\documentclass[3p,times,11pt,sort&compress]{elsarticle}

\usepackage[utf8]{inputenc}
\usepackage{graphicx}
\usepackage{enumerate}
\usepackage{wrapfig}
\usepackage{amsmath,amsthm}
\usepackage{caption,subcaption}

\usepackage{soul}
\usepackage{cancel}
\usepackage{ulem}
\usepackage{algorithm,algpseudocode,varwidth}
\usepackage{amsfonts}
\usepackage{mathtools}
\usepackage[version=4]{mhchem}
\usepackage{mathrsfs}
\usepackage{amssymb}
\usepackage{dsfont}
\usepackage{url}
\usepackage{hyperref}
\usepackage{xcolor}
\usepackage{tcolorbox}
\usepackage{colortbl}
\usepackage{booktabs}
\usepackage{multirow}
\allowdisplaybreaks

\makeatletter
\def\ps@pprintTitle{%
 \let\@oddhead\@empty
 \let\@evenhead\@empty
 \let\@oddfoot\@empty
 \let\@evenfoot\@empty
}
\makeatother



\newcommand{\vf}{\mathbf{f}}

\newcommand{\vq}{\mathbf{q}}

\newcommand{\vy}{\mathbf{y}}
\newcommand{\vz}{\mathbf{z}}

\newcommand{\vmu}{\boldsymbol{\mu}}

\newcommand{\vtheta}{\boldsymbol{\theta}}

\newcommand{\mD}{\mathbf{D}}

\newcommand{\mF}{\mathbf{F}}

\newcommand{\mU}{\mathbf{U}}

\newcommand{\mC}{\mathbf{C}}

\newcommand{\mS}{\mathbf{S}}

\newcommand{\vw}{\mathbf{w}}

\newcommand{\vOmega}{\boldsymbol{\omega}}

\newcommand{\phib}{\boldsymbol{\Phi}}
\newcommand{\psib}{\boldsymbol{\Psi}}

\newcommand{\phifb}{\phib\left(\psib^\intercal\phib\right)^{-1}}

\newcommand{\eqspace}{\,}

\usepackage{xcolor}

\newcommand\fulltitle{Petrov-Galerkin model reduction for collisional-radiative argon plasma}

\begin{document}

\begin{abstract}
    Accurate simulation of nonequilibrium plasmas—central to applications in hypersonic re-entry, fusion energy, and astrophysical flows—requires state-specific collisional–radiative (CR) kinetic models,  but the associated computational cost is often prohibitive.
While traditional approaches reduce this burden through empirical or physics-based simplifications, they frequently compromise accuracy in strongly nonequilibrium regimes.
To address these limitations, we develop a Petrov-Galerkin reduced-order model (ROM) for CR argon plasma based on oblique projections that optimally balance the covariance of full-order state trajectories with that of the system's output sensitivities.
This construction ensures that the ROM captures both the dominant energetic modes and the directions most relevant to input-output behavior.
After offline training in a zero-dimensional setting using nonlinear forward and adjoint simulations, the ROM is coupled to a finite-volume solver and applied to one- (1D) and two-dimensional (2D) ionizing shock-tube problems.
The ROM achieves a 3$\times$ reduction in state dimension and more than one order of magnitude savings in floating-point operations, while maintaining errors below 1\% for macroscopic quantities.
In both 1D and 2D, it robustly reproduces complex unsteady plasma flow features—such as periodic fluctuations, electron avalanches, triple points, and cellular ionization patterns—in contrast to standard ROM strategies, which become unstable or inaccurate under these challenging conditions.
These results demonstrate the potential of the proposed model reduction strategy to enable high-fidelity simulation of reactive plasma flows at reduced computational cost, offering new capabilities for exploring high-speed fluid systems governed by coupled transport, wave propagation, and kinetic nonequilibrium.

\end{abstract}
\begin{keyword}
    Reduced-order modeling \sep Thermochemistry \sep Plasma
\end{keyword}

\begin{frontmatter}

    \title{\fulltitle}
    \author{Ivan Zanardi}
    \ead{zanardi3@illinois.edu}
    \author{Alessandro Meini}
    \ead{ameini2@illinois.edu}
    \author{Alberto Padovan}
    \ead{padovan3@illinois.edu}
    \author{Daniel J. Bodony}
    \ead{bodony@illinois.edu}
    \author{Marco Panesi\corref{cor1}}
    \ead{mpanesi@uci.edu}
    
    \cortext[cor1]{Corresponding Author}
    
    \address{Center for Hypersonics and Entry Systems Studies, \\
	Department of Aerospace Engineering,
	University of Illinois Urbana-Champaign,
	Urbana, IL 61801, USA}

\end{frontmatter}

\section{Introduction}\label{sec:introduction}
Nonequilibrium plasmas play a crucial role in a wide range of scientific and engineering applications, including electric propulsion, plasma-assisted combustion, fusion energy, atmospheric re-entry, and astrophysical flows~\cite{Goebel2008FundamentalsThrusters,Pancheshnyi2006IgnitionDischarge,Park_Book_1990,Helber2016ExperimentalPlasmas,Goedbloed2004PrinciplesPlasmas}. In such regimes, the internal energy distributions of the constituent species—arising from electronic, vibrational, and rotational excitations—can deviate significantly from equilibrium, requiring detailed, state-specific collisional-radiative (CR) kinetic models to faithfully capture the underlying physics~\cite{Vlcek1989AData,Bultel2002InfluenceModel,Kapper2011IonizingStructure,Kapper2011IonizingEffects,Bultel2006Collisional-radiativeProblems,Bultel2013ElaborationAtmospheres,Panesi_Lani_PofF_2013,Capitelli2017CouplingConditions}. These models, typically formulated as state-to-state (StS) master equations~\cite{Nagnibeda_Book_2009,Panesi_JTHT_2011,Panesi_JCP_2013,Panesi_PR_2014,Macdonald_PRF_2016,Capitelli_Book_2016,Macdonald_JFC_2020}, track the population dynamics of individual internal energy levels through an extensive set of collisional and radiative processes. While offering an unprecedented level of physical accuracy in representing the thermochemical state of the plasma, the associated computational cost is prohibitive, particularly in multidimensional simulations, due to the large number of degrees of freedom (i.e., the molecules’ and atoms’ energy levels) and the extreme stiffness associated with the governing equations.

Historically, a number of model reduction strategies have been proposed to alleviate this cost. Multi-temperature (MT) models~\cite{Hammerling1959TheoryNitrogen,Heims1963MomentDissociation,Marrone1963ChemicalLevels,Treanor1962EffectRelaxation,Park_Book_1990,Park_JTHT_1989,Park_JTHT_1993,Park_JTHT_1994,Park_JTHT_2001}, for instance, simplify the problem by assuming thermal equilibrium distributions within each energy mode, introducing additional temperatures to describe vibrational, rotational, and electronic state populations. While computationally attractive, MT models rely on crude, semi-empirical assumptions and are known to break down in strongly nonequilibrium regimes—such as those found in ionizing shock layers or rapid relaxation zones~\cite{Panesi_JCP_2013,Panesi_PR_2014,Heritier_JCP_2014}. Alternatively, coarse-grained (CG) models~\cite{Magin2012Coarse-grainNitrogen,Panesi_Lani_PofF_2013,Munafo_PR_2014,Munafo_PF_2015,Liu_JCP_2015} provide a more effective solution compared to the traditional MT models by clustering energy levels into bins and reconstructing level populations using entropy-based closures. These approaches, while grounded in physical principles, require careful tuning and may not capture complex features of the internal state distribution function without significant prior analysis and optimization~\cite{Macdonald_I_JCP_2018,Macdonald_II_JCP_2018,Sahai_JCP_2017,Venturi_JCP_2020,Sharma_PR_2020,Zanardi2023AdaptiveFlows,Jacobsen_JComP_2024,Zanardi_SciTech_2022,Zanardi_SciTech_2023,Zanardi_SciTech_2024}.

More recently, the emergence of data-driven model reduction techniques has opened new avenues for compressing detailed kinetic systems without relying on physics-based closure assumptions. A simplistic approach involves orthogonally projecting the governing equations onto the span of proper orthogonal decomposition (POD) modes derived from high-fidelity simulation data~\cite{Rowley2004ModelProjection,Barone2009StableFlow,Sutherland2009CombustionAnalysis,Parente2013PrincipalSensitivity}. Bellemans et al.~\cite{Bellemans2015ReductionAnalysis,Bellemans2017Reduced-orderSensitivity} demonstrated the application of such orthogonal projections to reduce the dimensionality of CR models for argon plasmas, achieving good agreement with full-order results in steady 1D shock-tube configurations. 
However, POD constructs its basis purely from snapshot data, without accounting for sensitivity information encoded in the governing equations. 
As a result, the method often fails to capture dynamically important directions—particularly in systems exhibiting transient growth and high sensitivity to low-energy disturbances (e.g., low-population states in the distribution function)~\cite{Rowley2017ModelControl, Otto2022OptimizingTrajectories,Otto2023ModelCovariance,Padovan2024Data-DrivenDynamics}. 
Consequently, when deployed in predictive simulations—especially under highly nonlinear, unsteady, or extrapolative conditions—POD-based ROMs frequently suffer from instability, degraded accuracy, and poor generalization.

To overcome the limitations of these approaches, Zanardi et al.~\cite{Zanardi2025Petrov-GalerkinMixtures,Zanardi_SciTech_2025} proposed a reduced-order modeling framework based on Petrov-Galerkin projection, in which the full-order dynamics (i.e., the StS master equations) are \textit{obliquely} projected onto a low-dimensional linear subspace. 
This method builds upon the ``Covariance Balancing Reduction using Adjoint Snapshots'' (CoBRAS) framework~\cite{Otto2023ModelCovariance}, which constructs reduced subspaces by balancing the covariance of the full-order state trajectory with that of the system’s output sensitivities (i.e., gradients), similar to the balance truncation technique and its variants~\cite{Moore1981PrincipalReduction,Dullerud_Book_2000,Rowley2005ModelDecomposition,Varga2000BalancedSystems,Padovan2024Continuous-timeGramians}. 
This balance is defined in a statistical mean-square sense, ensuring that the reduced space captures both the dominant energetic modes and also the directions most influential to the system’s input-output behavior. 
Crucially, CoBRAS introduces a physically grounded alternative to POD by generating oblique projection operators specifically tailored to minimize output reconstruction error and preserve the system’s most relevant dynamical features. Zanardi et al.~\cite{Zanardi2025Petrov-GalerkinMixtures} demonstrated that CoBRAS-based Petrov-Galerkin ROMs yield stable and accurate reduced models even in highly stiff and nonlinear kinetic systems.
A key innovation in their framework is the use of linearized forward and adjoint equations to efficiently approximate the state and gradient covariances, enabling fast offline construction of the reduced basis while preserving fidelity.

In this work, we extend the CoBRAS-based reduced-order modeling framework—previously demonstrated only in zero-dimensional (0D) settings—to simulate multidimensional, unsteady, CR plasma flows, with a focus on ionizing shock dynamics.
These flows are particularly demanding due to their strong nonlinearity, multi-scale nature, and tight coupling between thermochemistry and hydrodynamics—conditions under which conventional ROMs often fail due to instability or loss of accuracy.
As a first step, we construct a CoBRAS-based Petrov-Galerkin ROM for detailed CR argon kinetics by computing projection spaces from nonlinear 0D forward and adjoint simulations of the full-order system.
Second, we integrate the resulting ROM into a finite-volume flow solver and apply it to one- (1D) and two-dimensional (2D) shock-tube configurations characterized by strongly unsteady plasma behavior.
Third, we systematically evaluate the performance of the proposed ROM against both the full-order model (FOM) and a standard POD-based ROM, demonstrating substantial improvements in numerical stability, accuracy, and physical fidelity. The CoBRAS approach proves robust even in regimes where traditional ROMs typically fail.
This marks a significant step forward in model reduction for reactive plasma flows, opening new avenues for both fundamental investigation and applied analysis of high-speed, nonequilibrium systems in aerospace, astrophysics, and energy science.

The remainder of this paper is organized as follows. Sections~\ref{sec:physics} and~\ref{sec:numerics} describe the physical modeling of CR argon plasma and the numerical framework used to solve multidimensional plasma flows.
Section~\ref{sec:rom} presents the reduced-order modeling methodology and its integration with a finite-volume solver. Section~\ref{sec:num_exp} reports on numerical results obtained in 0D, 1D, and 2D configurations, comparing CoBRAS performance against FOM and POD baselines.
Finally, section~\ref{sec:conclusions} summarizes the key findings and outlines directions for future work.

\section{Physical modeling}\label{sec:physics}
This section outlines the thermochemical and radiative modeling framework adopted in the present study, along with the governing equations describing the plasma flow. The formulation incorporates detailed kinetics and energy exchange processes relevant to weakly ionized argon. For a comprehensive description of the reaction rate coefficients, collisional-radiative source terms, and thermodynamic closures, the reader is referred to the methodology presented by Kapper and Cambier~\cite{Kapper2011IonizingStructure,Kapper2011IonizingEffects}.

\subsection{Collisional-radiative model}\label{sec:physics:crmodel}
To model the kinetics of weakly ionized argon plasmas, we employ a detailed collisional-radiative (CR) framework in which each electronic state is treated as a distinct pseudo-species. This enables a non-Boltzmann description of the atomic state distribution function (ASDF) and offers improved fidelity over traditional multi-temperature (MT) models. The CR mechanism used in this work was originally developed by Vl{\^c}ek~\cite{Vlcek1989AData} and Bultel~\cite{Bultel2002InfluenceModel}, and has been validated against experimental data from the UTIAS (University of Toronto, Institute for Aerospace Studies) shock tube~\cite{Glass1991OverWaves} by Kapper and Cambier~\cite{Kapper2011IonizingStructure,Kapper2011IonizingEffects}.

The plasma composition includes free electrons (e$^-$), neutral argon (Ar), and singly ionized argon (Ar$^+$), while higher-order ionization states (e.g., Ar$^{++}$ and Ar$_2^+$) are neglected due to their negligible contributions under the conditions considered. The argon species are resolved into their respective electronic states (31 for Ar and 2 for Ar$^+$) yielding a total of 34 pseudo-species.

The CR model accounts for five types of fundamental processes~\cite{Vlcek1989AData,Bultel2002InfluenceModel,Kapper2011IonizingStructure,Kapper2011IonizingEffects}:
\begin{enumerate}
  \item Excitation and ionization by electron impact:
    \begin{align}\label{eq:kin.e}
        \ce{
            Ar$(i)$ + e$^-$ &<=> Ar$(j)$ + e$^-$
        }\eqspace, \\
        \ce{
            Ar$(i)$ + e$^-$ &<=> Ar^+ + e$^-$ + e$^-$
        }\eqspace,
    \end{align}
  \item Excitation and ionization by heavy-particle impact:
    \begin{align}\label{eq:kin.h}
        \ce{
            Ar$(i)$ + Ar$(1)$ &<=> Ar$(j)$ + Ar$(1)$
        }\eqspace, \\
        \ce{
            Ar$(i)$ + Ar$(1)$ &<=> Ar^+ + e$^-$ + Ar$(1)$
        }\eqspace,
    \end{align}
  \item Spontaneous emission/absorption (bound-bound):
    \begin{equation}\label{eq:rad.bb}
        \ce{
            Ar$(i)$ + h $\nu_{ij}$ <=> Ar$(j)$
        }\eqspace,
    \end{equation}
  \item Photo-ionization/radiative recombination (free-bound and bound-free):
    \begin{equation}\label{eq:rad.bf}
        \ce{
            Ar$(i)$ + h $\nu_{i}$ <=> Ar^+ + e$^-$
        }\eqspace,
    \end{equation}
  \item Bremsstrahlung radiation (free-free):
    \begin{equation}\label{eq:rad.ff}
        \ce{
            e$^-$ + Ar^+ <=> e$^-$ + Ar^+ + h $\nu$
        }\eqspace.
    \end{equation}  
\end{enumerate}
Here, indices $i \in \{1,\dots,31\}$ and $j > i$ refer to the electronic levels of Ar; the two levels of Ar$^+$ are omitted for brevity. Superelastic collisions are neglected, as excited states remain sparsely populated under the conditions considered, rendering their contribution to heavy-particle collisions negligible. Accordingly, only the ground state, Ar$(1)$, is included in these interactions.
Radiative absorption is modeled using escape factors, $\Lambda$, which range from 0 to 1, corresponding to optically thick and optically thin plasma, respectively. A value of $\Lambda = 0$ indicates that all emitted photons are reabsorbed locally, while $\Lambda = 1$ implies complete photon escape without reabsorption. In this work, the plasma is assumed to be optically thin for all radiative transitions, with the exception of high-frequency bound–bound emissions to the ground state Ar(1), which are treated as fully reabsorbed by the gas~\cite{Kapper2011IonizingStructure,Kapper2011IonizingEffects}. Planck’s constant and the radiation frequency are denoted by $h$ and $\nu$, respectively.

\subsection{Governing equations}\label{sec:physics:gov_eqs}
The plasma is described using a single-fluid formulation, in which all species share a common velocity field, and thermal nonequilibrium is captured by distinguishing between electron and heavy-particle temperatures, $T_\mathrm{e}$ and $T_\mathrm{h}$, respectively. Charge neutrality is maintained by the short Debye length and the absence of applied electric fields. The assumption of strong collisional coupling justifies a single-fluid velocity. Viscous and diffusive transport effects are neglected. As a result, the dynamics are governed by the two-temperature Euler equations~\cite{Kapper2011IonizingEffects}, expressed in conservation form as:
\begin{equation}\label{eq:euler}
\frac{\partial \mU}{\partial t}+\frac{\partial \mF_\alpha}{\partial x_\alpha} = \mS
\end{equation}
for $\alpha \in \{1,2,3\}$, with $\mathbf{x}=[x_1,x_2,x_3]=[x,y,z]$ being the Cartesian coordinates and $t$ denoting time. Summation is implied over repeated indices. The conservative variables $\mU$, inviscid fluxes $\mF_\alpha$, and source terms $\mathbf{S}$ are defined as: 
\begin{equation}\label{eq:euler.terms}
\mU = \begin{bmatrix}
\rho_s^i \\
\phantom{;} \rho u_\beta \phantom{;} \\
\rho E \\
\rho e_\mathrm{e}
\end{bmatrix}\eqspace,\quad
\mF_\alpha = \begin{bmatrix}
\rho_s^i u_\alpha \\
\phantom{;} \rho u_\alpha u_\beta + p\delta_{\alpha\beta} \phantom{;} \\
\rho H u_\alpha \\
\rho e_\mathrm{e}u_\alpha
\end{bmatrix}\eqspace,\quad
\mS = \begin{bmatrix}
\omega_s^i \\
0 \\
\Omega_{\mathrm{bb}}^{\mathrm{R}}+\Omega_{\mathrm{ff}}^{\mathrm{R}} \\
\phantom{;}\Omega_{\mathrm{el}}^{\mathrm{C}}+\Omega_{\mathrm{in}}^{\mathrm{C}}+\Omega_{\mathrm{ion}}^{\mathrm{C}}+\Omega_{\mathrm{bf}}^{\mathrm{R}}+\Omega_{\mathrm{ff}}^{\mathrm{R}}-p_{\mathrm{e}} \partial u_\alpha / \partial x_\alpha\phantom{;}
\end{bmatrix}\eqspace,
\end{equation}
with $\beta \in \{1,2,3\}$, $s \in \mathcal{S}$, and $i \in \mathcal{I}_s$. Here, $\mathcal{S} = \{\mathrm{e}^-, \mathrm{Ar}, \mathrm{Ar}^+\}$ denotes the set of chemical species, and $\mathcal{I}_s$ the set of electronic states for species $s$. The symbols $\rho_s^i$ and $p_s^i$ represent the partial density and pressure of state $i$ of species $s$, while $\rho$ and $p$ denote the total plasma density and pressure. $u_\alpha$ corresponds to the $\alpha$-component of
the mass-averaged velocity with $\delta_{\alpha \beta}$ being the Kronecker delta, whereas $\rho E$, $\rho H$, and $\rho e_\mathrm{e}$ denote the total energy, total enthalpy, and electron internal energy densities, respectively. Mass source terms are denoted by $\omega_s^i$, while energy source terms include radiative contributions from bound-bound ($\Omega_{\mathrm{bb}}^{\mathrm{R}}$), bound-free ($\Omega_{\mathrm{bf}}^{\mathrm{R}}$), and free-free ($\Omega_{\mathrm{ff}}^{\mathrm{R}}$) transitions, as well as collisional energy transfer from elastic ($\Omega_{\mathrm{el}}^{\mathrm{C}}$), inelastic ($\Omega_{\mathrm{in}}^{\mathrm{C}}$), and ionizing ($\Omega_{\mathrm{ion}}^{\mathrm{C}}$) collisions. The bound-free radiative term is omitted from the total energy conservation equation, as its contribution is negligible compared to the free-free and especially the bound-bound terms~\cite{Kapper2011IonizingStructure}. The non-conservative term $-p_e \partial u_\alpha / \partial x_\alpha$ is related to the work of the self-consistent plasma electric field, which acts against charge separation, whose expression can be derived from the free-electron momentum equation when neglecting inertial and diffusive terms~\cite{Zeldovich1967PhysicsPhenomena}.

\section{Computational method}\label{sec:numerics}
The cell-centered finite volume discretization~\cite{Hirsch_Book_2007} is employed, with inviscid fluxes computed using the Van Leer flux-splitting scheme~\cite{vanLeer1982Flux-vectorEquations}. To achieve second-order spatial accuracy, left and right states at cell interfaces are reconstructed via the MUSCL approach~\cite{VanLeer_JComP_1979}, applied to primitive variables (partial densities, velocity components, and temperatures) using van Albada’s slope limiter~\cite{vanAlbada1997ADynamics}. To mitigate numerical instabilities associated with strong shock fronts—particularly when using reduced-order models—a reconstruction blending technique is implemented. In this approach, the reconstructed state at each cell interface, $\mathbf{U}_{f}$, is defined as a weighted combination of the high-order (MUSCL) reconstruction, $\mathbf{U}_{f}^{\mathrm{H}}$, and the low-order (first-order upwind) reconstruction, $\mathbf{U}_{f}^{\mathrm{L}}$:
\begin{equation}
\mathbf{U}_{f} = (1-\eta)\,\mathbf{U}_{f}^{\mathrm{H}}
    + \eta\,\mathbf{U}_{f}^{\mathrm{L}} \eqspace,
\end{equation}
where $\eta$ is a shock-sensing function that transitions smoothly between 1 (upstream and within the shock) and 0 (downstream of the shock) via a Gaussian smoothing profile. Shock locations are identified using the detection algorithm described in~\cite{Quirk1994ADebate}.

Time integration is performed using the operator-splitting technique proposed by Strang~\cite{Strang_SIAM_1968}. This method integrates the transport operator, $\boldsymbol{\mathcal{T}}\left(\mathbf{U}\right) = \partial \mF_\alpha / \partial x_\alpha$, and the reaction operator, $\boldsymbol{\mathcal{R}}\left(\mathbf{U}\right) = \mathbf{S}$, sequentially in a symmetric fashion:
\begin{align}
	\partial_t\mathbf{U}^{(1)} & = \boldsymbol{\mathcal{T}}\left(\mathbf{U}^{(1)}\right)\eqspace, & \hspace{15mm} &
	\hspace{-5cm}\mathbf{U}^{(1)}\left(t_n\right)=\mathbf{U}_n \label{eq:split.01}\\
	\partial_t\mathbf{U}^{(2)} & = \boldsymbol{\mathcal{R}}\left(\mathbf{U}^{(2)}\right)\eqspace, & \hspace{15mm} &
    \hspace{-5cm}\mathbf{U}^{(2)}\left(t_n\right)=\mathbf{U}^{(1)}\left(t_n+\Delta t/2\right) \label{eq:split.02} \\
	\partial_t\mathbf{U}^{(3)} & = \boldsymbol{\mathcal{T}}\left(\mathbf{U}^{(3)}\right)\eqspace, & \hspace{15mm} &
	\hspace{-5cm}\mathbf{U}^{(3)}\left(t_n+\Delta t/ 2\right)=\mathbf{U}^{(2)}\left(t_n+\Delta t\right) \label{eq:split.03} \\
	\mathbf{U}_{n+1} &=\mathbf{U}^{(3)}\left(t_n+\Delta t\right)\eqspace, & & \label{eq:split.04}
\end{align}
where $\Delta t$ is the time step. The splitting formulation is second-order accurate, strongly stable, and symplectic for nonlinear equations. Its convergence and stability properties have been extensively studied for reacting flow simulations~\cite{Knio_JComP_1999,Singer_CTM_2006,Ren_JComP_2014,Wu_CPC_2019}.
The transport operator is advanced in time using a four-stage, third-order strong-stability-preserving Runge-Kutta (SSP-RK3) scheme~\cite{Durran2010NumericalDynamics}, which allows a Courant-Friedrichs-Lewy (CFL) number of up to 2. The reaction operator is integrated implicitly via a second-order backward differentiation formula (BDF) solver provided by the LSODE library~\cite{Radhakrishnan_LSODE_1993}.

\section{Projection-based model reduction}\label{sec:rom}
The numerical solution of the nonlinear governing equations~\eqref{eq:euler} is often computationally expensive, due to the large number of pseudo-species involved and the extremely fast time scales introduced by the CR model, which impose severe restrictions on temporal integration schemes. These challenges are typical when solving thermochemical state-to-state models. To alleviate these issues, reduced-order models (ROMs) are employed to decrease the number of pseudo-species integrated in time, resulting in a reduction of computational cost compared to the full-order model (FOM). In this work, ROMs are constructed using Petrov-Galerkin projections, following the procedure outlined in Zanardi et al.~\cite{Zanardi2025Petrov-GalerkinMixtures}.

\subsection{0D state-space representation}\label{ssec:rom:state}
To reduce the number of pseudo-species in the simulation, we adopt a zero-dimensional (0D) formulation of the governing equations~\eqref{eq:euler}, neglecting transport terms and expressing the system in terms of primitive variables, namely species mass densities and temperatures. Exploiting mass conservation, where $d\rho/dt = 0$, the system can be expressed compactly as
\begin{equation} \label{eq:fom.param}
    \frac{d}{dt}\vq(t) = \vf(\vq(t);\rho)\eqspace,
    \quad \vq(0) = \vq_0\left(\vmu\right)\eqspace,
\end{equation}
where the thermochemical state vector $\vq(t) = [w_s^i, T_\mathrm{h}, T_\mathrm{e}] \in \mathbb{R}^N$ includes the mass fractions $w_s^i$ associated with each electronic state $i \in \mathcal{I}_s$ of species $s \in \mathcal{S}$, along with the heavy-particle and electron temperatures, $T_\mathrm{h}$ and $T_\mathrm{e}$.
Here, $N=2+\sum_{s\in\mathcal{S}}\left|\mathcal{I}_s\right|$ denotes the total number of variables, accounting for the two temperatures and all internal states across all species, while $|\mathcal{I}_s|$ represents the cardinality of the set $\mathcal{I}_s$.
The initial condition $\vq(0)$ is prescribed by the function $\vq_0: \mathbb{R}^{N_0} \rightarrow \mathbb{R}^N$, which depends on a set of parameters $\vmu \in \mathbb{R}^{N_0}$ including the initial heavy-particle and electron temperatures ($T_{\mathrm{h}_0}$ and $T_{\mathrm{e}_0}$) and the initial species mass fractions $w_{s_0}$. The ASDFs at $t=0$ are assumed to follow Boltzmann distributions evaluated at $T_{\mathrm{e}_0}$.

In addition, we define an output vector $\vy(t)$ as
\begin{equation} \label{eq:fom.output}
    \vy(t) = \mC \vq(t)\eqspace, 
\end{equation}
where $\mC$ is an output matrix, and $\vy(t)$ collects physical quantities of interest. Specifically, $\vy(t)$ contains the moments of orders $0$ through $m$ of the ASDF for heavy species only. The output dimension is therefore, $\mathrm{dim}(\vy)=(m+1)(|\mathcal{S}|-1)$. The exclusion of electrons is justified by the quasi-neutrality of the plasma: the zeroth-order moment of Ar$^+$, which is included in $\vy(t)$, directly corresponds to that of e$^-$. The $j$-th moment corresponding to species $s$ is defined as~\cite{Zanardi2025Petrov-GalerkinMixtures} 
\begin{equation} \label{eq:output_y}
    \vy_{s,j} = \left(\mC \vq\right)_{s,j} = \frac{1}{j!}\sum_{i\in\mathcal{I}_s} \frac{1}{M_s}\left(\epsilon_s^i\right)^j w_s^i \eqspace ,
\end{equation}
where $M_s$ is the molar mass of species $s$, and $\epsilon_s^i$ denotes the energy of state $i$ (expressed in electron-volts, eV). In this work, temperature-related outputs are not included, as the focus is solely on reducing the number of pseudo-species.

\subsection{Normalization}\label{sec:rom:scale}
Model reduction techniques based on singular value decomposition (SVD), such as the one employed in this work, are inherently sensitive to the relative magnitudes of the state variables. This sensitivity arises from the impact of variable scaling on the sample covariance matrix, which the algorithm seeks to diagonalize. In practice, the components of the state vector may span several orders of magnitude, making proper normalization a critical preprocessing step~\cite{Noda2008ScalingSpectra,Parente2013PrincipalSensitivity}. To address this, we apply a linear transformation of the form
\begin{equation}\label{eq:norm}
    \vq(t) = \mD \tilde{\vq}(t) + \bar{\vq}\eqspace,
\end{equation}
where $\mD \in \mathbb{R}^{N \times N}$ is a diagonal matrix of scaling weights and $\bar{\vq} \in \mathbb{R}^N$ is a centering vector. We adopt Pareto scaling, in which each variable is centered by its sample mean and scaled by the square root of its standard deviation, both computed from a representative ensemble of 0D trajectories by solving equation~\eqref{eq:fom.param}. Unlike standard normalization—which enforces unit variance—Pareto scaling preserves the relative variance structure by assigning each variable a variance equal to its standard deviation~\cite{Noda2008ScalingSpectra}. Substituting equation~\eqref{eq:norm} into the governing equations~\eqref{eq:fom.param} and the output map~\eqref{eq:fom.output}, we obtain
\begin{equation}\label{eq:fom.norm}
    \begin{aligned}
        \frac{d}{dt} \tilde{\vq}(t) 
            & = \mD^{-1}\vf(\mD \tilde{\vq}(t) + \bar{\vq};\rho)\eqspace,
            \quad \tilde{\vq}(0) = \mD^{-1}\vq_0\left(\vmu\right)-\bar{\vq}\eqspace, \\
        \vy(t) & = \mC\mD\tilde{\vq}(t) + \mC\bar{\vq}\eqspace, 
    \end{aligned}
\end{equation}
which defines the normalized system upon which the reduced-order model is constructed. For notational convenience, we rewrite system~\eqref{eq:fom.norm} in a more compact form as
\begin{equation}\label{eq:fom.norm.compact}
    \begin{aligned}
        \frac{d}{dt} \tilde{\vq}(t) 
            & = \tilde\vf(\tilde{\vq}(t);\rho)\eqspace,
            \quad \tilde{\vq}(0) = \tilde{\vq}_0\left(\vmu\right)\eqspace, \\
        \vy(t) & = \tilde\mC\tilde\vq(t) + \bar\vy\eqspace, 
    \end{aligned}
\end{equation}
where the transformed terms $\tilde{\vf}$, $\tilde{\vq}_0$, $\tilde{\mC}$, and $\bar{\vy}$ are defined by matching terms between equations~\eqref{eq:fom.norm} and~\eqref{eq:fom.norm.compact}.

\subsection{Petrov-Galerkin model reduction}\label{sec:rom:pg}
In (linear) Petrov-Galerkin model reduction, we replace the normalized full state $\tilde{\vq}(t) \in \mathbb{R}^N$ by its projection onto a lower-dimensional space, $\hat\vq(t) = \mathbb{P}\tilde{\vq}(t) \in \mathbb{R}^N$, where $\mathbb{P} \in \mathbb{R}^{N \times N}$ is a rank-$p$ projector operator (with $p \ll N$) whose range defines the approximation subspace.
Inserting this ansatz into \eqref{eq:fom.norm}, it is easy to see that the dynamics of $\hat\vq$ are governed by
\begin{equation} \label{eq:pg_rom}
    \begin{aligned}
        \frac{d}{dt}\hat\vq(t) &= \mathbb{P}\tilde\vf(\mathbb{P}\hat\vq(t);\rho)\eqspace,
        \quad \hat\vq(0) = \mathbb{P}\tilde{\vq}_0\left(\vmu\right)\eqspace, \\
        \hat\vy(t) &= \tilde\mC\mathbb{P}\hat\vq(t) + \bar\vy\eqspace.
    \end{aligned}
\end{equation}
Although $\hat\vq(t)$ remains $N$-dimensional, its dynamics evolve within the $p$-dimensional range of $\mathbb{P}$. To make this explicit, we can factor the projector as $\mathbb{P} = \phifb\psib^\intercal$, where $\phib, \psib \in \mathbb{R}^{N \times p}$ are the trial and test basis matrices, respectively. Defining the reduced coordinate vector $\hat\vz(t) = \psib^\intercal \hat{\vq}(t)$ and left-multiplying the first equation in \eqref{eq:pg_rom} by $\psib^\intercal$ yields the final $p$-dimensional Petrov-Galerkin ROM:
\begin{equation} \label{eq:rom}
    \begin{aligned}
        \frac{d}{dt}\hat\vz(t) &= \psib^\intercal\tilde\vf(\phifb \hat\vz(t);\rho)\eqspace,
        \quad \hat\vz(0) = \psib^\intercal\tilde{\vq}_0\left(\vmu\right)\eqspace, \\
        \hat\vy(t) &= \tilde\mC\phifb\hat\vz(t) + \bar\vy\eqspace.
    \end{aligned}
\end{equation}
The accuracy of the reduced-order model is determined entirely by the choice of trial and test subspaces spanned by $\phib$ and $\psib$, respectively. In this work, these subspaces are constructed following the methodology proposed by Zanardi et al.~\cite{Zanardi2025Petrov-GalerkinMixtures}, which builds upon the CoBRAS framework recently introduced by Otto et al.~\cite{Otto2023ModelCovariance}.
Throughout the paper, the reduced-order state $\hat\vz(t) \in \mathbb{R}^p$ is taken to be $\hat\vz(t) = [\hat\vz_\mathrm{h}, w_\mathrm{e}, T_\mathrm{h}, T_\mathrm{e}]$,
where $\hat\vz_\mathrm{h}\in\mathbb{R}^r$ denotes the latent variables associated with the normalized ASDF of the heavy-particle species Ar and Ar$^+$, $T_h$ and $T_e$ are the two temperatures and $w_e$ is the electron mass fraction.
The electron mass fraction is retained exactly because it corresponds to a single-state species, and its evolution equation can be used to enforce the quasi-neutrality condition of the plasma, i.e., $dn_\mathrm{e}/dt = dn_{\mathrm{Ar}^+}/dt$.
The two temperatures ($T_\mathrm{h}$ and $T_\mathrm{e}$) are similarly excluded from the reduction, following previous findings~\cite{Isaac2014Reduced-orderAnalysis} showing that applying model reduction solely to species mass fractions yields better accuracy and stability.


\subsection{Computing the optimal projector}\label{sec:rom:proj}
We aim to approximate the full-order input-output map defined by
\begin{equation} \label{eq:F}
    F: \mathbb{R}^N \to \mathbb{R}^{\mathrm{dim}(\vy)}: \tilde\vq(t_0;\rho,\vmu)\mapsto \left(\vy(t_0;\rho,\vmu),\vy(t_1;\rho,\vmu),\ldots,\vy(t_{L-1};\rho,\vmu)\right)\eqspace,
\end{equation}
which maps an initial state $\vq(t_0;\rho,\vmu)$ to a sequence of $L$ output observations along a trajectory of~\eqref{eq:fom.norm}, using its Petrov-Galerkin surrogate:
\begin{equation} \label{eq:Fhat}
    \hat{F}: \mathbb{R}^N \to \mathbb{R}^{\mathrm{dim}(\vy)}: \mathbb{P}\tilde\vq(t_0;\rho,\vmu)\mapsto \left(\hat\vy(t_0;\rho,\vmu),\hat\vy(t_1;\rho,\vmu),\ldots,\hat\vy(t_{L-1};\rho,\vmu)\right)\eqspace.
\end{equation}
Here, the state $\tilde\vq$ is a function of the temporal variable $t_0$ and of the physical parameters $\rho$ and $\vmu$. 
Rather than directly minimizing the mean-square error $\mathbb{E}\left[F(\tilde\vq)-\hat{F}\left(\mathbb{P}\tilde\vq\right)\right]$, we follow Zahm et al.~\cite{Zahm2020Gradient-BasedFunctions}, who show this error is bounded by
\begin{equation} \label{eq:bound}
    \text{trace}\left(W_g \left(I-\mathbb{P}\right)W_s\left(I-\mathbb{P}\right)^\intercal\right)\eqspace,
\end{equation}
where $W_s \coloneqq \mathbb{E}\left[\tilde\vq\tilde\vq^\intercal\right]\approx XX^\intercal$ and $W_g \coloneqq \mathbb{E}\left[\nabla F\nabla F^\intercal\right]\approx YY^\intercal$ are the state and gradient covariance matrices associated with the system, respectively. The matrices $X$ and $Y$ are numerical-quadrature factors defined as in equations (20) and (21) of \cite{Rowley2005ModelDecomposition}. Here, $W_s$ captures the variance of state snapshots, while $W_g$ encodes the sensitivity of the output sequence to the initial state, with $\nabla F = D_{\tilde\vq} F(\tilde\vq)^\intercal \in \mathbb{R}^{N\times L\, \mathrm{dim}(\vy)}$ and $D_{\tilde\vq}$ denoting the differential (or Jacobian) operator with respect to $\tilde\vq$. 
The minimizer $\mathbb{P}$ of equation~\eqref{eq:bound} may be written as $\mathbb{P} = \phib \psib^\intercal$, where 
\begin{equation} \label{eq:phi_psi}
    \phib = X V_r \Sigma_r^{-1/2}\eqspace,
    \quad \psib = Y U_r \Sigma_r^{-1/2}\eqspace,
\end{equation}
and $U_r$, $\Sigma_r$ and $V_r$ are the first $r$ singular triplets of $Y^\intercal X$. Rows corresponding to the retained variables $(w_\mathrm{e}, T_\mathrm{h}, T_\mathrm{e})$ are removed from both $X$ and $Y$ prior to computing the singular value decomposition of $Y^\intercal X$. This ensures that the resulting projection operator $\mathbb{P}$ is constructed exclusively for the normalized ASDF associated with the heavy-particle species Ar and Ar$^+$ (see section \ref{sec:rom:pg}).

In contrast to the approach of Zanardi et al.~\cite{Zanardi2025Petrov-GalerkinMixtures}, which relies on linearizing the dynamics to approximate the covariance factors $X$ and $Y$, we obtain these directly from the full nonlinear system \eqref{eq:fom.norm} and its associated adjoint (see Proposition 1 in~\cite{Zanardi2025Petrov-GalerkinMixtures}), as in the original CoBRAS formulation \cite{Otto2023ModelCovariance}.
Linearization was found to inadequately capture the dominant subspaces for the CR model considered in this work. Accordingly, for each sampled training instance $(t_0, \rho, \vmu)$, we (i) integrate the nonlinear forward model to generate state snapshots for $X$, and (ii) solve the corresponding nonlinear adjoint to obtain gradient snapshots for $Y$. While this approach preserves the system's intrinsic nonlinearity, it significantly increases computational cost due to repeated adjoint integrations. To mitigate this, we adopt the subsampling strategy of Otto et al.~\cite{Otto2023ModelCovariance}, selecting an uniformly sparse set of $t_0$ values (approximately 10\% of the total) along extended trajectories to estimate the gradient covariance with far fewer adjoint integrations.

\subsection{Multidimensional reduced governing equations}\label{ssec:rom:nd_gov_eqs}
Once the zero-dimensional ROM is constructed, it can be embedded in a multidimensional flow solver by projecting the full Euler species‐continuity equations onto the reduced basis. Recall that only the heavy-particle species mass fractions are subject to dimensionality reduction, while temperatures and the electron mass fraction are retained exactly. Therefore, we restrict our attention to the species-continuity equations from the Euler system~\eqref{eq:euler}, which are written as:
\begin{equation}\label{eq:fom.species_cont}
    \frac{\partial}{\partial t} \rho\vw 
        + \frac{\partial}{\partial x_\alpha} \rho u_\alpha\vw 
        = \vOmega(\vw;\vtheta) \eqspace,
\end{equation}
where $\vOmega$ is the source term associated with species mass densities and depends on the thermochemical state vector, expressed here as the mass fraction vector $\vw$ and the thermodynamic parameters $\vtheta = (\rho, T_\mathrm{h}, T_\mathrm{e})$. The parameters in $\vtheta$ are obtained from the mass and energy conservation equations, which are solved concurrently within \eqref{eq:euler}.

Let $\mD_w$ and $\bar{\vw}$ denote the scaling matrix and centering vector for $\vw$, computed as described in section~\ref{sec:rom:scale}, and let $\phib_w$ and $\psib_w$ denote the optimal trial and test basis matrices for the normalized species mass fractions, $\tilde{\vw}$, constructed following section~\ref{sec:rom:proj}.
Applying the normalization and projection procedures from sections~\ref{sec:rom:scale} and~\ref{sec:rom:pg} to equation~\eqref{eq:fom.species_cont}, we obtain the reduced species-continuity equations:
\begin{equation}\label{eq:rom.species_cont}
    \frac{\partial}{\partial t} \rho\hat{\vz}
        + \frac{\partial}{\partial x_\alpha} \rho u_\alpha\hat{\vz}
        = \psib_w^\intercal\mD_w^{-1}\vOmega(\mD_w\phib_w\hat{\vz} + \bar{\vw};\vtheta) \eqspace,
\end{equation}
where the reduced coordinate vector is given by $\hat{\vz}(t) = [\hat\vz_\mathrm{h}, w_\mathrm{e}]\in\mathbb{R}^{r+1}$.
Equations \eqref{eq:rom.species_cont} are then integrated in space and time together with the unmodified momentum and energy equations to simulate the full multidimensional plasma flow. Since the computation of thermodynamic and chemical production terms $\mS$ (see \eqref{eq:euler.terms}) requires the complete set of species mass fractions, we reconstruct $\hat{\vw}$ from the reduced variables $\hat{\vz}$ after each time step~\cite{Bellemans2017Reduced-orderSensitivity}.

\section{Numerical results}\label{sec:num_exp}
In this section, we assess the performance of the proposed CoBRAS-based model reduction strategy for CR argon plasma. Results are compared against both the FOM and the conventional POD-Galerkin approach~\cite{Sutherland2009CombustionAnalysis, Parente2013PrincipalSensitivity, Bellemans2015ReductionAnalysis, Bellemans2017Reduced-orderSensitivity}. The ROMs are first constructed in a zero-dimensional setting, as described in section~\ref{sec:rom}, and subsequently evaluated in one- and two-dimensional shock tube simulations, following the configuration described in~\cite{Kapper2011IonizingEffects}.

The implementation used to build the 0D ROMs is available at
\url{https://github.com/ivanZanardi/romar}. All 1D and 2D simulations were performed using \textsc{hegel}~\cite{Munafo2024HEGEL:Simulations}, a parallel, multi-block, structured solver for plasma hydrodynamics, coupled with \textsc{plato}~\cite{Munafo2025PlatoPlasmas}, a physico-chemical library for computing thermodynamic properties and evaluating source terms associated with nonequilibrium collisional and radiative processes. The ROMs were integrated into \textsc{plato} and seamlessly coupled with \textsc{hegel} to enable reduced-order simulations within the plasma flow solver.

\subsection{Physical model validation}\label{sec:num_exp:valid}
We first validate the CR model using a steady, one-dimensional inviscid flow behind a normal shock. The analysis is conducted in the shock reference frame, where the shock front is modeled as a discontinuity located at $x = 0$, across which flow properties exhibit a sharp jump from their freestream values. Given the upstream velocity, pressure, and temperature, post-shock conditions are computed using the Rankine-Hugoniot jump relations, assuming frozen chemistry across the shock. The electron temperature is also held constant through the discontinuity, and any precursor ionization upstream of the shock is neglected. Under these assumptions, the governing equations reduce to a steady, one-dimensional form of the Euler equations \eqref{eq:euler}, which results in a system of ordinary differential equations. These are integrated using the LSODE (``Livermore Solver for Ordinary Differential Equations'') library~\cite{Radhakrishnan_LSODE_1993} within \textsc{plato}.

\begin{figure}[htb!]
\centering
\begin{subfigure}[htb!]{0.32\textwidth}
    \includegraphics[width=\textwidth]{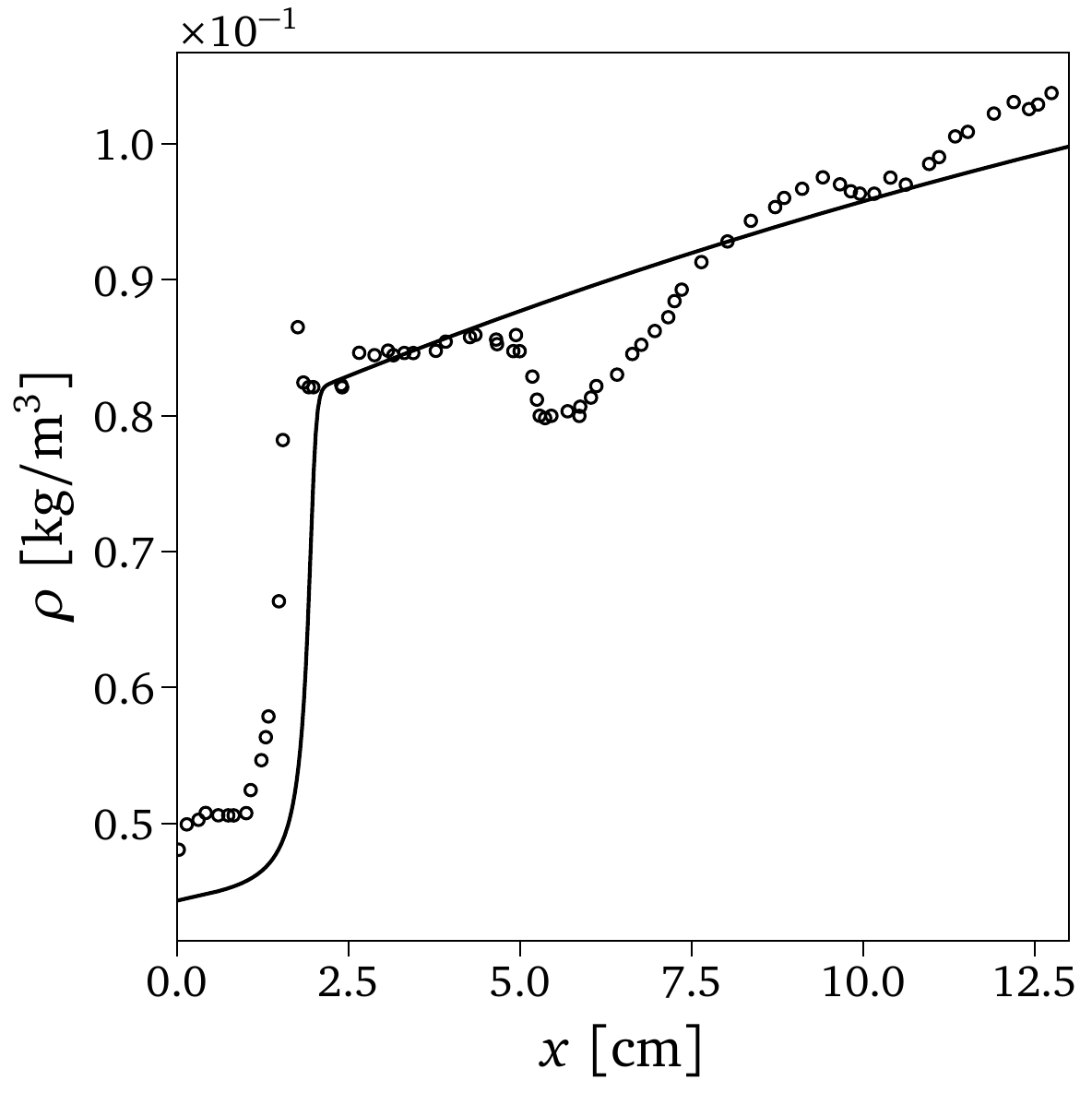}
\end{subfigure}
\quad
\begin{subfigure}[htb!]{0.32\textwidth}
    \includegraphics[width=\textwidth]{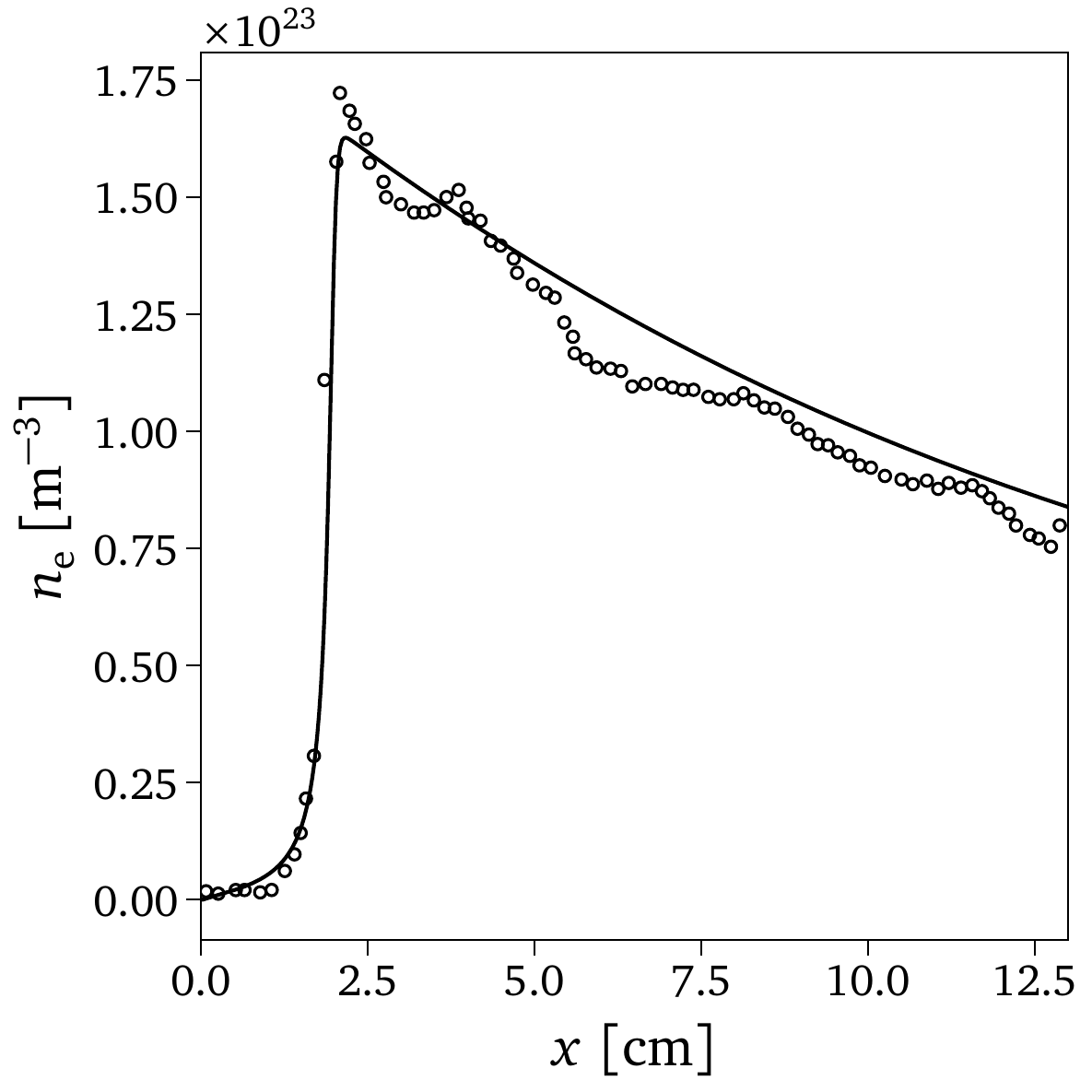}
\end{subfigure}
\caption{\textit{Steady post-shock relaxation simulation.} Comparison of the ionizing shock structure between experimental measurements (circle markers) and numerical solutions (solid line) under the following equilibrium freestream conditions: $p_\infty = 685.2$~Pa, $T_\infty = 293.6$~K, and $\text{Ma}_\infty = 15.9$ ($u_\infty = 5\,074.2$~m/s). The left panel shows the total mass density $\rho$, while the right panel presents the electron number density $n_{\mathrm{e}}$.}
\label{fig:shock.exp}
\end{figure}

Figure~\ref{fig:shock.exp} presents a comparison between the simulated results and experimental measurements from shock tube tests conducted at UTIAS~\cite{Glass1991OverWaves}. Specifically, we compare electron number density and total mass density profiles under freestream conditions of $p_\infty = 685.2$~Pa, $T_\infty = 293.6$~K, and $\text{Ma}_\infty = 15.9$ ($u_\infty = 5\,074.2$~m/s). The close agreement with experimental data demonstrates that the CR model accurately captures the steady-state relaxation behavior behind the shock.
Furthermore, the steady-state solution enables calibration of the radiation model against experimental measurements, supporting the assumption that transitions to the ground state are fully reabsorbed.
Although unsteady simulations are capable of reproducing experimentally observed shock-structure fluctuations driven by kinetic-wave interactions~\cite{Kapper2011IonizingEffects}, as will be shown in subsequent sections, we focus here on the steady case for model validation, as it provides a controlled and well-characterized benchmark.

\subsection{Zero‐dimensional simulations}\label{ssec:0d}
To build our ROM and reduce the number of pseudo-species solved, we follow the procedure described in section~\ref{sec:rom}, using the zero-dimensional formulation of the governing equations. Training and testing trajectories are generated by imposing a sudden temperature increase on the argon mixture confined within an ideal, constant-volume reactor. In this setup, the initial heavy-particle temperature \(T_{\mathrm{h}_0}\) is varied across simulations, while the initial electron temperature \(T_{\mathrm{e}_0}\) is held fixed at 293.6~K. The total mass density $\rho$ also varies between trajectories but remains constant within each simulation, consistent with the closed, fixed-volume nature of the reactor (i.e., no mass exchange with the surroundings). At the beginning of each trajectory, the initial chemical composition and the ASDF of Ar and Ar$^+$ are assumed to be in thermochemical equilibrium, following a Boltzmann distribution at 293.6~K. Since our primary interest lies in predicting the time evolution of macroscopic observables such as mass and energy, we define the output vector $\vy \in \mathbb{R}^4$ in equation~\eqref{eq:fom.output} to include the zeroth- and first-order moments of the ASDF for both Ar and Ar$^+$.
 
The sampling bounds and distributions used to generate the training dataset are summarized in table~\ref{table:train.dset}. Specifically, we generate a uniform grid of 40 values for \(T_{\mathrm{h}_0}\) and 8 logarithmically spaced values for \(\rho\), yielding a total of 320 distinct training trajectories.
\begin{table}[!htb]
    \centering
    \begin{tabular}{c|cc}
        \arrayrulecolor{black}\cmidrule{2-3}
        & $T_{\mathrm{h}_0}$ [K] & $\rho$ [kg/m$^3$] \\
        \arrayrulecolor{black}\midrule
        Minimum & 20\,000 & 0.005 \\
        Maximum & 50\,000 & 0.1 \\
        Distribution & Uniform & Log-uniform \\
		Samples & 40 & 8 \\
        \arrayrulecolor{black}\midrule
    \end{tabular}
    \caption{\textit{Training dataset.} Sampling bounds, distributions, and number of samples for each initial parameter used to generate the training trajectories.}
    \label{table:train.dset}
\end{table}
To assess the generalization performance of the ROM, we define three testing datasets, summarized in table~\ref{table:test.dsets}. The first dataset samples within the training domain and is used to evaluate interpolation accuracy. The second and third datasets represent extrapolation scenarios: the second targets a high-temperature regime (\(T_{\mathrm{h}_0} > 50\,000\)~K), corresponding to more extreme shock conditions, while the third focuses on a low-temperature regime (\(T_{\mathrm{h}_0} < 20\,000\)~K).
\begin{table}[!htb]
    \centering
    \begin{tabular}{c|c|c|c|c}
        \arrayrulecolor{black}\midrule
        Dataset & $T_{\mathrm{h}_0}$ [K] & $\rho$ [kg/m$^3$] & Samples & Type \\
        \arrayrulecolor{black}\midrule
        1 & $[2,5]\times10^4$ & & 100 & Interpolation \\
        2 & $[5,7]\times10^4$ & $[0.005,0.1]$ & 50 & Extrapolation \\
        3 & $[1,2]\times10^4$ & & 50 & Extrapolation \\
        \arrayrulecolor{black}\midrule
    \end{tabular}
    \caption{\textit{Testing datasets}. Each testing dataset is defined by the lower and upper bounds of the sampled initial parameters: initial heavy-particle temperature ($T_{\mathrm{h}_0}$) and density ($\rho$), along with the number of samples and whether the dataset is used for interpolation or extrapolation.}
    \label{table:test.dsets}
\end{table}

Table \ref{table:err.dset1} compares the mean relative error of CoBRAS‐ and POD‐based ROMs across the test trajectories of the first dataset. This comparison highlights the effectiveness of the proposed model reduction technique relative to a more conventional approach. Errors are reported for key quantities of interest—including the mixture number density $n$, mass fractions $\vw$, species zeroth- and first-order moments (molar fractions and internal energies, respectively), and the two temperatures—across different values of the reduced dimension $r$. Here, $r$ denotes the number of latent variables required to represent the normalized ASDF of Ar and Ar$^+$ combined (see section~\ref{sec:rom:pg}).
Table~\ref{table:err.dset1} reports the lower error in bold for each quantity, consistently showing that CoBRAS outperforms POD across all reduced dimensions and target variables. In particular, CoBRAS provides significantly more accurate predictions of the electron molar fraction $x_\mathrm{e}$—and, by quasi-neutrality, the molar fraction of Ar$^+$—which is crucial for capturing the degree of ionization of the plasma. Furthermore, the CoBRAS model achieves high accuracy in reconstructing the mass fraction vector, despite the fact that the ROM is trained only on macroscopic outputs (chemical species total masses and internal energies). This demonstrates the model’s ability to reconstruct fine-scale features from low-order moment targets, while also reducing adjoint computations by limiting the number of output quantities.
\begin{table}[!htb]
    \centering
    \begin{tabular}{c|ccc|ccc}
        \arrayrulecolor{black}\cmidrule{2-7}
        & \multicolumn{6}{c}{Testing Error [\%] - Dataset 1} \\
        \arrayrulecolor{black}\cmidrule{2-7}
        & \multicolumn{3}{c|}{CoBRAS} & \multicolumn{3}{c}{POD} \\
        \arrayrulecolor{black}\midrule
        $r$ & 8 & 9 & 10 & 8 & 9 & 10 \\
        \arrayrulecolor{black}\midrule
        $n$ & \textbf{0.013} & \textbf{0.002} & \textbf{0.000} & 0.032 & 0.089 & 0.084 \\
        $\vw$ & \textbf{2.005} & \textbf{1.205} & \textbf{1.397} & 3.638 & 2.684 & 2.773 \\
        $x_{\mathrm{e}}$ & \textbf{0.239} & \textbf{0.037} & \textbf{0.009} & 0.827 & 2.385 & 2.098 \\
        $x_{\text{Ar}}$ & \textbf{0.021} & \textbf{0.003} & \textbf{0.000} & 0.101 & 0.092 & 0.088 \\
        $x_{\text{Ar}^+}$ & \textbf{0.101} & \textbf{0.028} & \textbf{0.014} & 2.496 & 0.671 & 0.574 \\
        $e_{\text{Ar}}$ & \textbf{0.270} & \textbf{0.061} & \textbf{0.018} & 1.003 & 1.295 & 1.232 \\
        $e_{\text{Ar}^+}$ & \textbf{0.010} & \textbf{0.003} & \textbf{0.003} & 0.178 & 0.037 & 0.028 \\
        $T_\mathrm{h}$ & \textbf{0.019} & \textbf{0.004} & \textbf{0.001} & 0.085 & 0.077 & 0.076 \\
        $T_\mathrm{e}$ & \textbf{0.129} & \textbf{0.056} & \textbf{0.047} & 0.241 & 0.119 & 0.108 \\
        \arrayrulecolor{black}\midrule
    \end{tabular}
    \caption{\textit{Testing error for dataset 1 of table \ref{table:test.dsets}.} Mean relative errors (\%) for the quantities of interest, computed using CoBRAS and POD ROMs with varying reduced dimensions $r$ (total reduced state dimension $d = r + 3$, including the electron mass fraction and the two temperatures). Results are compared against the FOM solutions. Bold values indicate the lower error between CoBRAS and POD for each entry.}
    \label{table:err.dset1}
\end{table}
Similar trends are observed in tables~\ref{table:err.dset2} and~\ref{table:err.dset3} (see Supplementary Material), which present results for the two extrapolation datasets. While both ROMs exhibit increased errors in the low-temperature regime, CoBRAS maintains competitive accuracy for $r \geq 9$, even under extrapolative conditions. The performance degradation at lower temperatures likely stems from discrepancies between the subspaces required to accurately capture low-temperature dynamics and those spanned by the training data (see table~\ref{table:train.dset}). This is consistent with the physics of the system: the reaction operators governing species evolution are highly temperature-dependent, with rate constants governed by Arrhenius-type laws. Due to their exponential form, $\exp(-1/T)$, these rates exhibit greater sensitivity at lower temperatures, resulting in sharper gradients in the ASDFs and increased difficulty for the reduced basis to generalize accurately.

Figure~\ref{fig:0d.testcase3}, together with figures~\ref{fig:0d.testcase1} and~\ref{fig:0d.testcase2} in the Supplementary Material, presents a side-by-side comparison of CoBRAS and POD ROMs predictions with reduced dimension $r = 8$, against FOM results for three test trajectories spanning mild (case 1) to harsh (case 3) conditions, as summarized in table~\ref{table:testcases}.
\begin{table}[!htb]
    \centering
    \begin{tabular}{c|c|c}
        \arrayrulecolor{black}\midrule
        Case & $T_{\mathrm{h}_0}$ [K] & $\rho$ [kg/m$^3$] \\
        \arrayrulecolor{black}\midrule
        1 & 20\,000 & 0.005 \\
        2 & 35\,000 & 0.01 \\
        3 & 50\,000 & 0.1 \\
        \arrayrulecolor{black}\midrule
    \end{tabular}
    \caption{\textit{Test cases.} Initial parameter values used to generate the three testing trajectories for visual comparison between the ROMs and FOM solutions.}
    \label{table:testcases}
\end{table}
In all figures, the top row shows the time evolution of the zeroth-order moments (i.e., molar fractions) of e$^-$, Ar, and Ar$^+$, while the bottom row displays the first-order moments (internal energies) of Ar and Ar$^+$, along with the evolution of heavy-particle and electron temperatures. 
For the first two test cases, both ROMs show good agreement with the FOM, as illustrated in the Supplementary Material, except for the Ar$^+$ internal energy ($e_{\mathrm{Ar}^+}$), where CoBRAS consistently outperforms POD. In contrast, under the more demanding conditions of case 3 (figure~\ref{fig:0d.testcase3}), CoBRAS shows superior performance across all quantities, corroborating the error trends reported in tables~\ref{table:err.dset1}, \ref{table:err.dset2}, and~\ref{table:err.dset3}. 
Notably, the reduced accuracy of POD in predicting internal energies leads to visible discrepancies in the evolution of both heavy-particle and electron temperatures, particularly evident in the last panel of figure~\ref{fig:0d.testcase3}. Immediately following the electron peak and within the thermal equilibrium region (see \cite{Kapper2011IonizingStructure}), POD fails to track the FOM temperature evolution, whereas CoBRAS remains in close agreement.
It is also worth noting that, as previously reported by Zanardi et al.~\cite{Zanardi2025Petrov-GalerkinMixtures}, neither CoBRAS nor POD enforces positivity of the mass fractions. In our simulations, this limitation led to small non-physical negative values at early times ($t < 10^{-8}$~s).
\begin{figure}[!htb]
\centering
\begin{subfigure}[htb!]{0.32\textwidth}
    \includegraphics[width=\textwidth]{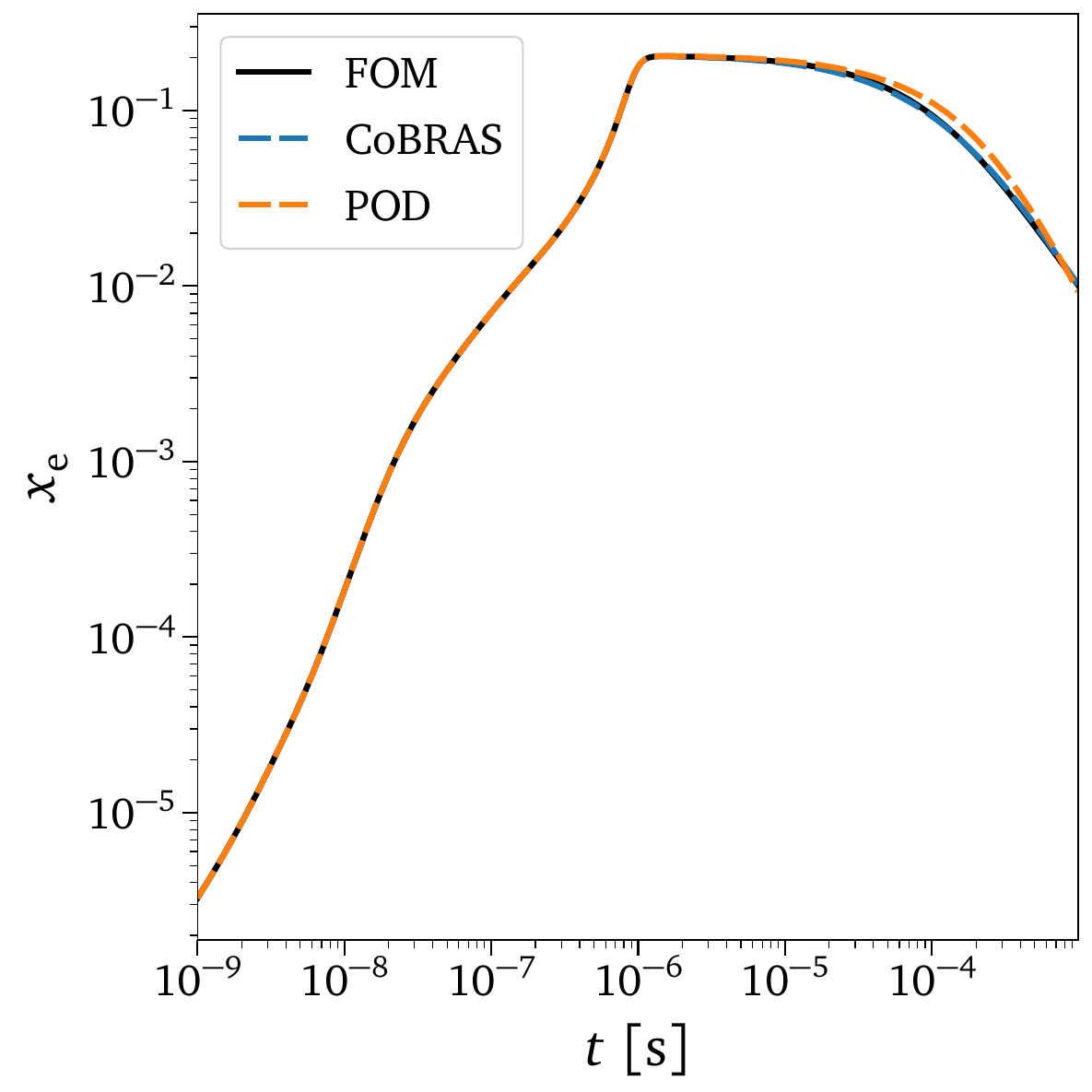}
\end{subfigure}
\begin{subfigure}[htb!]{0.32\textwidth}
    \includegraphics[width=\textwidth]{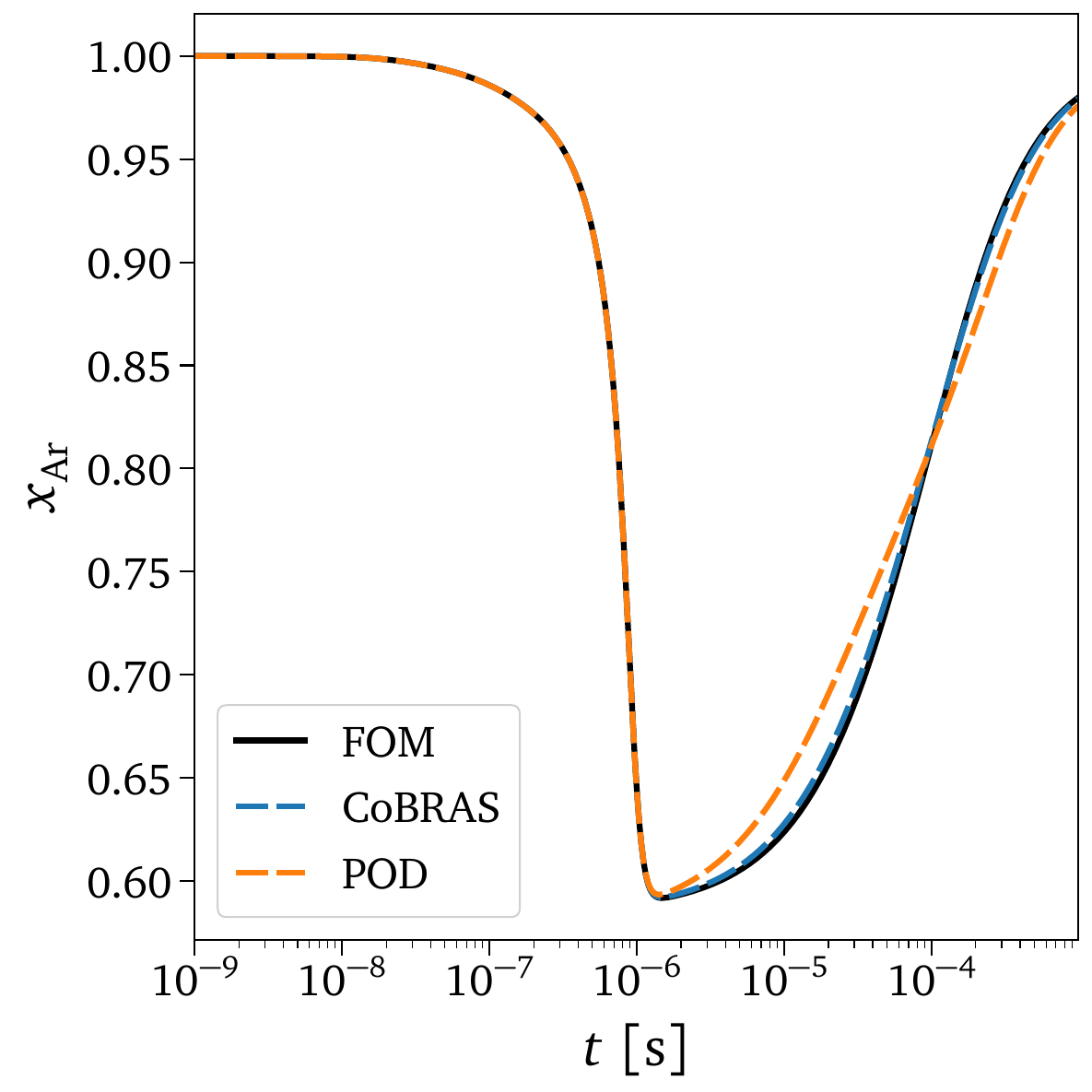}
\end{subfigure}
\begin{subfigure}[htb!]{0.32\textwidth}
    \includegraphics[width=\textwidth]{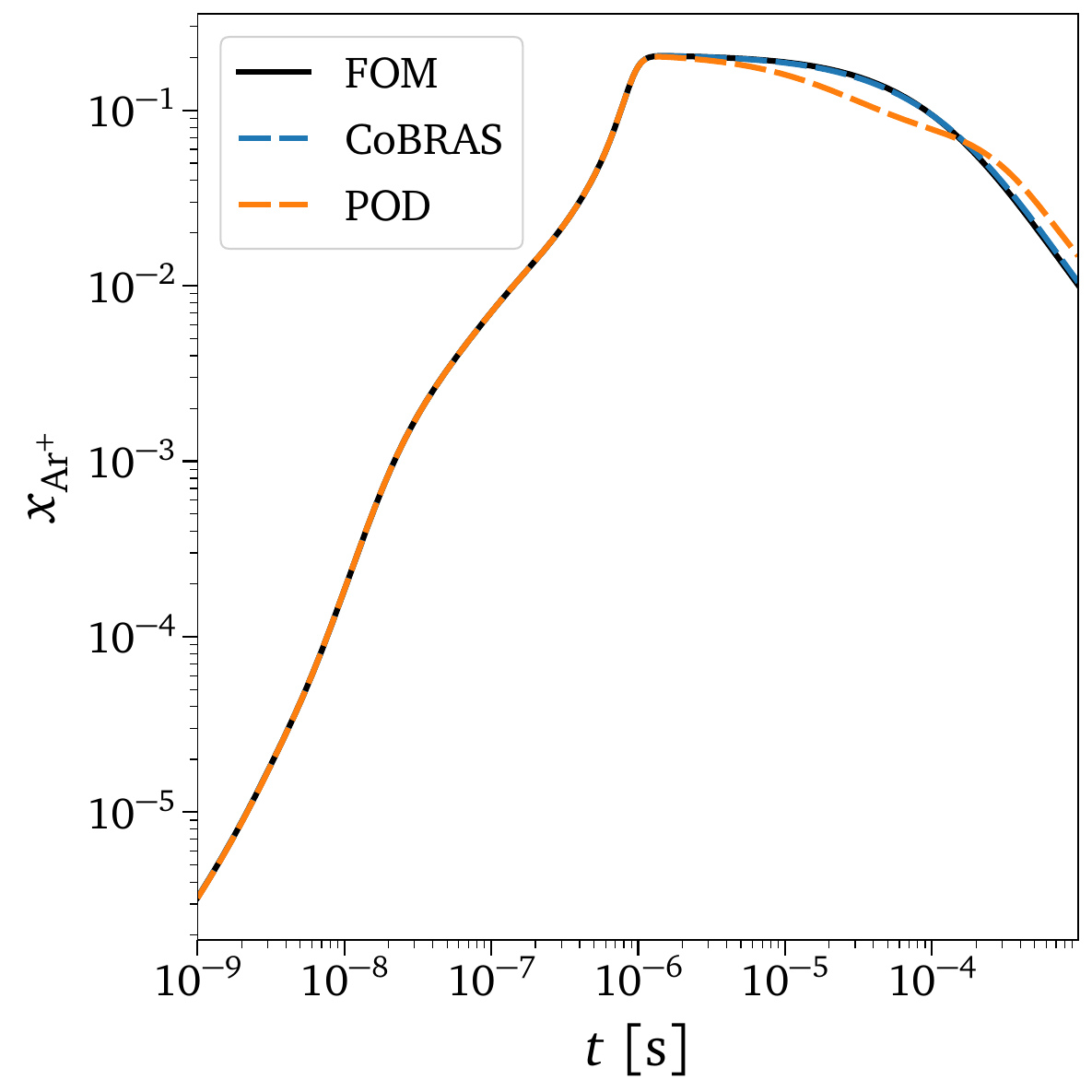}
\end{subfigure}
\\[4pt]
\begin{subfigure}[htb!]{0.32\textwidth}
    \includegraphics[width=\textwidth]{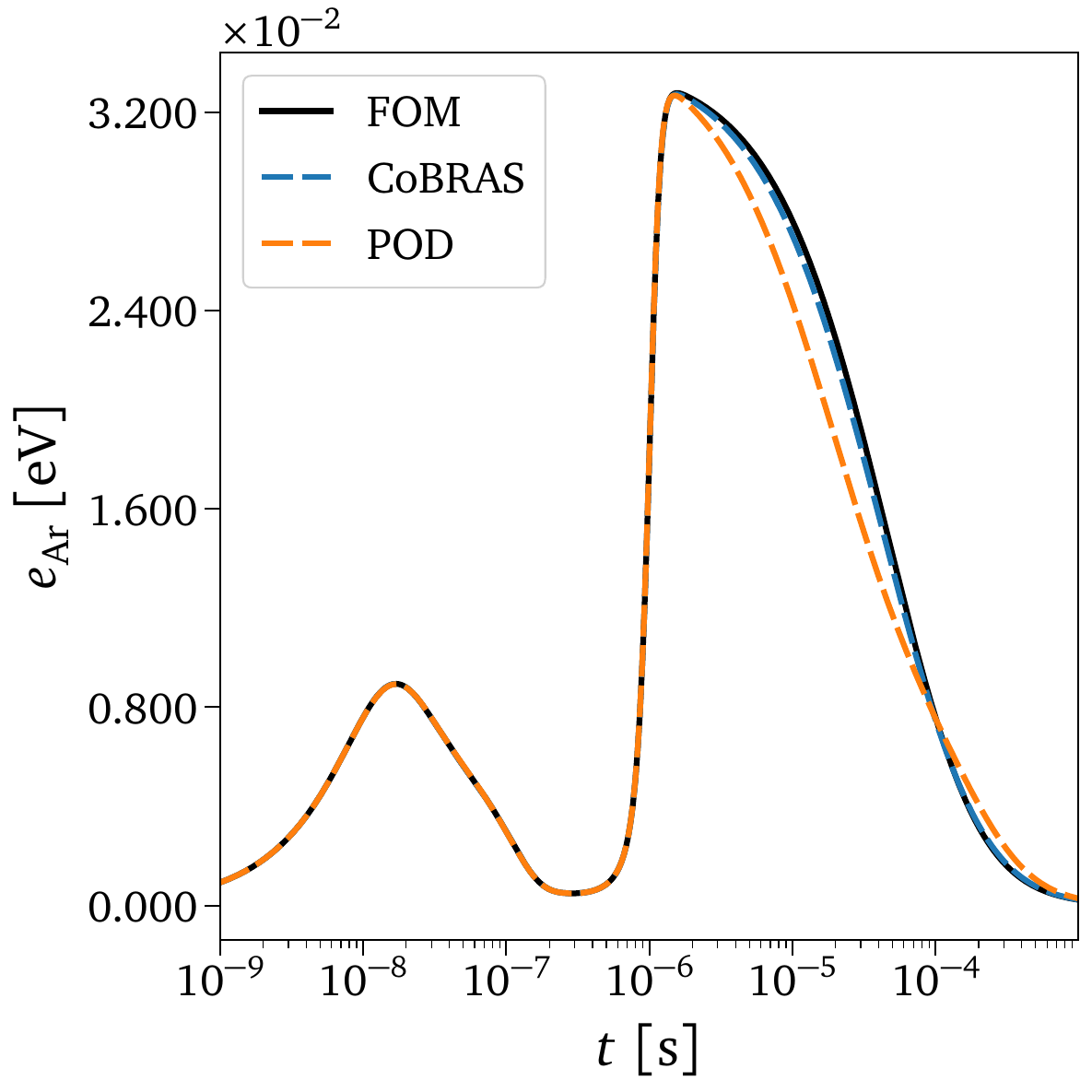}
\end{subfigure}
\begin{subfigure}[htb!]{0.32\textwidth}
    \includegraphics[width=\textwidth]{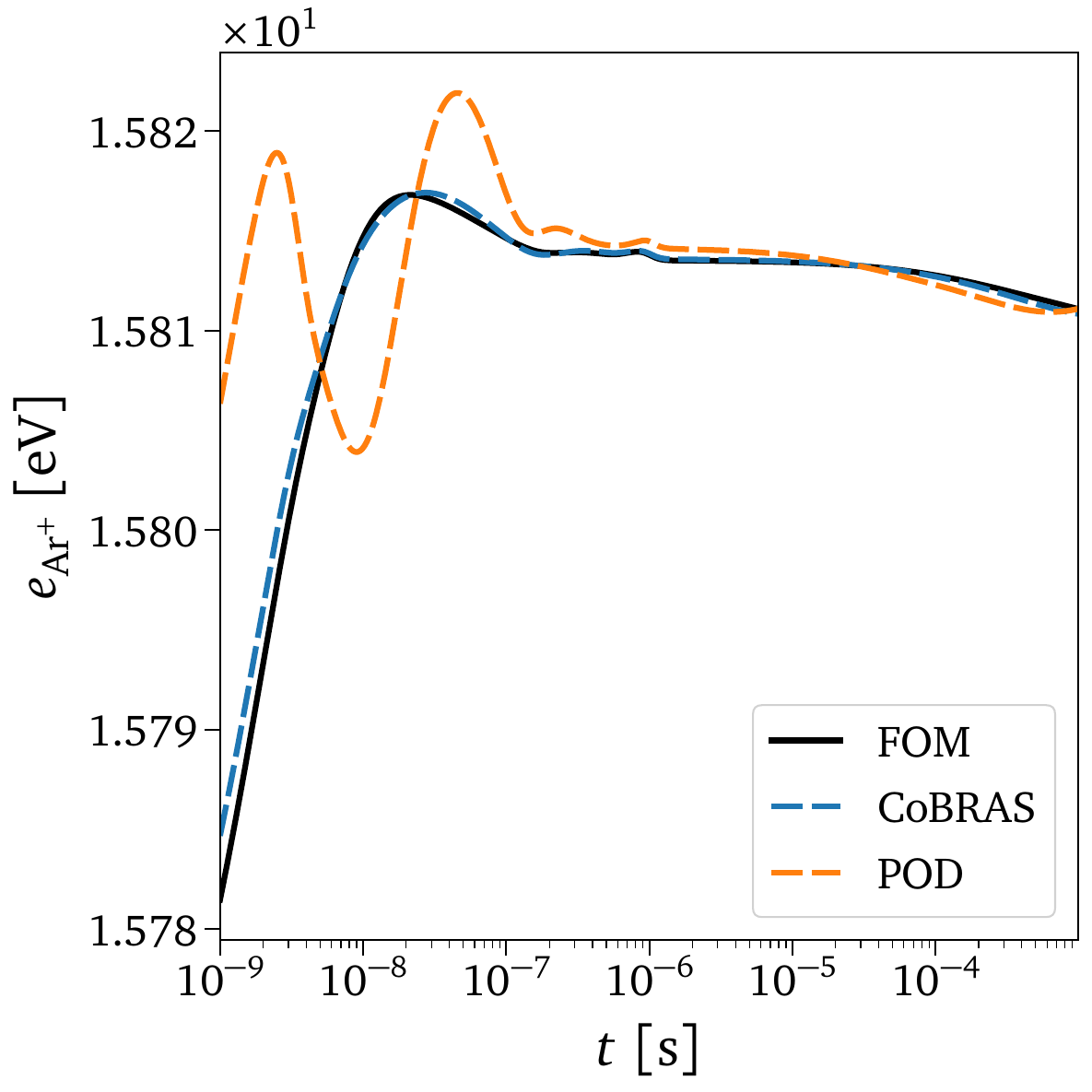}
\end{subfigure}
\begin{subfigure}[htb!]{0.32\textwidth}
    \includegraphics[width=\textwidth]{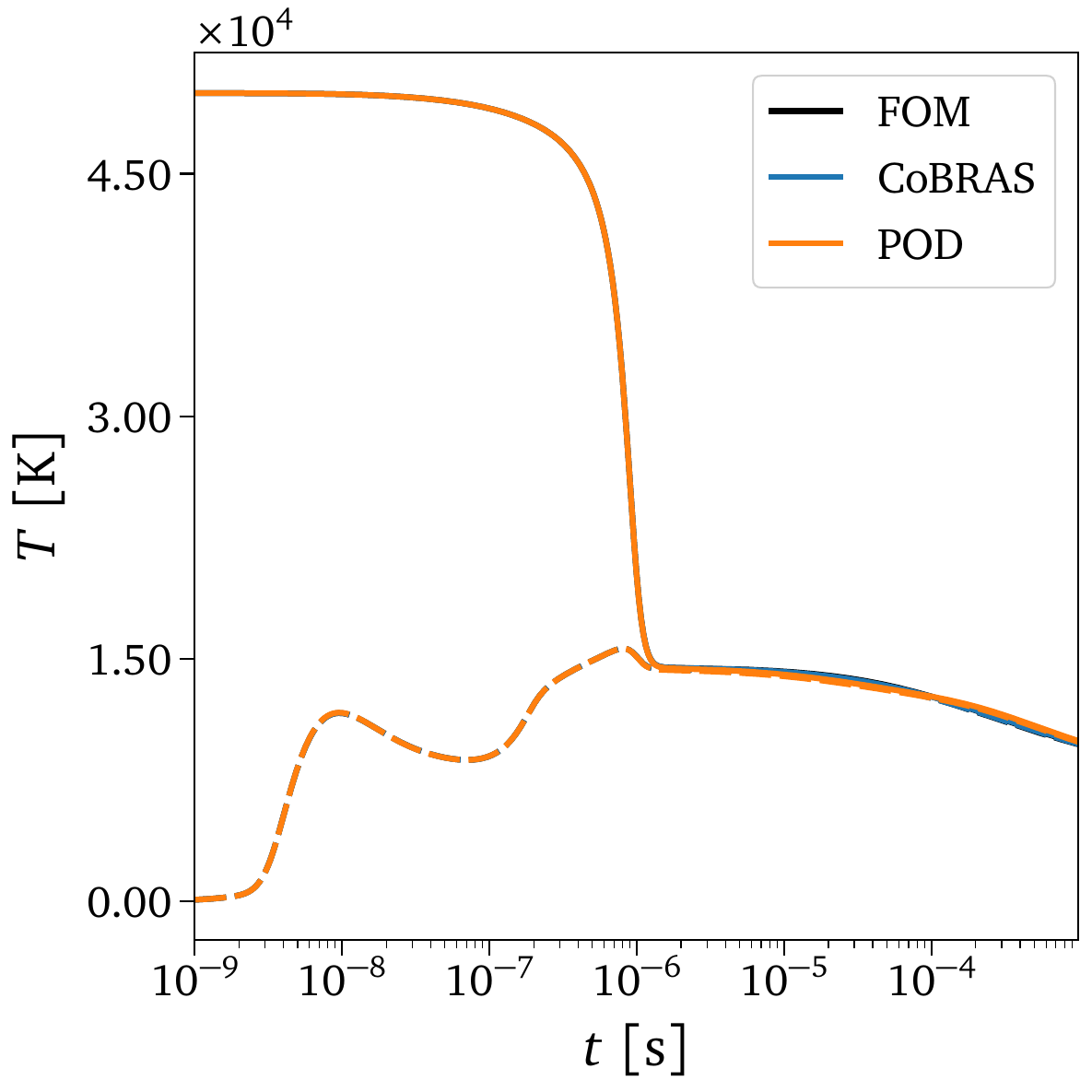}
\end{subfigure}
\caption{\textit{FOM vs. ROMs for 0D simulations: Test case 3.} Time evolution of species zeroth-order moments (molar fractions), first-order moments (internal energies), and temperatures, as predicted by the FOM and by CoBRAS and POD reduced-order models with dimension $r = 8$. Results correspond to test case 3 described in table~\ref{table:testcases}.}
\label{fig:0d.testcase3}
\end{figure}

\subsubsection{Computational performance}\label{sssec:0d.cost}
In this section, we assess computational efficiency by comparing the total floating-point operations (FLOPs) required by the FOM and the CoBRAS-based ROM as the reduced dimension $r$ varies. In addition, we report the \textit{arithmetic intensity}—defined as the ratio of FLOPs to memory access in bytes (FLOP/Byte), following the formulation in~\cite{Zanardi2025Petrov-GalerkinMixtures}—for key numerical kernels. While actual runtimes are influenced by hardware-specific features such as memory bandwidth, processor speed, and cache hierarchy, FLOP counts and arithmetic intensity offer consistent, architecture-agnostic comparisons.
\par
As shown in table~\ref{table:0d.costs}, where $d$ is the system dimension ($d = r + 3$ for the ROM), the CoBRAS-based ROM yields significant reductions in computational cost. In particular, the number of FLOPs required for right-hand side (RHS) evaluations—dominated by dense matrix-vector multiplications that scale as $\mathcal{O}(d^2)$ (see equations (52)-(55) in~\cite{Kapper2011IonizingStructure})—is reduced by more than an order of magnitude. In our current implementation, these savings are estimated rather than fully realized, as we do not assemble reduced operators offline. Instead, the full-order state is reconstructed at each time step to evaluate the RHS of the FOM, which is subsequently projected to compute the evolution of the reduced coordinates. While this approach avoids the need for precomputing and storing reduced operators, it limits the practical benefits of the theoretical FLOP reductions. Nonetheless, in contexts where reduced operators are assembled offline—as demonstrated in~\cite{Zanardi2025Petrov-GalerkinMixtures}—the presented estimates are directly applicable. The cost of solving the linear system (LSS) associated with the implicit BDF integrator, which scales as $\mathcal{O}(d^3)$, is similarly reduced by over a factor of 30. However, the actual efficiency of both RHS and LSS operations is ultimately constrained by their \textit{arithmetic intensity}. In the ROM setting, these operations exhibit relatively low FLOP/Byte ratios, suggesting that performance is likely to be memory-bound rather than compute-bound. Consequently, the theoretical FLOP reductions may not translate linearly into runtime gains due to limitations in memory access speed.
\begin{table}[!htb]
    \centering
    \begin{tabular}{c|c||c|c|c|c}
        \arrayrulecolor{black}\cmidrule{3-6}
        \multicolumn{2}{c||}{} & \multicolumn{2}{c|}{FLOPs} & \multicolumn{2}{c}{FLOP/Byte} \\
        \arrayrulecolor{black}\midrule
        Model & $d$ & RHS - \(\mathcal{O}(d^2)\) & LSS - \(\mathcal{O}(d^3)\) & RHS - \(\mathcal{O}(1)\) & LSS - \(\mathcal{O}(d)\) \\
        \arrayrulecolor{black}\midrule
        FOM & 36 & \(2.556\times 10^3\) & \(3.370\times 10^{4}\) & 0.234 & 3.079 \\
        \arrayrulecolor{gray}\midrule
        \multirow{3}*{CoBRAS} & 13 & \(3.250\times 10^2\) & \(1.803\times 10^3\) & 0.208 & 1.156 \\
        & 12 & \(2.760\times 10^2\) & \(1.440\times 10^3\) & 0.205 & 1.071 \\
        & 11 & \(2.310\times 10^2\) & \(1.129\times 10^3\) & 0.202 & 0.987 \\
        \arrayrulecolor{black}\midrule
    \end{tabular}
    \caption{\textit{FOM vs. CoBRAS for 0D simulations: Computational cost}. The table reports total FLOPs and FLOP/Byte ratios for key numerical operations, including right-hand side (RHS) evaluation and linear system solve (LSS). Results are shown for the FOM and three CoBRAS ROM configurations.}
    \label{table:0d.costs}
\end{table}
\par
\begin{figure}[htb!]
\centering
\begin{subfigure}[htb!]{0.32\textwidth}
    \includegraphics[width=\textwidth]{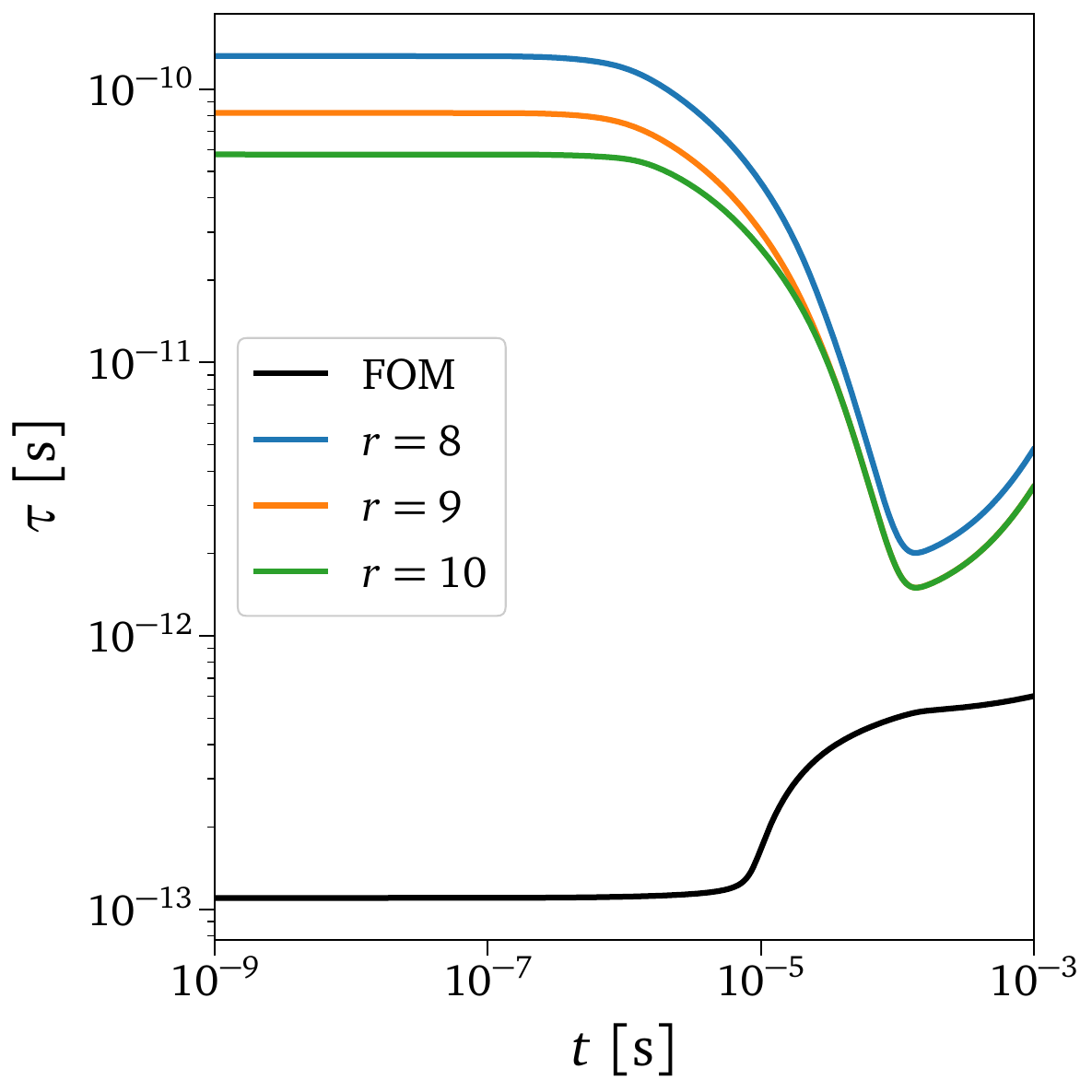}
\end{subfigure}
\quad
\begin{subfigure}[htb!]{0.32\textwidth}
    \includegraphics[width=\textwidth]{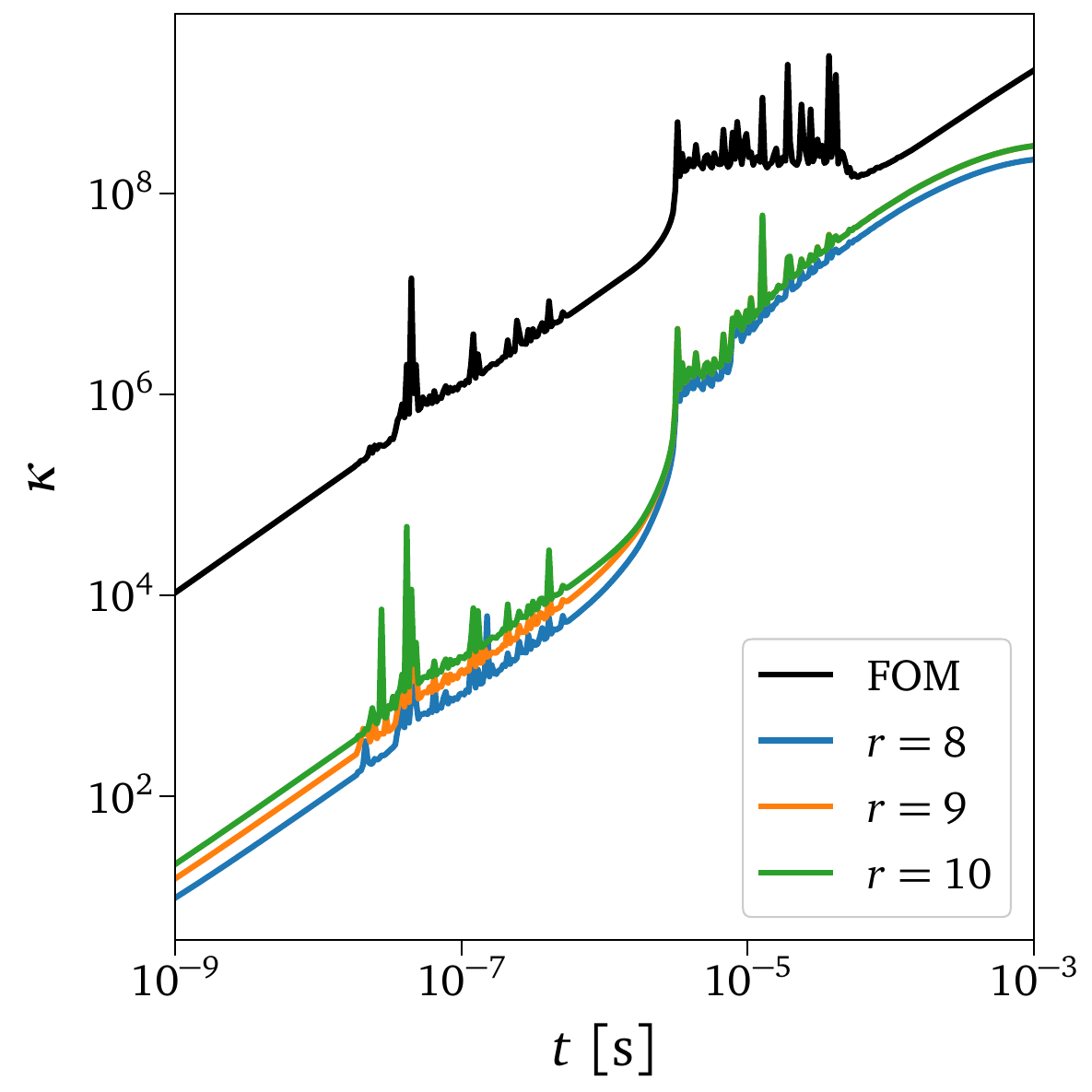}
\end{subfigure}
\caption{\textit{FOM vs. CoBRAS for 0D simulations: Integration scheme parameters.} Comparison between the FOM and CoBRAS ROM across different reduced dimensions $r$, showing the smallest timescale (left panel) and condition number (right panel) of the implicit integration scheme. These quantities are computed using the Jacobian of the respective system.}
\label{fig:0d.tmin.knum}
\end{figure}

Two additional performance metrics are examined to assess numerical stiffness and solver sensitivity. The first is the fastest timescale, approximated as the inverse of the smallest eigenvalue of the Jacobian, \(\tau=1/|\lambda_{\min}|\). The second is the condition number $\kappa$ of the matrix $\mathbf{P}$ defined in equation \eqref{eq:p.mat.time} in \ref{app:cond.num} (with $\beta_0 = 1$, i.e. backward Euler method) used in the linear solve of the BDF scheme. Both metrics are computed as averaged values over 100 test trajectories with fixed density $\rho=10^{-2}$ kg/m$^3$, while varying the initial heavy-particle temperature $T_{\mathrm{h}_0}$ (within the training regime).
As illustrated in figure~\ref{fig:0d.tmin.knum} (left panel), the CoBRAS-based ROM exhibits reduced stiffness, as indicated by larger timescales, which enables bigger time steps during integration and contributes to further computational savings. Similarly, the right panel of figure~\ref{fig:0d.tmin.knum} shows that the ROM achieves a lower condition number, indicating improved numerical conditioning and a more stable and efficient linear solve. Together, these results highlight that CoBRAS not only reduces dimensional complexity but also improves the numerical properties of the system.

We emphasize that the computational savings achieved by the proposed ROM are not limited to the 0D configuration. In multidimensional simulations, the reduced number of state variables translates directly into fewer scalar quantities to be transported, leading to additional performance gains. Specifically, when using explicit time integration for the transport operator, a linear speedup proportional to the reduction in state dimension can be expected. For steady-state simulations employing implicit schemes, the computational advantage becomes even more pronounced—potentially scaling cubically—due to the reduced cost of solving smaller linear systems.

\subsection{One‐dimensional simulations}\label{ssec:1d}
We evaluate the performance of CoBRAS in a canonical one-dimensional (1D) shock tube configuration, where the coupling between transport and collisional-radiative kinetics plays a critical role in determining the plasma dynamics. This test case serves as a benchmark for assessing the ROM’s ability to accurately capture highly unsteady and nonlinear behavior of the flow. The numerical discretization and time integration schemes are detailed in section~\ref{sec:numerics}, while the governing equations for the reduced latent variables are derived in section~\ref{ssec:rom:nd_gov_eqs}. The ROM used here employs a latent dimension of \(r = 9\), which, when combined with the retained components (electron mass fraction and two temperatures), results in a reduced state vector of size \(d = r + 3 = 12\). This represents a 3$\times$ reduction compared to the full thermochemical state vector of 36 variables.
\begin{figure}[!htb]
\centering
\includegraphics[width=0.74\textwidth]{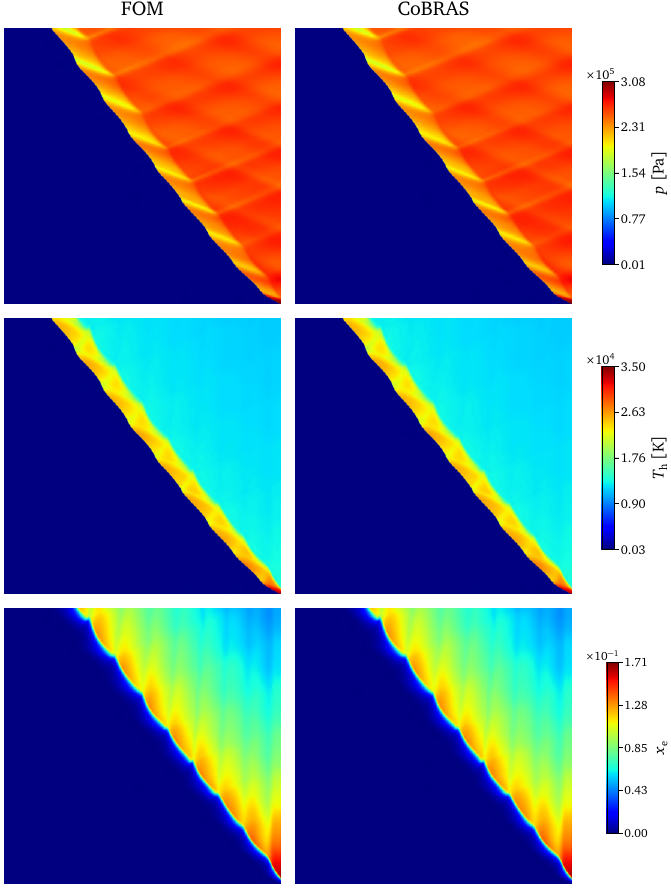}
\caption{\textit{FOM vs. CoBRAS for 1D simulations: Space-time diagrams.} Comparison of FOM and CoBRAS ROM ($r = 9$) for pressure $p$, heavy-particle temperature $T_\mathrm{h}$, and electron molar fraction $x_{\mathrm{e}}$ in space-time ($x$-$y$ plane) diagrams obtained from the 1D simulation.}
\label{fig:1d.v1.timespace}
\end{figure}
\par
The simulation setup follows the shock-reflection experiments of Kapper and Cambier~\cite{Kapper2011IonizingEffects}. In this configuration, a supersonic stream of argon gas impinges upon an ideally reflecting wall, generating a strong shock that propagates upstream against the inflow. The selected freestream conditions are: $p_\infty = 685.2$~Pa, $T_\infty = 293.6$~K, $u_\infty = 4\,535$~m/s, chosen to reproduce a Mach 15.9 shock in the laboratory frame, consistent with UTIAS shock tube measurements~\cite{Glass1991OverWaves}. Although the initial conditions correspond to a stronger (Mach 19) shock, the reflected shock rapidly weakens as thermal energy is converted into electronic excitation and ionization, yielding the lower effective Mach number observed experimentally. The computational domain spans \(x \in [0, 0.2]\) m, discretized with a spatial resolution \(\Delta x = 10^{-4}\) m. Time integration is carried out over \(t \in [0, 2.5 \times 10^{-4}]\) s with time step \(\Delta t = 5 \times 10^{-8}\) s. Boundary conditions consist of a supersonic inflow on the left and a perfectly reflecting wall on the right. This setup enables direct comparison with experimental data and isolates the unsteady interaction between plasma kinetics and hydrodynamics.
\par
Figure~\ref{fig:1d.v1.timespace} presents space-time (\(x\)-\(t\)) diagrams for three key quantities: pressure \(p\), heavy-particle temperature \(T_\mathrm{h}\), and electron molar fraction \(x_\mathrm{e}\). For each variable, the left column shows the FOM solution and the right column shows the CoBRAS prediction. Both models capture the characteristic undamped, periodic fluctuations observed in such flows. The shock front is visible as the leftmost propagating feature, followed by the electron avalanche, defined by the first peak in electron density. These structures travel from the bottom-right to the top-left of each panel, with the induction region lying between them.
The dominant physical feature of this flow is the presence of periodic disturbances within the induction zone.
The electron avalanche generates a pressure wave that travels back to the shock. Upon interacting with the shock front, this wave alters the local conditions, strengthening the shock and generating a new entropy wave that travels downstream.
This entropy wave accelerates excitation and ionization processes, triggering the formation of a subsequent electron avalanche. The new avalanche, in turn, emits another pressure wave, thus repeating the cycle of fluctuations.
These mechanisms were analyzed in detail by Kapper and Cambier~\cite{Kapper2011IonizingEffects}. The ROM accurately captures this cyclic behavior, preserving both the phase and amplitude of the fluctuations and faithfully reproducing the underlying physical processes.
\par
\begin{figure}[htb!]
\centering
\begin{subfigure}[htb!]{0.4\textwidth}
    \caption*{\hspace{-3mm}\fontfamily{mdbch}\normalsize CoBRAS}
    \includegraphics[width=\textwidth]{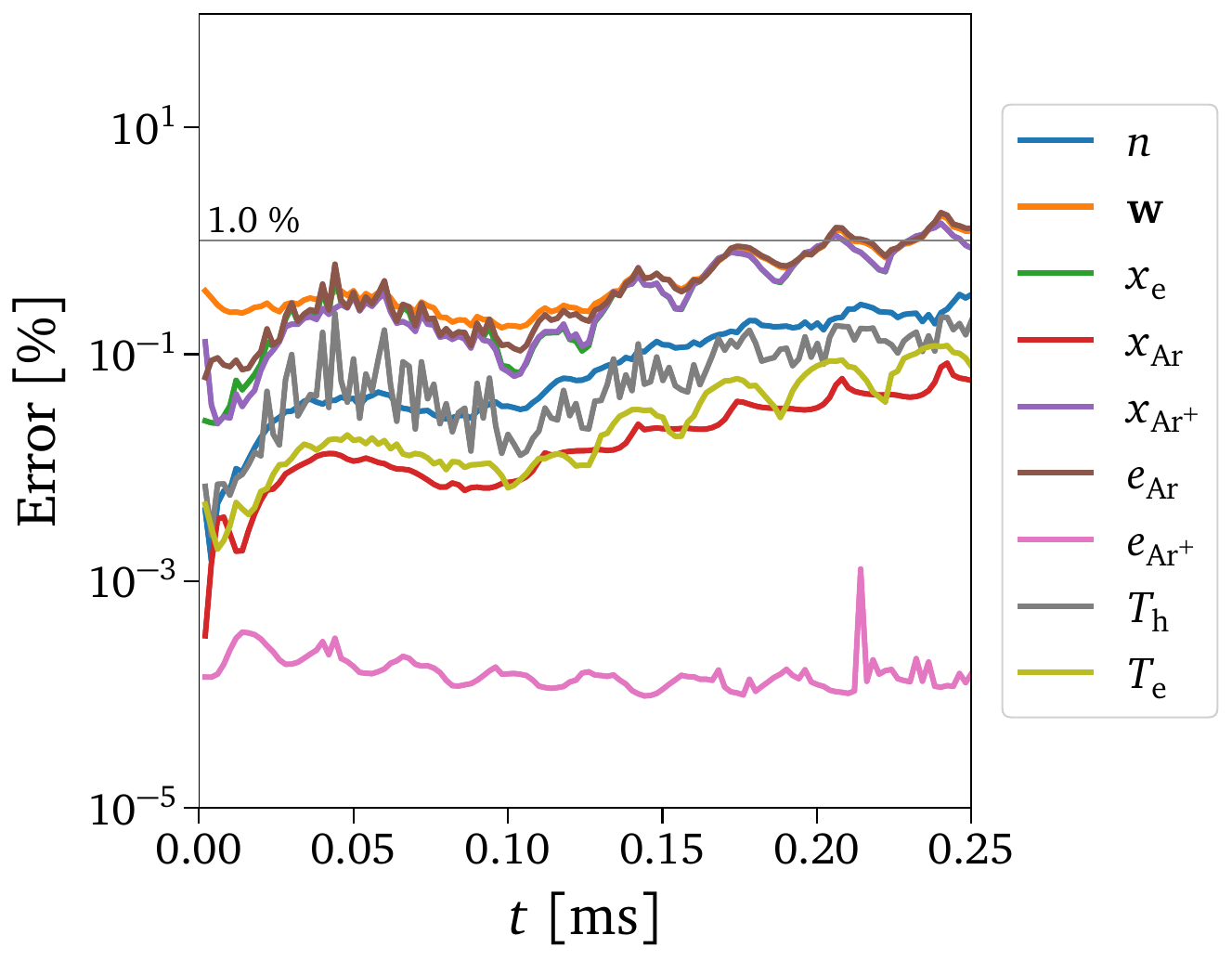}
\end{subfigure}
\quad
\begin{subfigure}[htb!]{0.4\textwidth}
    \caption*{\hspace{-3mm}\fontfamily{mdbch}\normalsize POD}
    \includegraphics[width=\textwidth]{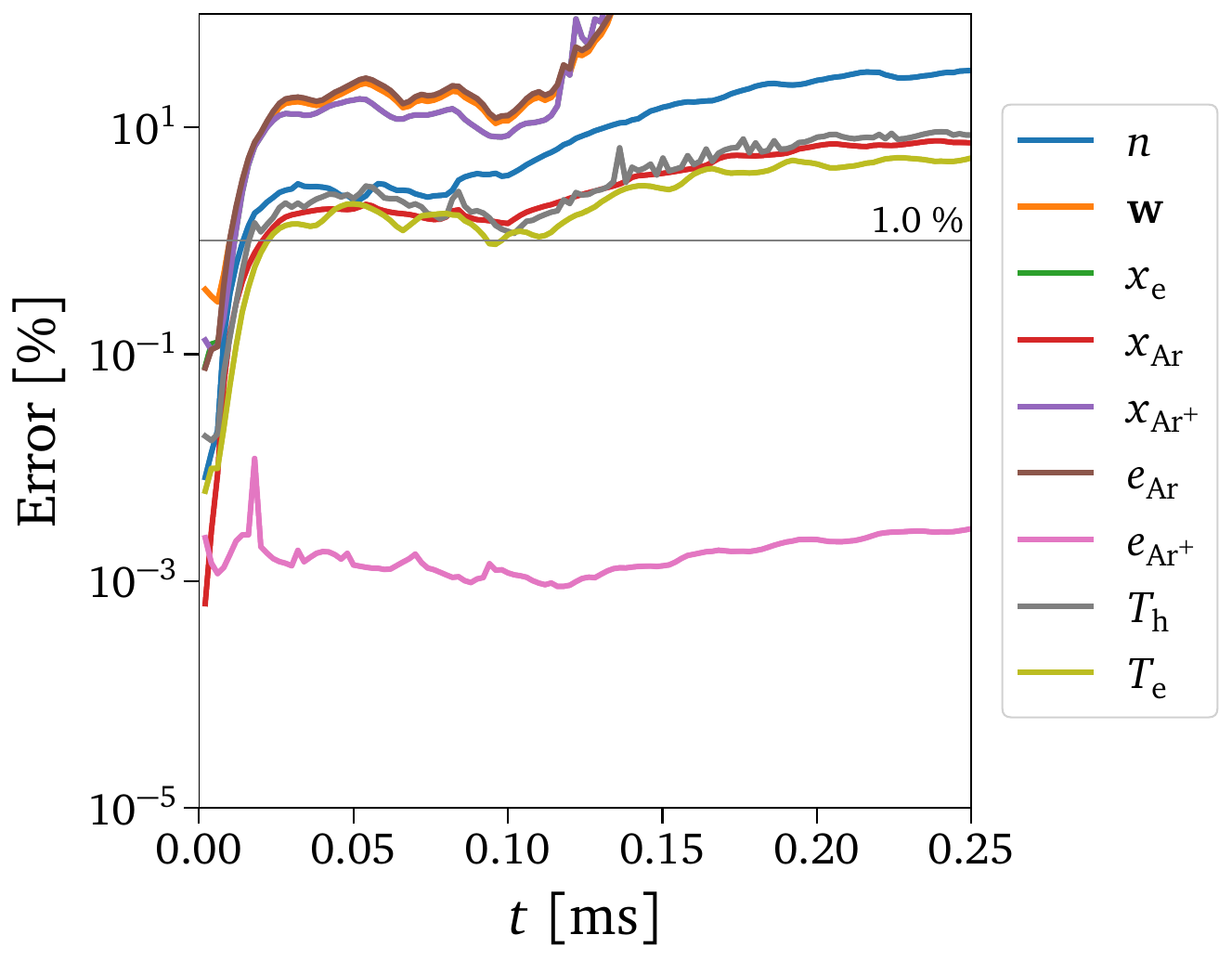}
\end{subfigure}
\caption{\textit{Error evolution of ROMs in 1D simulations.}
Space‐averaged relative errors (\%) for CoBRAS (left panel) and POD (right panel) ROMs with $r = 9$, computed for key quantities of interest.}
\label{fig:1d.v1.err_evol}
\end{figure}
A quantitative comparison is shown in figure~\ref{fig:1d.v1.err_evol}, which reports the evolution of space-averaged relative errors for a set of thermochemical variables: the mixture number density \(n\), species mass fractions \(\vw\), zeroth- and first-order moments of ASDFs, and the heavy-particle and electron temperatures. To ensure the error metric reflects meaningful dynamics, points within the freestream region—where values remain constant—are excluded from the computation, preventing artificial deflation of the error. The CoBRAS-based ROM consistently maintains relative errors below 1\% throughout the simulation, demonstrating excellent agreement with the FOM. In contrast, a standard POD-based ROM exhibits instability, with errors growing rapidly after approximately \(t = 0.1\) ms. This highlights the robustness of the CoBRAS projection framework in stiff, nonequilibrium kinetic regimes, where naive dimensionality reduction techniques may fail.
\par
\begin{figure}[!htb]
\centering
\includegraphics[width=0.96\textwidth]{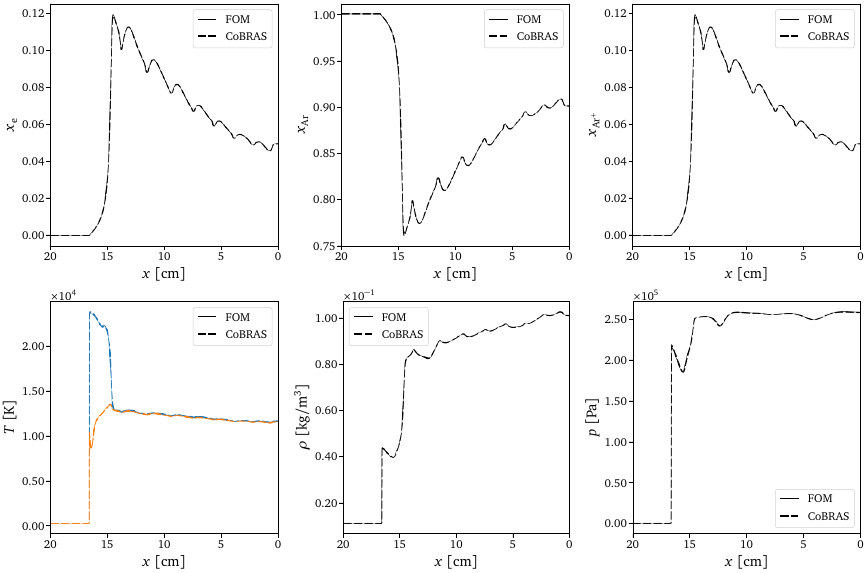}
\caption{\textit{Final-time comparison of FOM and CoBRAS for 1D simulations.} Comparison of FOM and CoBRAS ($r=9$) at the final time $t=2.5\times10^{-4}$ s for the 1D simulation presented in figure~\ref{fig:1d.v1.timespace}. Top row: species molar fractions of e$^-$, Ar, and Ar$^+$. Bottom row: heavy-particle temperature $T_{\mathrm{h}}$ (blue), electron temperature $T_{\mathrm{e}}$ (orange), density $\rho$, and pressure $p$.}
\label{fig:1d.v1.snapshots}
\end{figure}
To further evaluate the ROM's spatial accuracy, figure~\ref{fig:1d.v1.snapshots} compares final-time (\(t = 2.5 \times 10^{-4}\) s) snapshots of macroscopic fields. The top row shows molar fractions of e\(^-\), Ar, and Ar\(^+\), while the bottom row displays heavy-particle temperature \(T_\mathrm{h}\), electron temperature \(T_\mathrm{e}\), density \(\rho\), and pressure \(p\). The CoBRAS-based ROM closely matches the FOM across all variables. In particular, the sharp transitions and oscillatory features in the molar fractions and temperature fields are accurately retained, confirming the ROM's ability to resolve fine-scale plasma dynamics without sacrificing stability or fidelity.
\par
\begin{figure}[htb!]
\centering
\begin{subfigure}[htb!]{0.32\textwidth}
    \includegraphics[width=\textwidth]{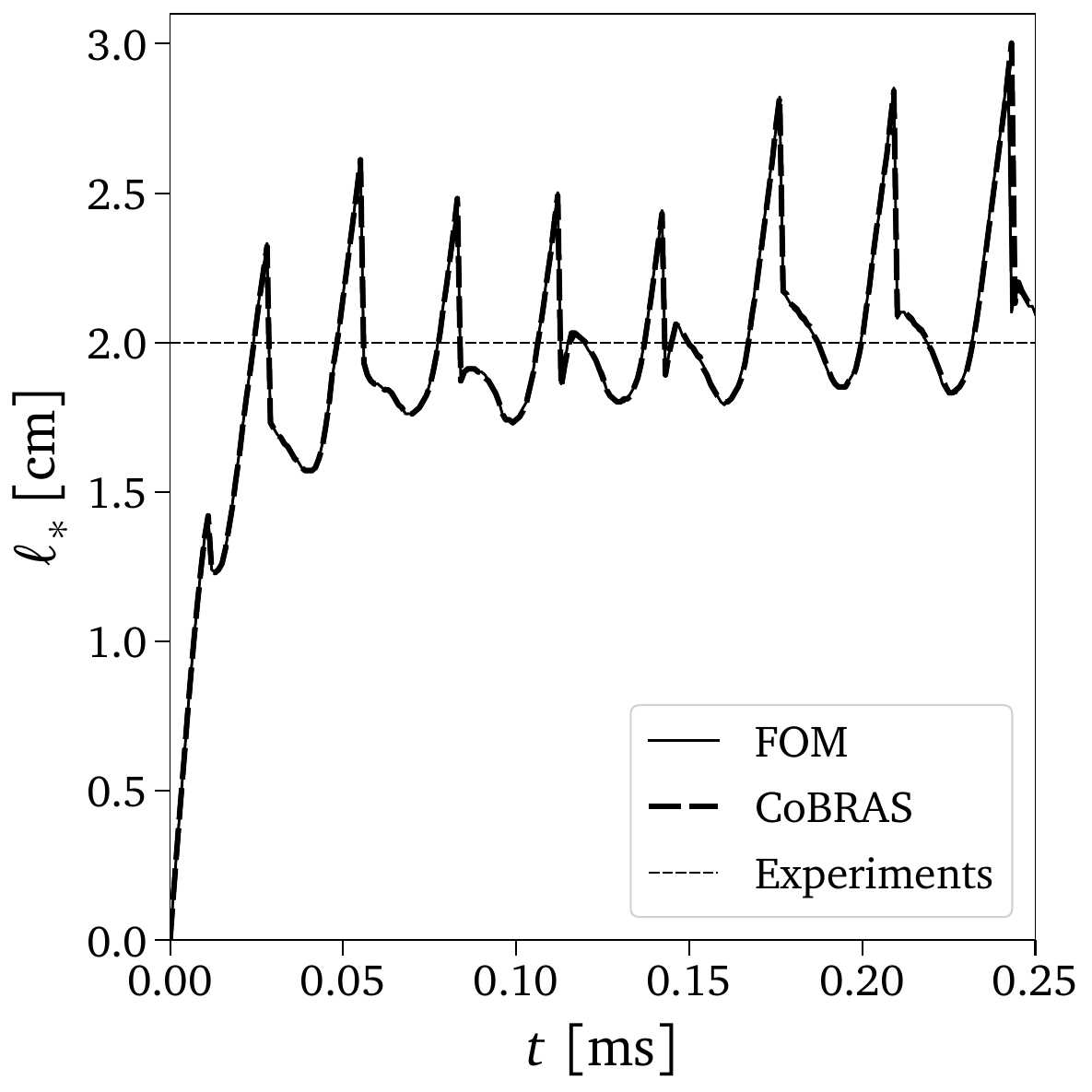}
\end{subfigure}
\quad
\begin{subfigure}[htb!]{0.32\textwidth}
    \includegraphics[width=\textwidth]{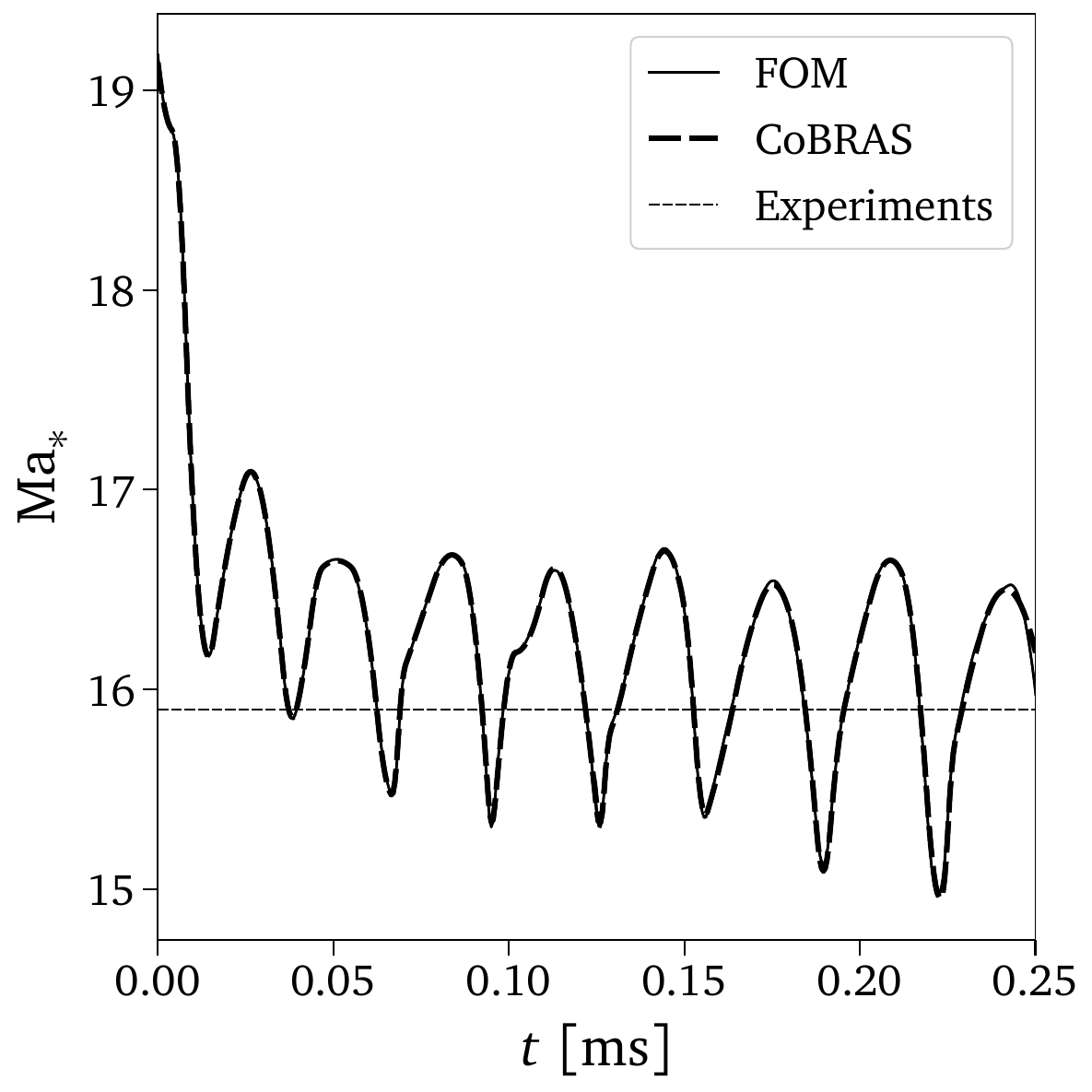}
\end{subfigure}
\caption{\textit{Comparison against average experimental values for 1D simulations.} The FOM and CoBRAS ROM ($r = 9$) are compared against average experimental values for the induction length (left panel) and the shock Mach number (right panel).}
\label{fig:1d.exp}
\end{figure}
In addition to local field comparisons, we also track global quantities that are relevant to experimental observables. Figure~\ref{fig:1d.exp} shows the evolution of the instantaneous shock Mach number and the induction length, defined by the distance between the shock front and the first peak electron density. Both quantities exhibit undamped, nonlinear oscillations, consistent with prior observations. The oscillation period is approximately 30.57~\(\mu\)s for the FOM and 30.71~\(\mu\)s for the ROM, indicating strong agreement. The induction length undergoes discontinuous shifts, further highlighting the nonlinear character of the system. The CoBRAS model captures these behaviors with high fidelity, further validating its predictive capability.
\par
Additional test cases are provided in the Supplementary Material for freestream velocities of $u_\infty=4\,000$ m/s and $u_\infty=5\,000$ m/s. Figures \ref{fig:1d.v3.timespace} and \ref{fig:1d.v3.snapshots}, as well as figures \ref{fig:1d.v2.timespace} and \ref{fig:1d.v2.snapshots}, show the corresponding space-time evolutions and final-time comparisons for the lower and higher velocity cases, respectively. In both cases, CoBRAS continues to match the FOM closely across all major variables. Moreover, figures \ref{fig:1d.v3.err_evol} and \ref{fig:1d.v2.err_evol} confirm that CoBRAS consistently outperforms POD in terms of accuracy, reinforcing its effectiveness across a wide range of conditions.

\subsection{Two-dimensional simulations}\label{ssec:2d}
Building on the insights gained from the 1D simulations—particularly the identification of oscillation mechanisms and the resulting wave patterns—we extend our analysis to a more complex two-dimensional (2D) shock-reflection problem. In the 2D setting, disturbances are no longer confined to the longitudinal direction, allowing for the development of transverse instabilities and rich wave interactions. This presents a rigorous test for the CoBRAS-based ROM and its ability to capture multidimensional plasma flow features.

\begin{figure}[!htb]
\centering
\includegraphics[width=0.98\textwidth]{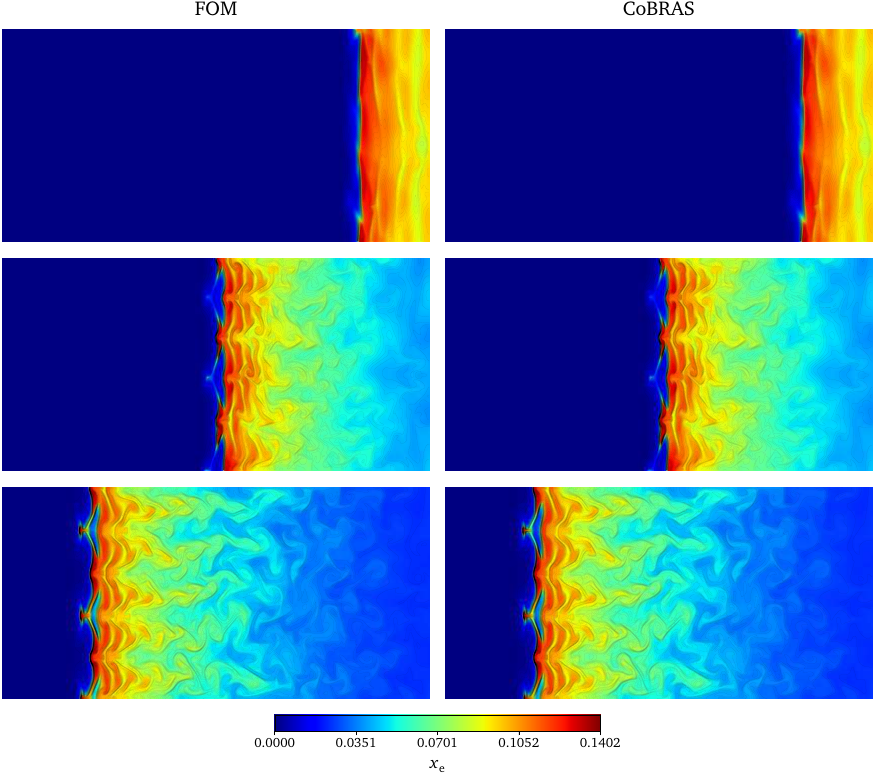}
\vspace{-2mm}
\caption{\textit{FOM vs. CoBRAS for 2D simulations: Electron molar fraction snapshots.} Comparison of FOM and CoBRAS ROM ($r = 9$) at three time instants: $t_i=[1,3,5]\times 10^{-4}$~s for the electron molar fraction, $x_{\mathrm{e}}$.}
\label{fig:2d.xe}
\end{figure}
The computational domain spans $(x, y) \in [0, 0.36] \times [0, 0.18]$~m and is discretized using uniform grid spacing $\Delta x = \Delta y = 3 \times 10^{-4}$~m. Time integration proceeds over the interval $t \in [0, 5] \times 10^{-4}$~s with a time step of $\Delta t = 10^{-7}$~s. As in the 1D case, a high-speed argon flow impinges on a perfectly reflecting wall, initiating a strong shock wave that propagates upstream. The freestream conditions are identical to those used in the previous section: $p_\infty = 685.2$~Pa, $T_\infty = 293.6$~K, $u_\infty = 4\,535$~m/s. To promote the growth of transverse instabilities and facilitate the emergence of multidimensional structures, we introduce a weak Gaussian pressure perturbation near the right wall. The perturbation has an amplitude three times the freestream pressure and density, a spatial extent of $3 \times 10^{-3}$~m in the $x$-direction, and is centered at 40\% of the shock tube height in the $y$-direction (see figure~\ref{fig:init_pert} in the Supplementary Material). This perturbation accelerates the development of instabilities while remaining weak enough to avoid overwhelming the flow physics.

Figure~\ref{fig:2d.xe} illustrates the evolution of the electron molar fraction $x_\mathrm{e}$ at three representative times, $t_i=[1,3,5]\times 10^{-4}$~s, comparing the FOM (left) with CoBRAS (right). Initially disordered transverse waves emerge from the perturbation. Over time, these waves self-organize into a repeating structure consisting of incident and reflected shocks, Mach stems, and triple points. These triple points form a regular cellular pattern in both the longitudinal and transverse directions, indicating a resonant wave interaction. The vorticity generated at these intersections gives rise to ionization cells—highly structured regions of increased temperature, electron density, and vorticity—resembling detonation cells in chemical reactive flows. These localized peaks radiate pressure waves outward, reinforcing the oscillatory motion and maintaining the flow instability~\cite{Kapper2011IonizingEffects}. The CoBRAS-based ROM accurately captures these dynamics. It reproduces not only the global structure and timing of wave interactions, but also the fine-scale features such as vorticity generation and ionization cell formation. Additional field comparisons are provided in the Supplementary Material for mass density $\rho$ (figure~\ref{fig:2d.rho}), pressure $p$ (figure~\ref{fig:2d.p}), heavy-particle temperature $T_\mathrm{h}$ (figure~\ref{fig:2d.Th}), and electron temperature $T_\mathrm{e}$ (figure~\ref{fig:2d.Te}), all of which show excellent agreement between the FOM and CoBRAS.

\begin{figure}[htb!]
\centering
\begin{subfigure}[htb!]{0.4\textwidth}
    \caption*{\hspace{-3mm}\fontfamily{mdbch}\normalsize CoBRAS}
    \includegraphics[width=\textwidth]{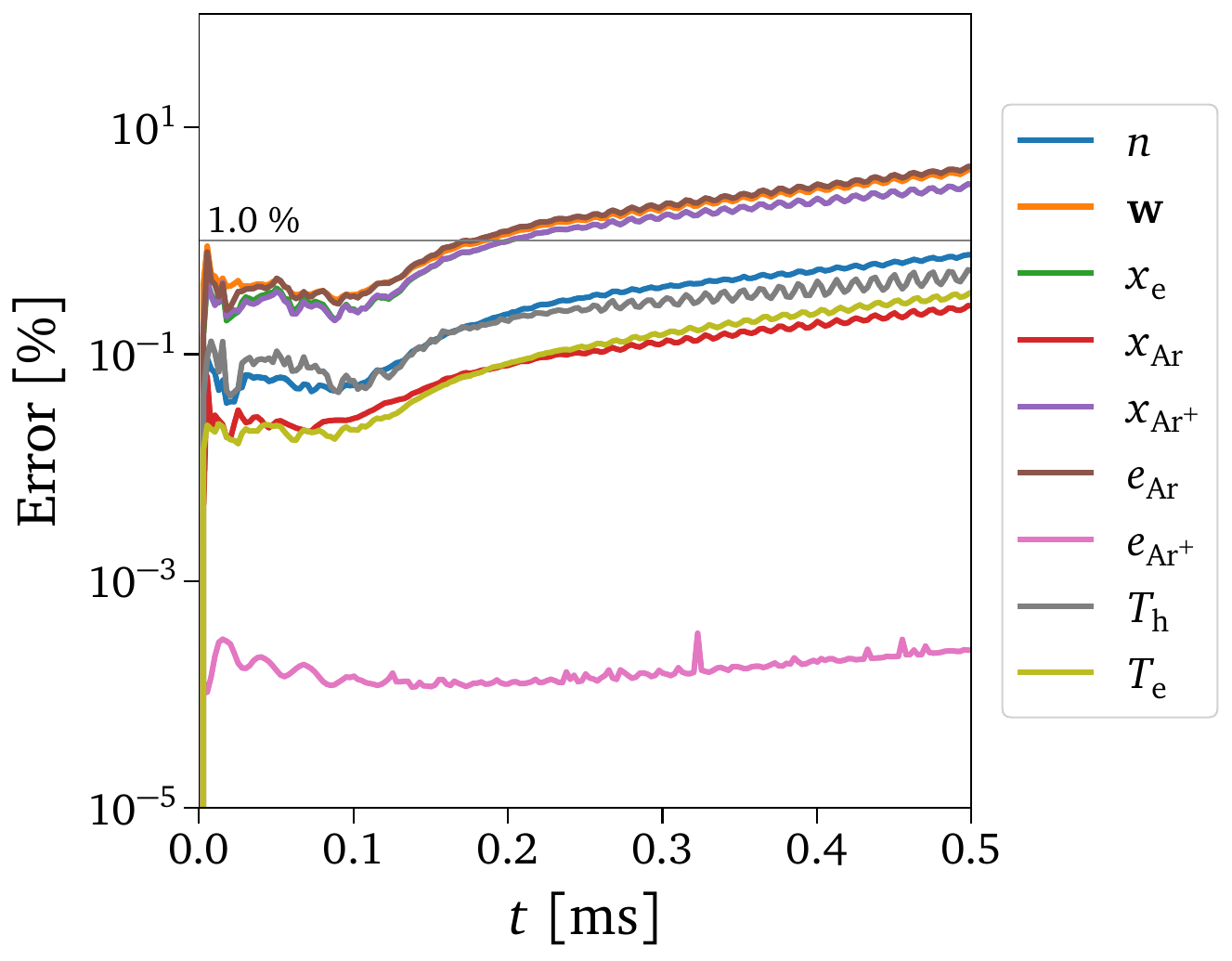}
\end{subfigure}
\quad
\begin{subfigure}[htb!]{0.4\textwidth}
    \caption*{\hspace{-3mm}\fontfamily{mdbch}\normalsize POD}
    \includegraphics[width=\textwidth]{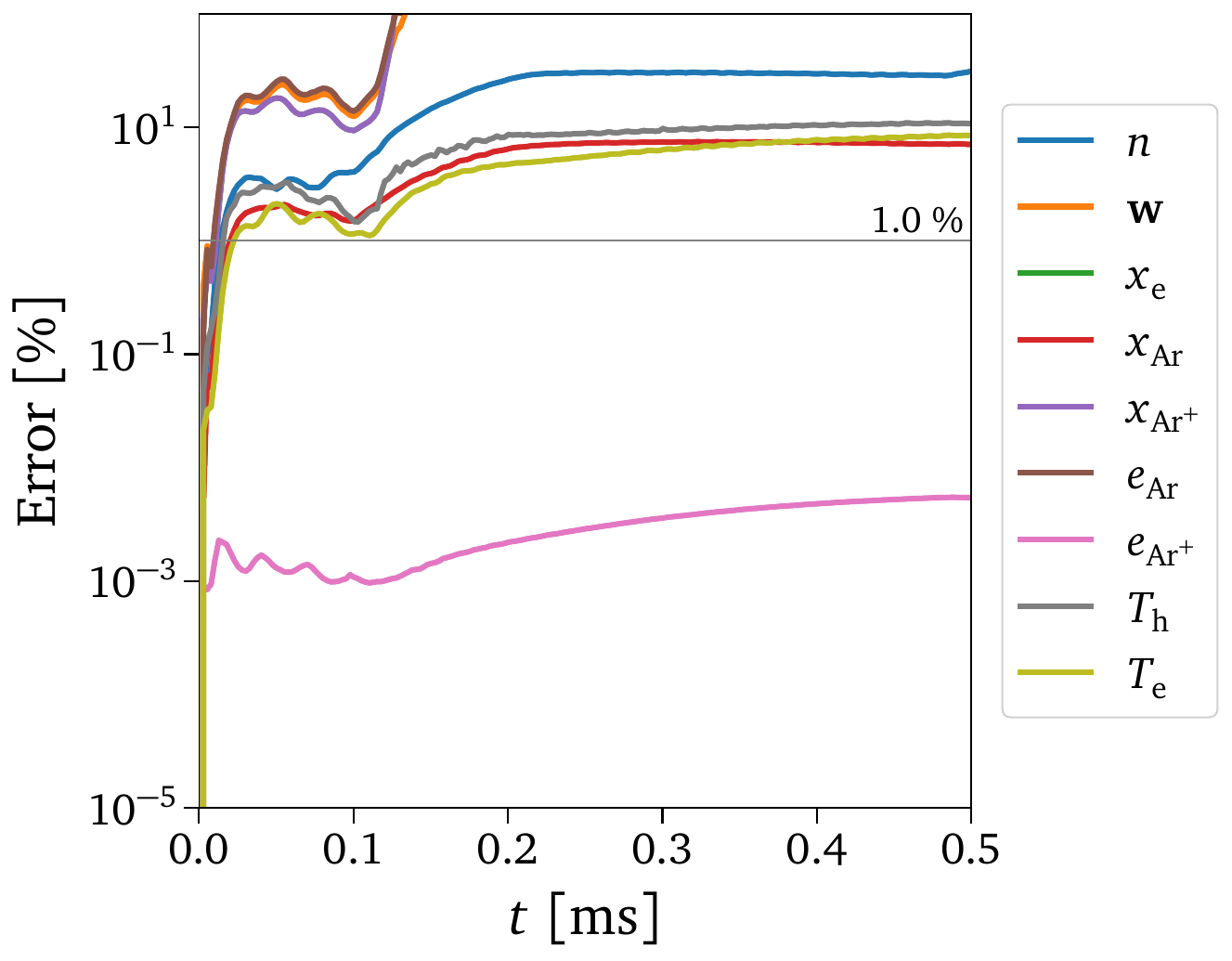}
\end{subfigure}
\caption{\textit{Error evolution of ROMs in 2D simulations.}
Space‐averaged relative errors (\%) for CoBRAS (left panel) and POD (right panel) ROMs with $r = 9$, computed for key quantities of interest.}
\label{fig:2d.err_evol}
\end{figure}

As in the 1D case, a quantitative assessment is presented in figure~\ref{fig:2d.err_evol}, which shows the space-averaged relative error for the same set of thermochemical variables. The CoBRAS-based ROM maintains relative errors below 10\%—and below 1\% for most quantities—throughout the simulation. In contrast, POD becomes unstable, with rapidly growing errors beyond $t = 0.15$~ms. This comparison highlights the robustness and accuracy of the CoBRAS framework in handling highly unsteady, nonlinear, and multidimensional plasma flows, regimes in which conventional projection-based ROMs like POD often fail due to the lack of physically consistent basis selection.

\section{Conclusions}\label{sec:conclusions}
In this work, we developed and assessed a Petrov-Galerkin reduced-order model (ROM) for collisional-radiative (CR) argon plasma kinetics. The framework builds upon the recent work of Zanardi et al.~\cite{Zanardi2025Petrov-GalerkinMixtures}, who applied the ``Covariance Balancing Reduction using Adjoint Snapshots'' (CoBRAS) method~\cite{Otto2023ModelCovariance} to detailed state-to-state kinetics in thermochemical nonequilibrium. ROM construction was performed in a zero-dimensional (0D) setting by computing a linear projection operator through the balancing of state and gradient covariances associated with the full-order nonlinear system. The resulting ROM was then coupled with a finite-volume solver to simulate multidimensional, unsteady, ionizing shock flows in both one- and two-dimensional domains. We benchmarked the CoBRAS-based ROM against full-order model (FOM) simulations and compared its performance with standard proper orthogonal decomposition (POD) methods across a wide range of conditions.

The 0D results demonstrated that CoBRAS delivers substantial reductions in computational complexity—exceeding an order of magnitude in FLOP savings—while maintaining high accuracy. Relative errors remained consistently below 1\% for macroscopic quantities and around 2\% for the fine-scale atomic state distribution functions (ASDFs). In the 1D simulations, the ROM accurately captured periodic shock-induced oscillations and spatially localized features such as electron avalanches and induction zones, achieving a 3$\times$ reduction in state dimension with space-averaged errors consistently below 1\%. The method remained robust in highly unsteady regimes where POD-based ROMs exhibited instability and rapid error growth. In the more challenging 2D simulations, CoBRAS maintained its accuracy, successfully reproducing complex multidimensional structures, including shock-shock interactions, triple points, and ionization-driven cellular patterns. These results confirm that the CoBRAS projection framework enables efficient model reduction while preserving the essential physical mechanisms required for predictive, high-fidelity plasma simulations.

While the proposed ROM framework already demonstrates substantial gains in efficiency and accuracy, several avenues for improvement remain.
One limitation is the lack of a built-in positivity constraint on the ASDF, which can result in small, nonphysical negative populations during early transients (e.g., \(t < 10^{-8}\) s), as previously observed by Zanardi et al.~\cite{Zanardi2025Petrov-GalerkinMixtures}.
Future developments could incorporate positivity-preserving projection techniques or constrained optimization formulations to enhance robustness and physical consistency.
Moreover, integrating the ROM into uncertainty quantification (UQ) pipelines and Bayesian inference frameworks would enable efficient exploration of large parametric spaces, especially in extrapolative regimes.

More broadly, this work contributes to the growing need for physically grounded, numerically stable reduced-order models capable of capturing multiscale, nonlinear interactions in reactive, radiating flows. Such flows arise in a wide range of fluid mechanics contexts—from plasma–shock interactions in hypersonic boundary layers to radiative magnetohydrodynamics in astrophysical jets—where high-fidelity full-order models remain prohibitively expensive. The proposed methodology offers a systematic approach to accelerating these simulations without sacrificing physical fidelity, enabling new opportunities in both fundamental fluid dynamics research and high-impact engineering applications.

\section*{Acknowledgements}
I.\ Zanardi, A.\ Meini, and M.\ Panesi acknowledge support from the Vannevar Bush Faculty Fellowship OUSD(RE), under Grant \#N00014-21-1-295. A. Padovan was supported by the Center for Hypersonics and Entry Systems Studies through internal funding provided by M. Panesi and K. Stephani via The Grainger College of Engineering at the University of Illinois Urbana-Champaign. The views and conclusions expressed in this work are those of the authors and do not necessarily reflect the official policies or endorsements, either expressed or implied, of the United States Government.

\appendix
\section{Condition number of the BDF scheme}\label{app:cond.num}
We consider a general system of ordinary differential equations (ODEs) of the form:
\begin{equation} \label{eq:ode}
    \frac{d}{dt}\vq(t) = \vf(t,\vq(t))\eqspace,
    \quad \vq(0) = \vq_0\eqspace,
\end{equation}
where $\vq(t) \in \mathbb{R}^N$ is the state vector and $\vf: \mathbb{R} \times \mathbb{R}^N \to \mathbb{R}^N$ is a smooth nonlinear function defining the system dynamics. To numerically solve equation~\eqref{eq:ode}, we employ linear multistep methods~\cite{Radhakrishnan_LSODE_1993}. At time step $n$, the general $k$-step formulation is given by:
\begin{equation}
\vq_n = \sum_{j=1}^{k_1} \alpha_j \vq_{n-j} + h_n \sum_{j=0}^{k_2} \beta_j \mathbf{f}\left(t_{n-j}, \vq_{n-j}\right)\eqspace,
\end{equation}
where $h_n$ is the time step size at iteration $n$, and $\alpha_j$, $\beta_j$ are method-specific coefficients. Choosing $k_1 = p$ and $k_2 = 0$ yields the $p$-th order backward differentiation formula (BDF), which takes the form:
\begin{equation}
\vq_n = \sum_{j=1}^p \alpha_j \vq_{n-j} + h_n \beta_0 \mathbf{f}\left(t_n, \vq_n\right)\eqspace.
\end{equation}
This is an implicit method, requiring the solution of a nonlinear system at each time step. We solve this system using the Newton-Raphson (NR) iteration. At iteration $m$ of time step $n$, the NR update solves the linear system:
\begin{align}
\mathbf{P}\left(\vq_n^{[m+1]} - \vq_n^{[m]}\right)
&= -\mathbf{r}\left(\vq_n^{[m]}\right) \nonumber \\
&= \sum_{j=1}^p \alpha_j \vq_{n-j} + h_n \beta_0 \mathbf{f}\left(\vq_n^{[m]}\right) - \vq_n^{[m]}\eqspace,
\end{align}
where $\vq_n^{[m]}$ is the $m$-th NR iterate, and $\mathbf{r}(\vq_n^{[m]})$ is the residual of the BDF equation evaluated at $\vq_n^{[m]}$. The Jacobian matrix $\mathbf{P}$ of the residual with respect to $\vq$ is given by:
\begin{equation}\label{eq:p.mat}
\mathbf{P} = \frac{\partial \mathbf{r}}{\partial \vq} = \mathbf{I} - h_n \beta_0 \mathbf{J}\eqspace,
\end{equation}
where $\mathbf{I}$ is the identity matrix, and $\mathbf{J} = \partial \vf / \partial \vq$ is the Jacobian of the system dynamics with respect to the state vector. The condition number of the matrix $\mathbf{P}$ is defined as:
\begin{equation}
\kappa(\mathbf{P}) = \frac{|\lambda_{\max}|}{|\lambda_{\min}|}\eqspace,
\end{equation}
where $\lambda_{\max}$ and $\lambda_{\min}$ denote the eigenvalues of $\mathbf{P}$ with the largest and smallest magnitudes, respectively. Since the time step $h_n$ varies over the course of the integration and only approximate condition number estimates are needed, it is convenient to define a continuous, time-dependent version of $\mathbf{P}$ as:
\begin{equation}\label{eq:p.mat.time}
\mathbf{P}(t, \vq) = \mathbf{I} - t \beta_0 \mathbf{J}(t, \vq),
\end{equation}
which mirrors the structure of equation~\eqref{eq:p.mat} by interpreting $t$ as a characteristic time scale of the integration step.


\bibliography{biblio/global,biblio/hypersonics,biblio/rom,biblio/ml_pde}

\clearpage

\renewcommand*{\thetable}{S\arabic{table}}
\renewcommand*{\thefigure}{S\arabic{figure}}
\renewcommand*{\thesection}{S.\arabic{section}}
\renewcommand*{\thesubsection}{\thesection.\arabic{subsection}}
\renewcommand*{\thesubsubsection}{\thesubsection.\arabic{subsubsection}}
\renewcommand*{\theequation}{\thesection.\arabic{equation}}

\setcounter{section}{0}
\setcounter{page}{1}
\setcounter{figure}{0}
\setcounter{equation}{0}


\makeatletter
\let\@title\@empty
\makeatother

\title{Supplementary Material to \\ ``\textit{\fulltitle}''}

\def\ps@pprintTitle{%
     \let\@oddhead\@empty
     \let\@evenhead\@empty
     \def\@oddfoot{\footnotesize\itshape
        Supplementary Data for \ifx\@journal\@empty Elsevier
       \else\@journal\fi\hfill\today}%
     \let\@evenfoot\@oddfoot}
\makeatother

\begin{abstract}
    Accurate simulation of nonequilibrium plasmas—central to applications in hypersonic re-entry, fusion energy, and astrophysical flows—requires state-specific collisional–radiative (CR) kinetic models,  but the associated computational cost is often prohibitive.
While traditional approaches reduce this burden through empirical or physics-based simplifications, they frequently compromise accuracy in strongly nonequilibrium regimes.
To address these limitations, we develop a Petrov-Galerkin reduced-order model (ROM) for CR argon plasma based on oblique projections that optimally balance the covariance of full-order state trajectories with that of the system's output sensitivities.
This construction ensures that the ROM captures both the dominant energetic modes and the directions most relevant to input-output behavior.
After offline training in a zero-dimensional setting using nonlinear forward and adjoint simulations, the ROM is coupled to a finite-volume solver and applied to one- (1D) and two-dimensional (2D) ionizing shock-tube problems.
The ROM achieves a 3$\times$ reduction in state dimension and more than one order of magnitude savings in floating-point operations, while maintaining errors below 1\% for macroscopic quantities.
In both 1D and 2D, it robustly reproduces complex unsteady plasma flow features—such as periodic fluctuations, electron avalanches, triple points, and cellular ionization patterns—in contrast to standard ROM strategies, which become unstable or inaccurate under these challenging conditions.
These results demonstrate the potential of the proposed model reduction strategy to enable high-fidelity simulation of reactive plasma flows at reduced computational cost, offering new capabilities for exploring high-speed fluid systems governed by coupled transport, wave propagation, and kinetic nonequilibrium.

\end{abstract}
\begin{keyword}
    Reduced-order modeling \sep Thermochemistry \sep Plasma
\end{keyword}

\maketitle

\clearpage
\section{Zero‐dimensional simulations}\label{suppl:0d}
\begin{table}[!htb]
    \centering
    \begin{tabular}{c|ccc|ccc}
        \arrayrulecolor{black}\cmidrule{2-7}
        & \multicolumn{6}{c}{Testing Error [\%] - Dataset 2} \\
        \arrayrulecolor{black}\cmidrule{2-7}
        & \multicolumn{3}{c|}{CoBRAS} & \multicolumn{3}{c}{POD} \\
        \arrayrulecolor{black}\midrule
        $r$ & 8 & 9 & 10 & 8 & 9 & 10 \\
        \arrayrulecolor{black}\midrule
        $n$ & \textbf{0.043} & \textbf{0.004} & \textbf{0.001} & 0.237 & 0.136 & 0.165 \\
        $\vw$ & \textbf{2.331} & \textbf{1.426} & \textbf{1.697} & 4.694 & 2.887 & 3.268 \\
        $x_{\mathrm{e}}$ & \textbf{0.311} & \textbf{0.030} & \textbf{0.011} & 2.643 & 1.167 & 1.399 \\
        $x_{\text{Ar}}$ & \textbf{0.087} & \textbf{0.008} & \textbf{0.002} & 0.552 & 0.230 & 0.233 \\
        $x_{\text{Ar}^+}$ & \textbf{0.268} & \textbf{0.036} & \textbf{0.014} & 2.805 & 0.582 & 1.126 \\
        $e_{\text{Ar}}$ & \textbf{0.400} & \textbf{0.069} & \textbf{0.028} & 2.956 & 1.354 & 1.270 \\
        $e_{\text{Ar}^+}$ & \textbf{0.004} & \textbf{0.002} & \textbf{0.002} & 0.028 & 0.015 & 0.016 \\
        $T_\mathrm{h}$ & \textbf{0.034} & \textbf{0.005} & \textbf{0.002} & 0.248 & 0.103 & 0.097 \\
        $T_\mathrm{e}$ & \textbf{0.11} & \textbf{0.047} & \textbf{0.051} & 0.325 & 0.116 & 0.051 \\
        \arrayrulecolor{black}\midrule
    \end{tabular}
    \caption{\textit{Testing error for dataset 2 of table \ref{table:test.dsets}.} Mean relative errors (\%) for the quantities of interest, computed using CoBRAS and POD ROMs with varying reduced dimensions $r$ (total reduced state dimension $d = r + 3$, including the electron mass fraction and the two temperatures). Results are compared against the FOM solutions. Bold values indicate the lower error between CoBRAS and POD for each entry.}
    \label{table:err.dset2}
\end{table}

\begin{table}[!htb]
    \centering
    \begin{tabular}{c|ccc|ccc}
        \arrayrulecolor{black}\cmidrule{2-7}
        & \multicolumn{6}{c}{Testing Error [\%] - Dataset 3} \\
        \arrayrulecolor{black}\cmidrule{2-7}
        & \multicolumn{3}{c|}{CoBRAS} & \multicolumn{3}{c}{POD} \\
        \arrayrulecolor{black}\midrule
        $r$ & 8 & 9 & 10 & 8 & 9 & 10 \\
        \arrayrulecolor{black}\midrule
        $n$ & \textbf{0.001} & \textbf{0.000} & \textbf{0.000} & 0.004 & 0.013 & 0.010 \\
        $\vw$ & \textbf{7.012} & \textbf{6.576} & \textbf{8.897} & 18.853 & 11.694 & 20.586 \\
        $x_{\mathrm{e}}$ & \textbf{0.183} & \textbf{0.14} & \textbf{0.182} & 2.767 & 3.546 & 1.651 \\
        $x_{\text{Ar}}$ & \textbf{0.002} & \textbf{0.001} & \textbf{0.000} & 0.011 & 0.014 & 0.011 \\
        $x_{\text{Ar}^+}$ & \textbf{0.424} & \textbf{0.422} & \textbf{0.589} & 4.642 & 3.909 & 5.354 \\
        $e_{\text{Ar}}$ & \textbf{0.752} & \textbf{0.275} & \textbf{0.321} & 4.128 & 3.155 & 2.367 \\
        $e_{\text{Ar}^+}$ & \textbf{32.844} & \textbf{3.439} & \textbf{2.033} & 34.969 & 17.771 & 22.201 \\
        $T_\mathrm{h}$ & \textbf{0.005} & \textbf{0.002} & \textbf{0.001} & 0.030 & 0.028 & 0.019 \\
        $T_\mathrm{e}$ & \textbf{2.931} & \textbf{3.686} & \textbf{3.286} & 19.962 & 31.797 & 26.803 \\
        \arrayrulecolor{black}\midrule
    \end{tabular}
    \caption{\textit{Testing error for dataset 3 of table \ref{table:test.dsets}.} Mean relative errors (\%) for the quantities of interest, computed using CoBRAS and POD ROMs with varying reduced dimensions $r$ (total reduced state dimension $d = r + 3$, including the electron mass fraction and the two temperatures). Results are compared against the FOM solutions. Bold values indicate the lower error between CoBRAS and POD for each entry.}
    \label{table:err.dset3}
\end{table}

\begin{figure}[!htb]
\centering
\begin{subfigure}[htb!]{0.32\textwidth}
    \includegraphics[width=\textwidth]{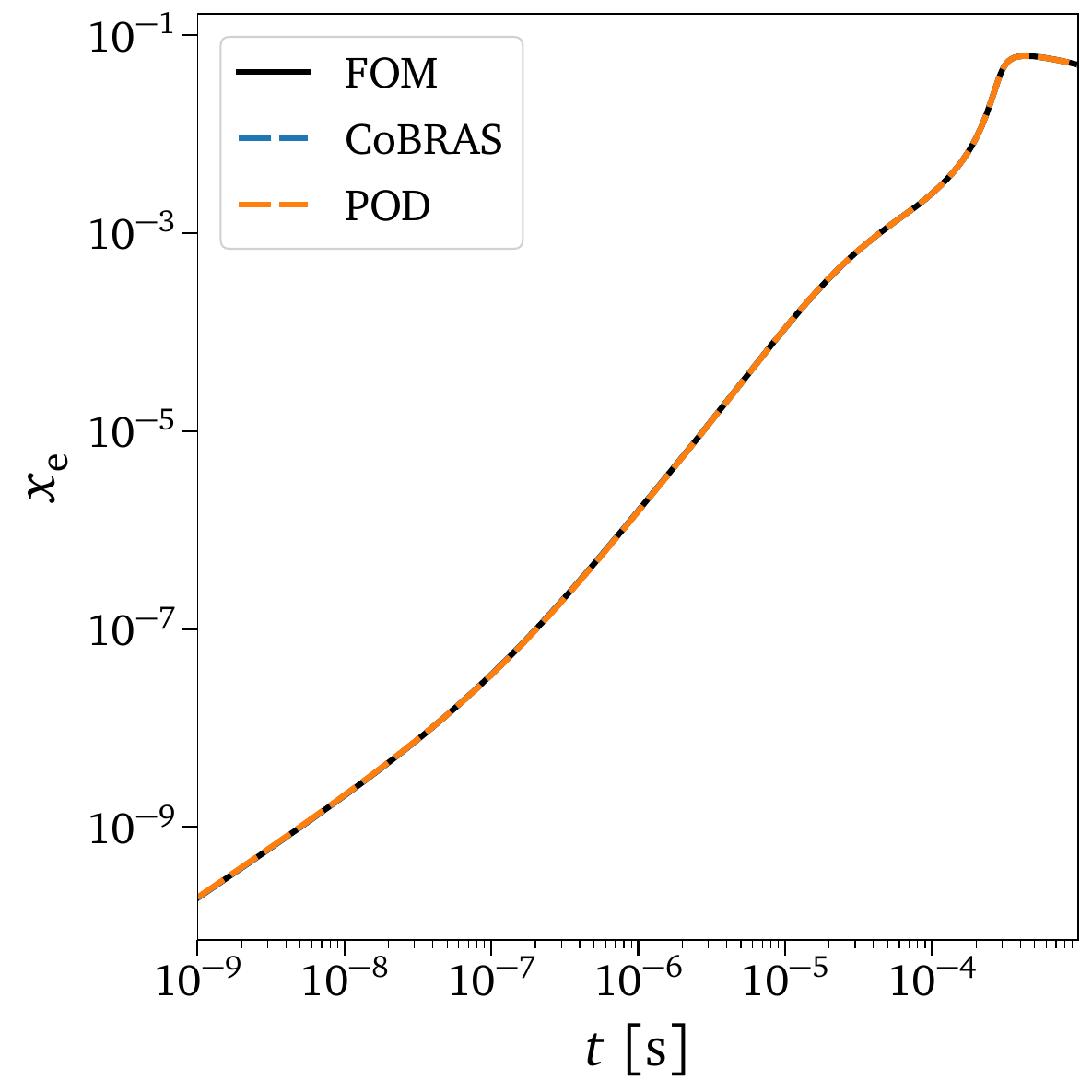}
\end{subfigure}
\begin{subfigure}[htb!]{0.32\textwidth}
    \includegraphics[width=\textwidth]{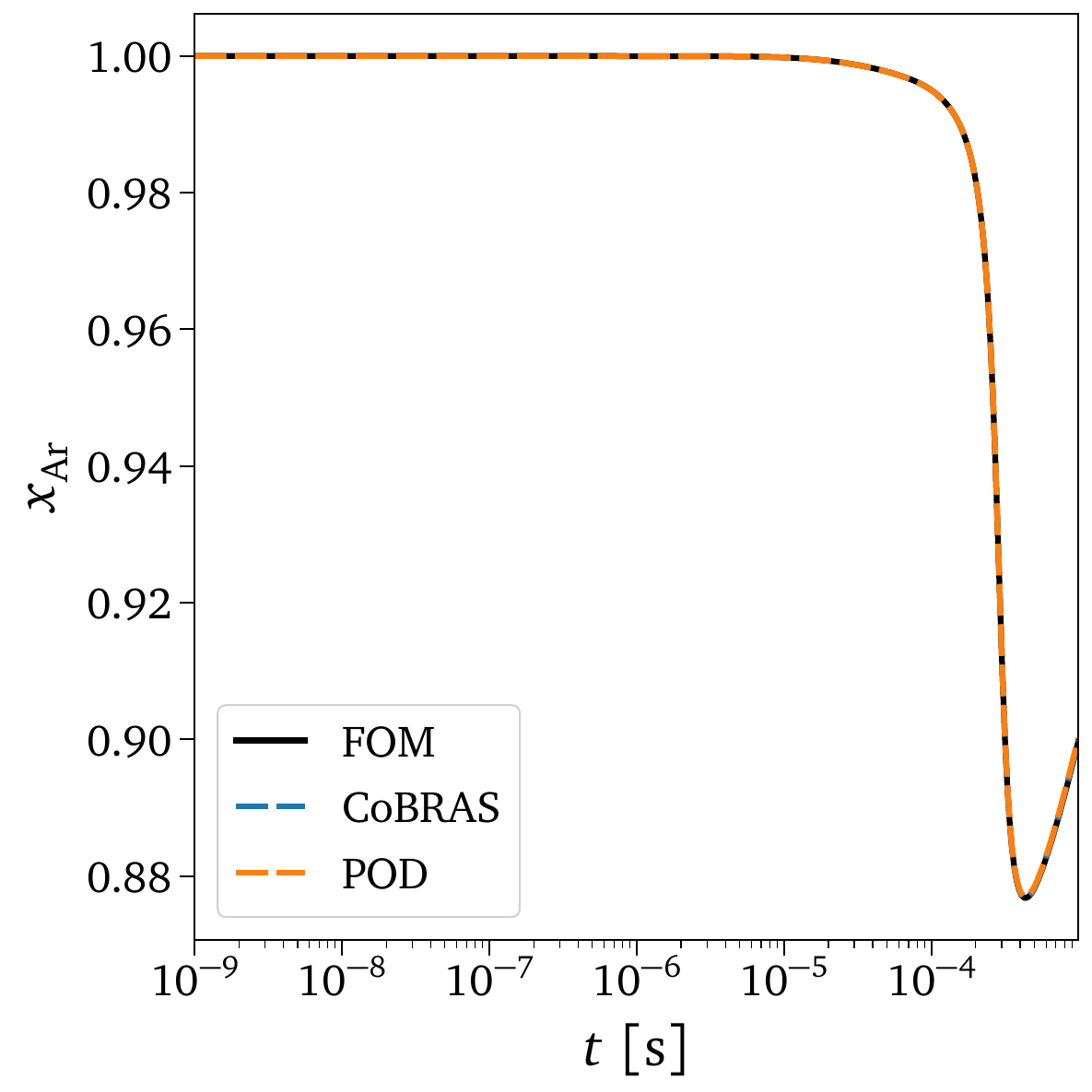}
\end{subfigure}
\begin{subfigure}[htb!]{0.32\textwidth}
    \includegraphics[width=\textwidth]{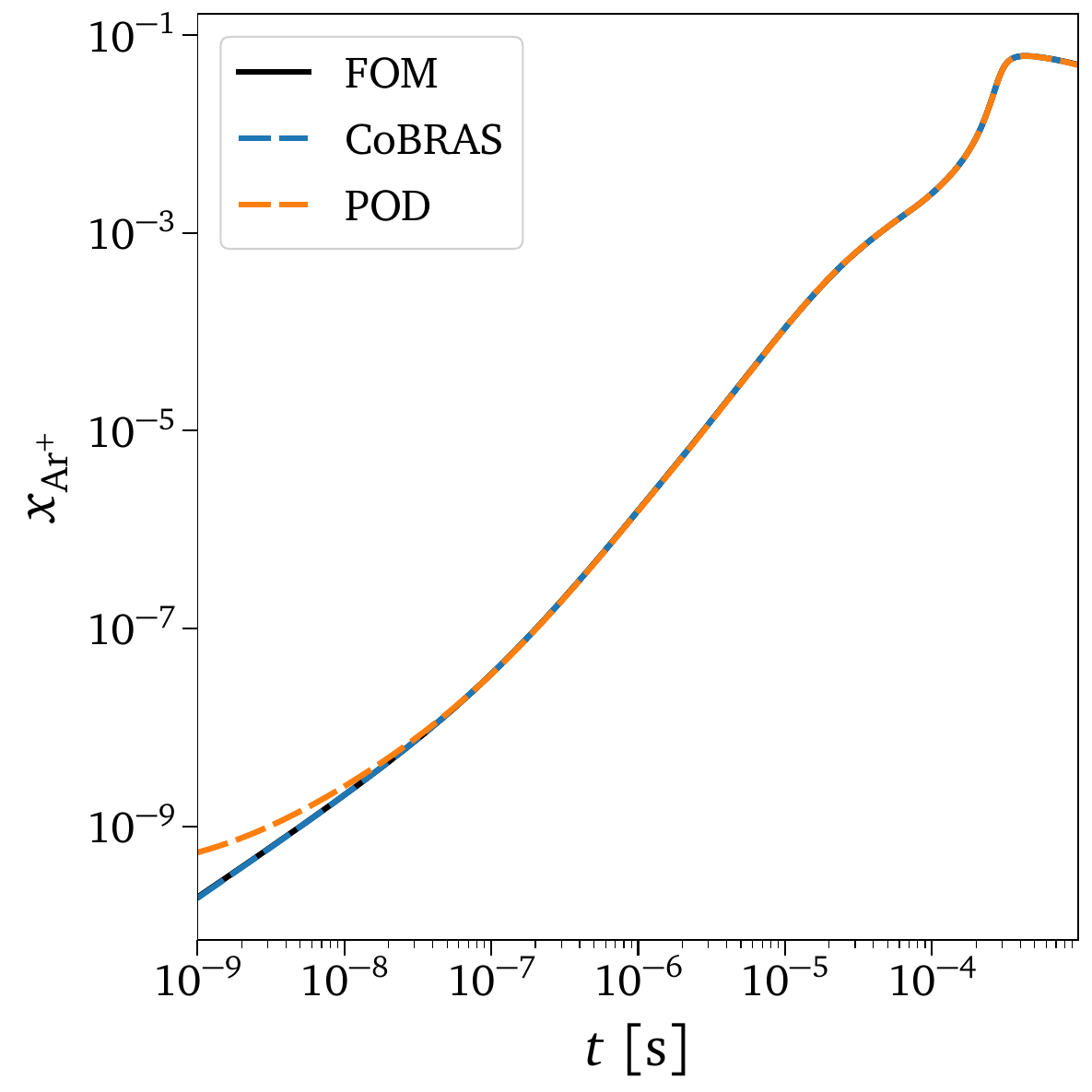}
\end{subfigure}
\\[4pt]
\begin{subfigure}[htb!]{0.32\textwidth}
    \includegraphics[width=\textwidth]{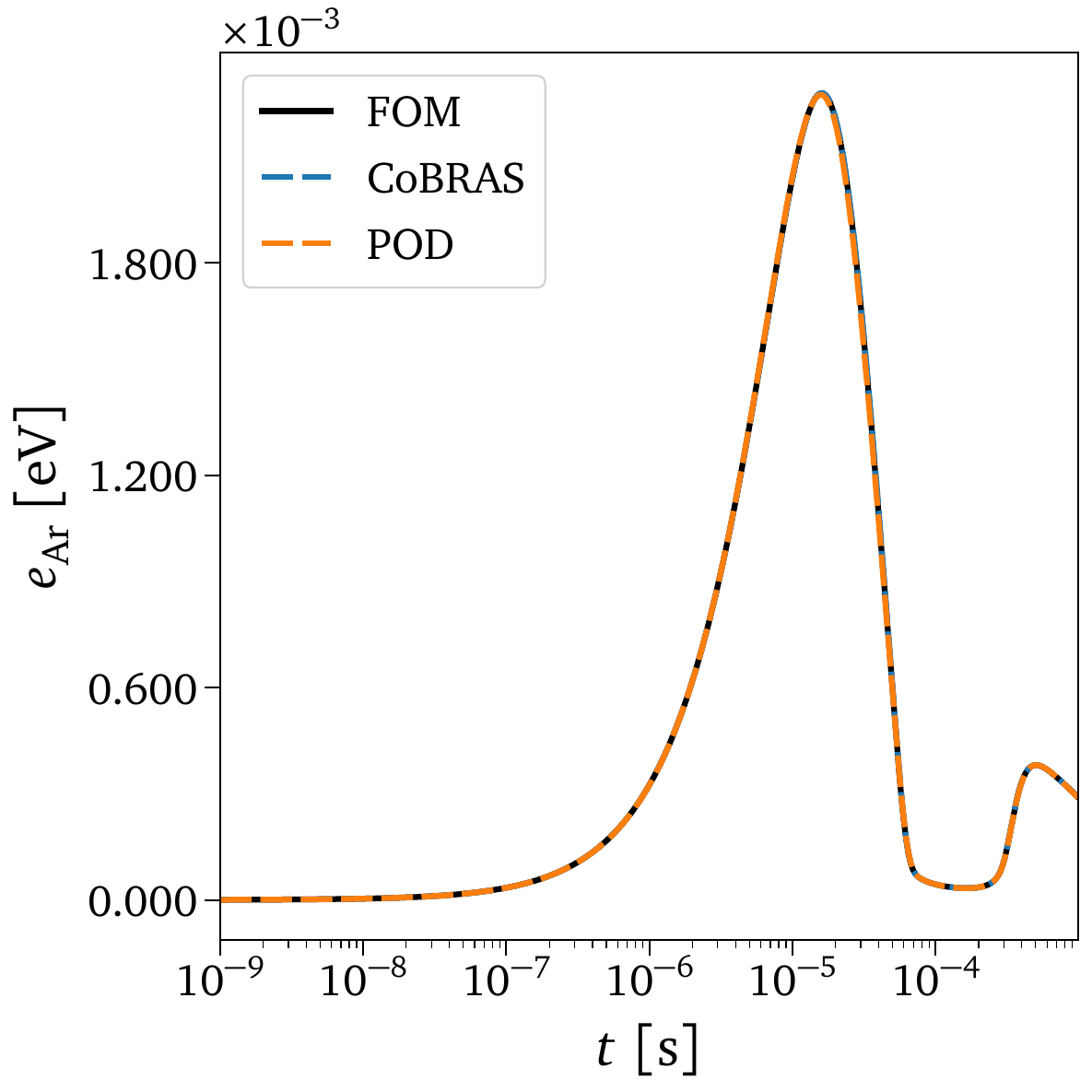}
\end{subfigure}
\begin{subfigure}[htb!]{0.32\textwidth}
    \includegraphics[width=\textwidth]{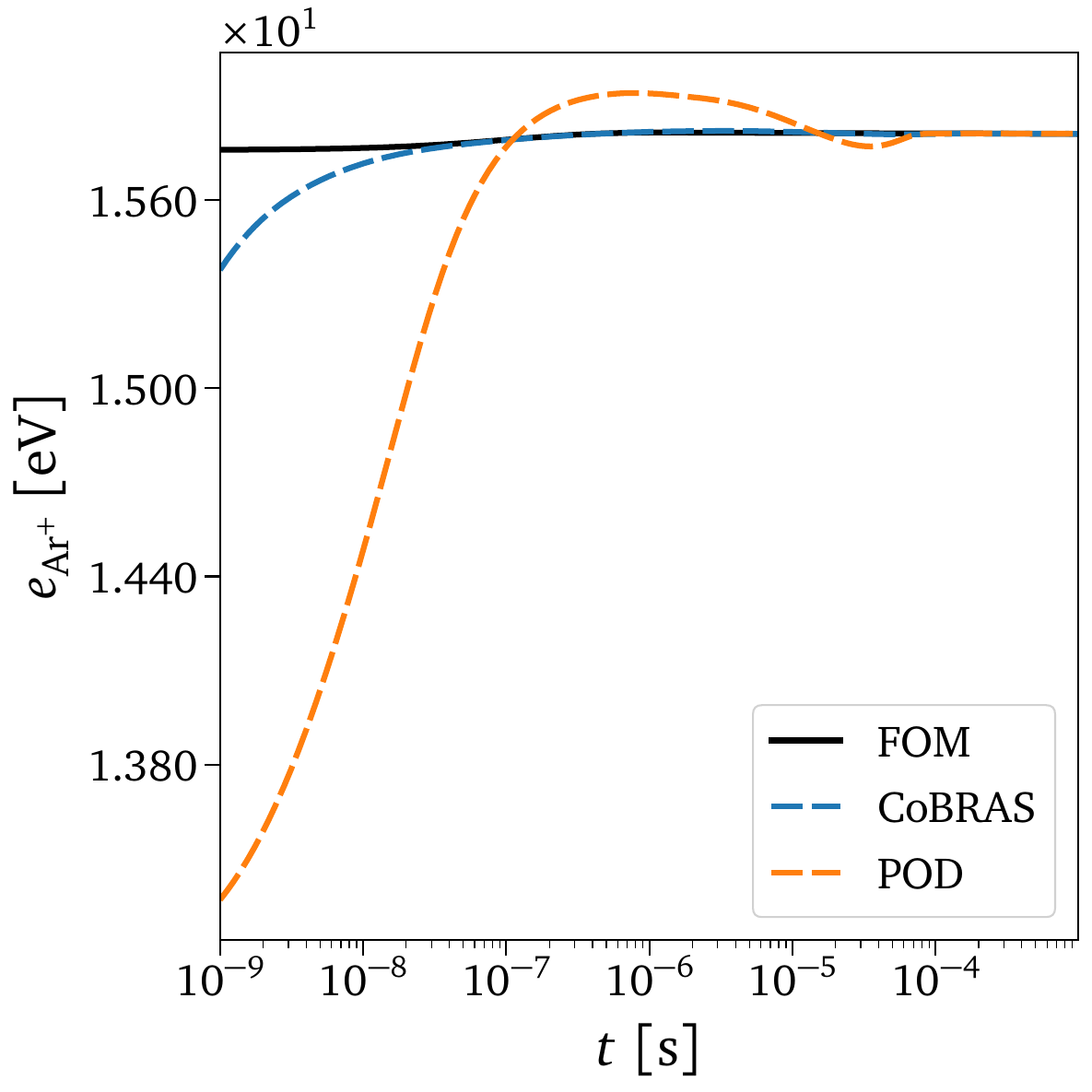}
\end{subfigure}
\begin{subfigure}[htb!]{0.32\textwidth}
    \includegraphics[width=\textwidth]{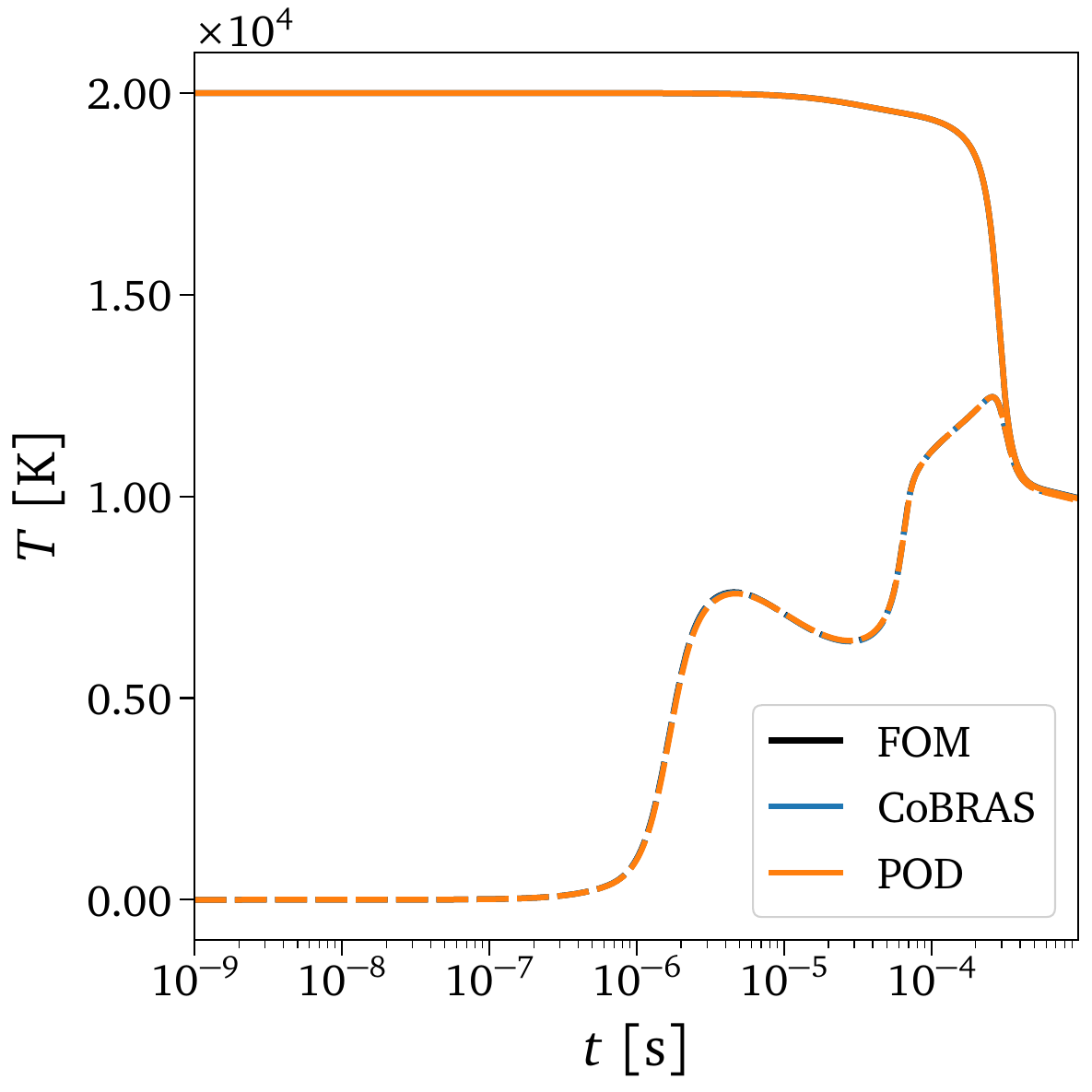}
\end{subfigure}
\caption{\textit{FOM vs. ROMs for 0D simulations: Test case 1.} Time evolution of species zeroth-order moments (molar fractions), first-order moments (internal energies), and temperatures, as predicted by the FOM and by CoBRAS and POD reduced-order models with dimension $r = 8$. Results correspond to test case 1 described in table~\ref{table:testcases}.}
\label{fig:0d.testcase1}
\end{figure}

\begin{figure}[!htb]
\centering
\begin{subfigure}[htb!]{0.32\textwidth}
    \includegraphics[width=\textwidth]{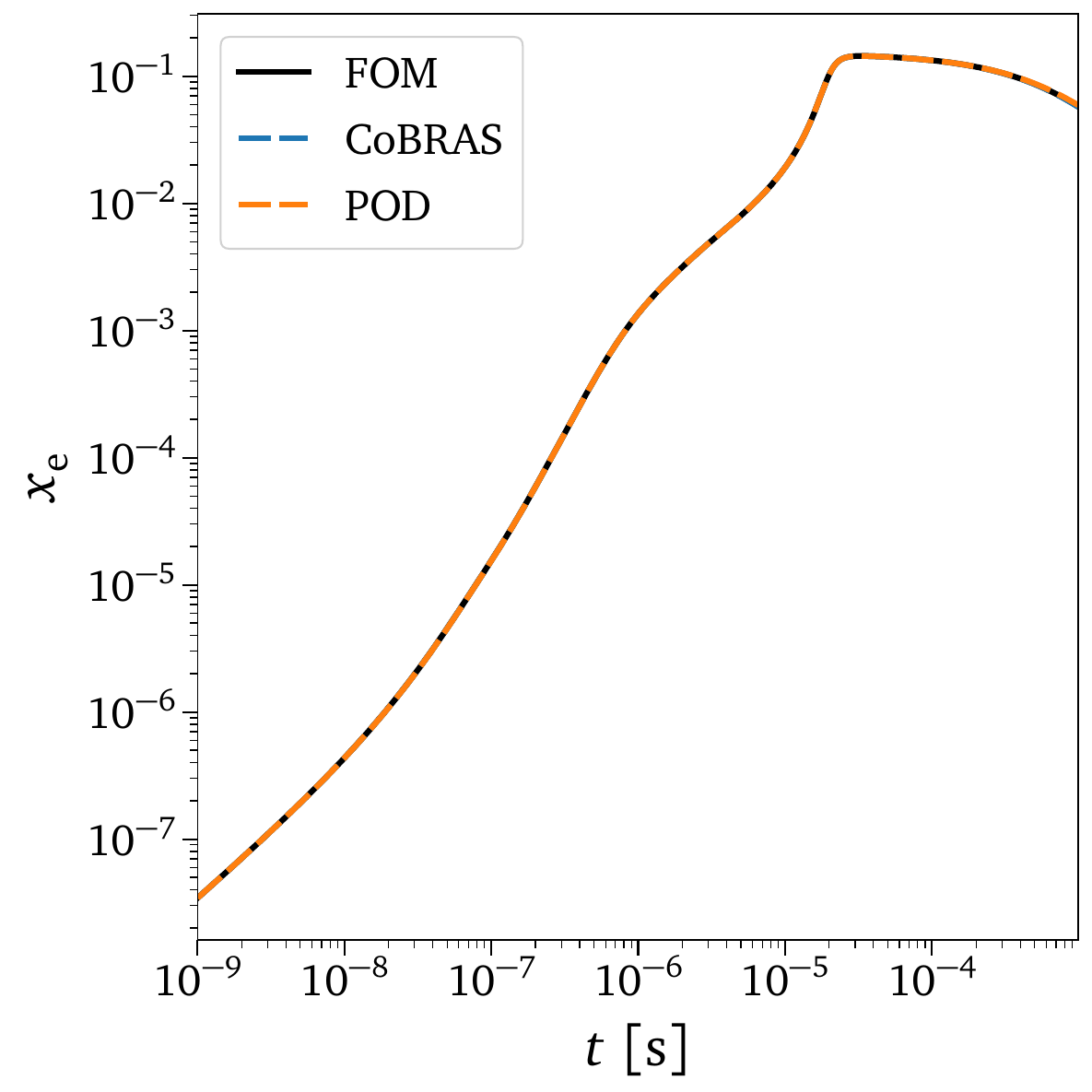}
\end{subfigure}
\begin{subfigure}[htb!]{0.32\textwidth}
    \includegraphics[width=\textwidth]{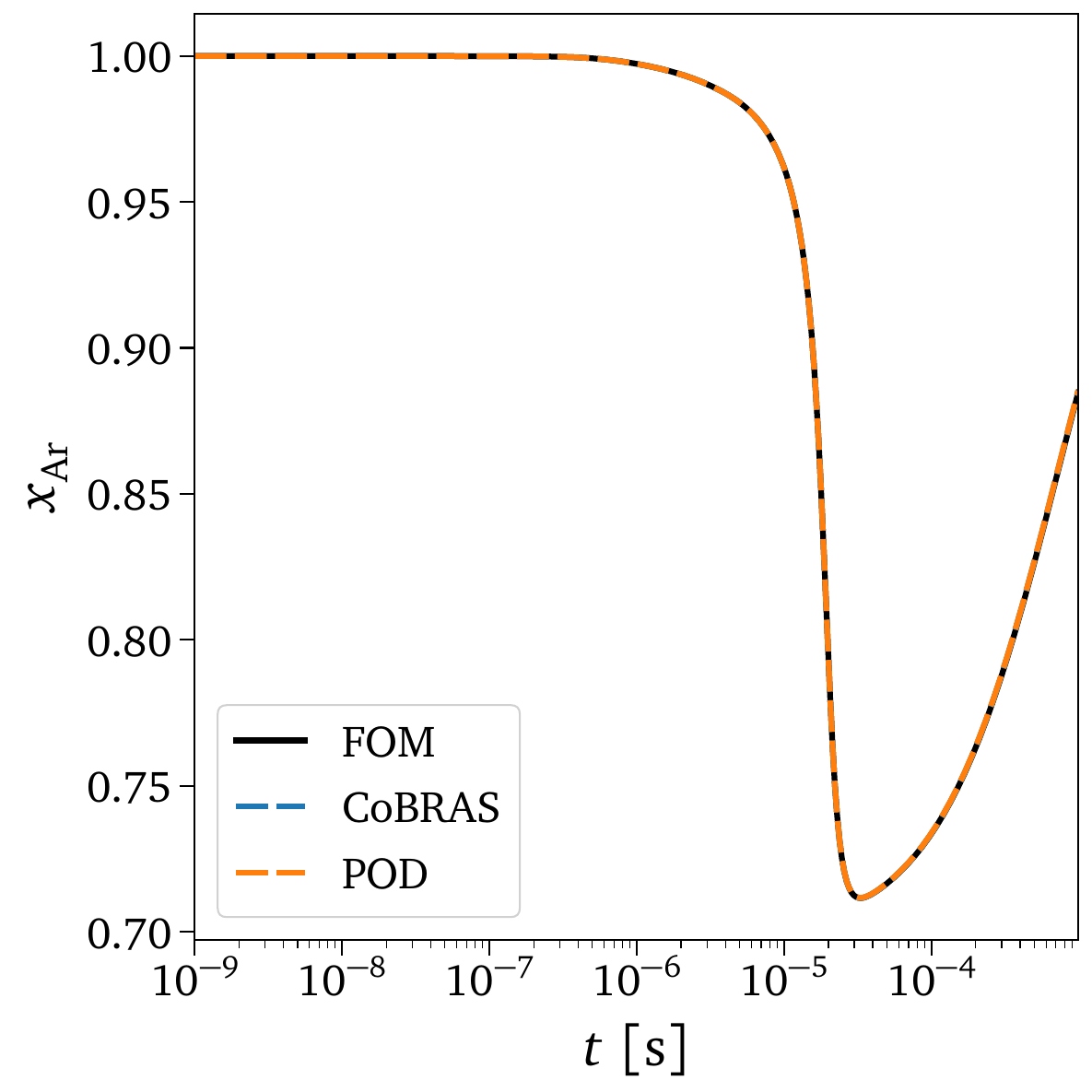}
\end{subfigure}
\begin{subfigure}[htb!]{0.32\textwidth}
    \includegraphics[width=\textwidth]{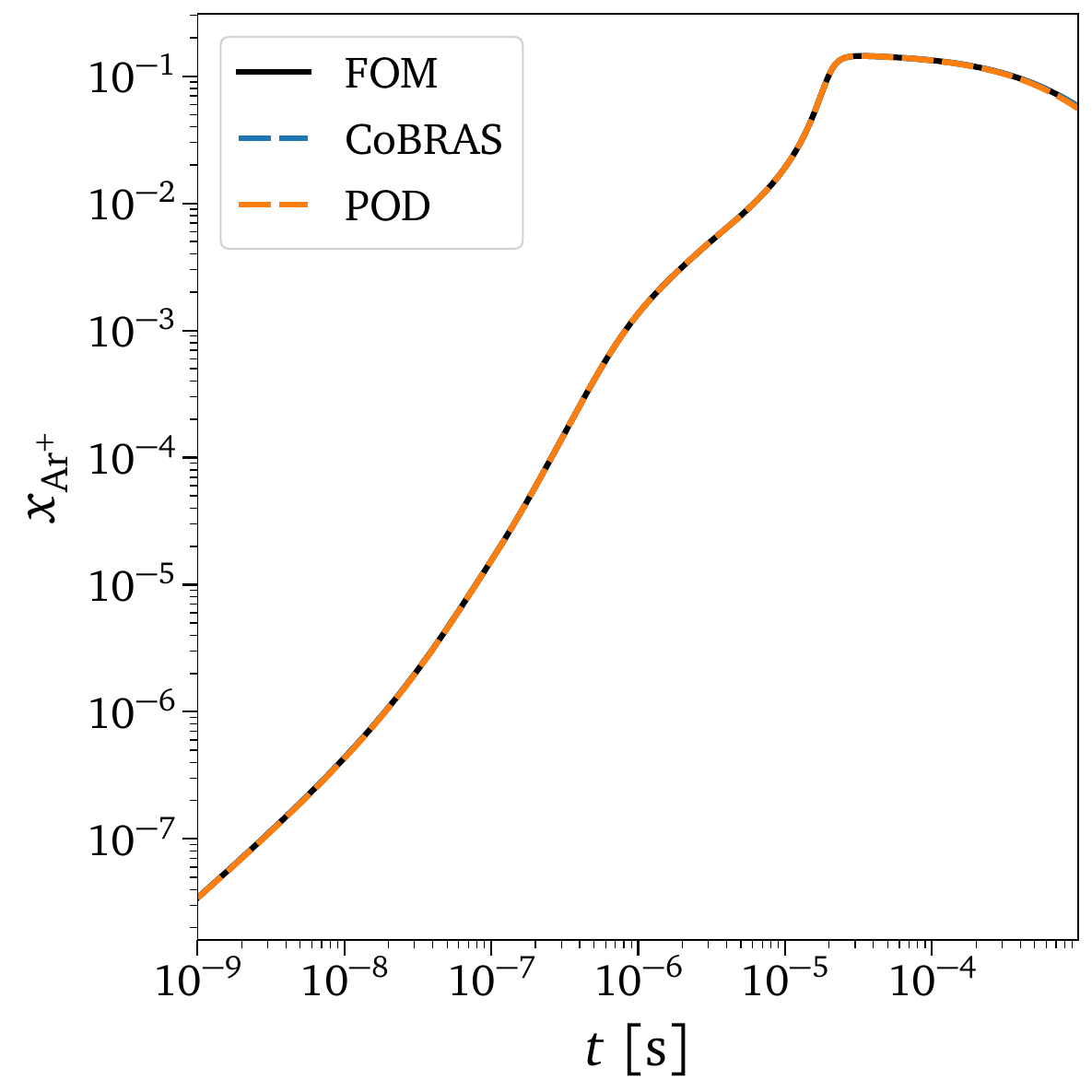}
\end{subfigure}
\\[4pt]
\begin{subfigure}[htb!]{0.32\textwidth}
    \includegraphics[width=\textwidth]{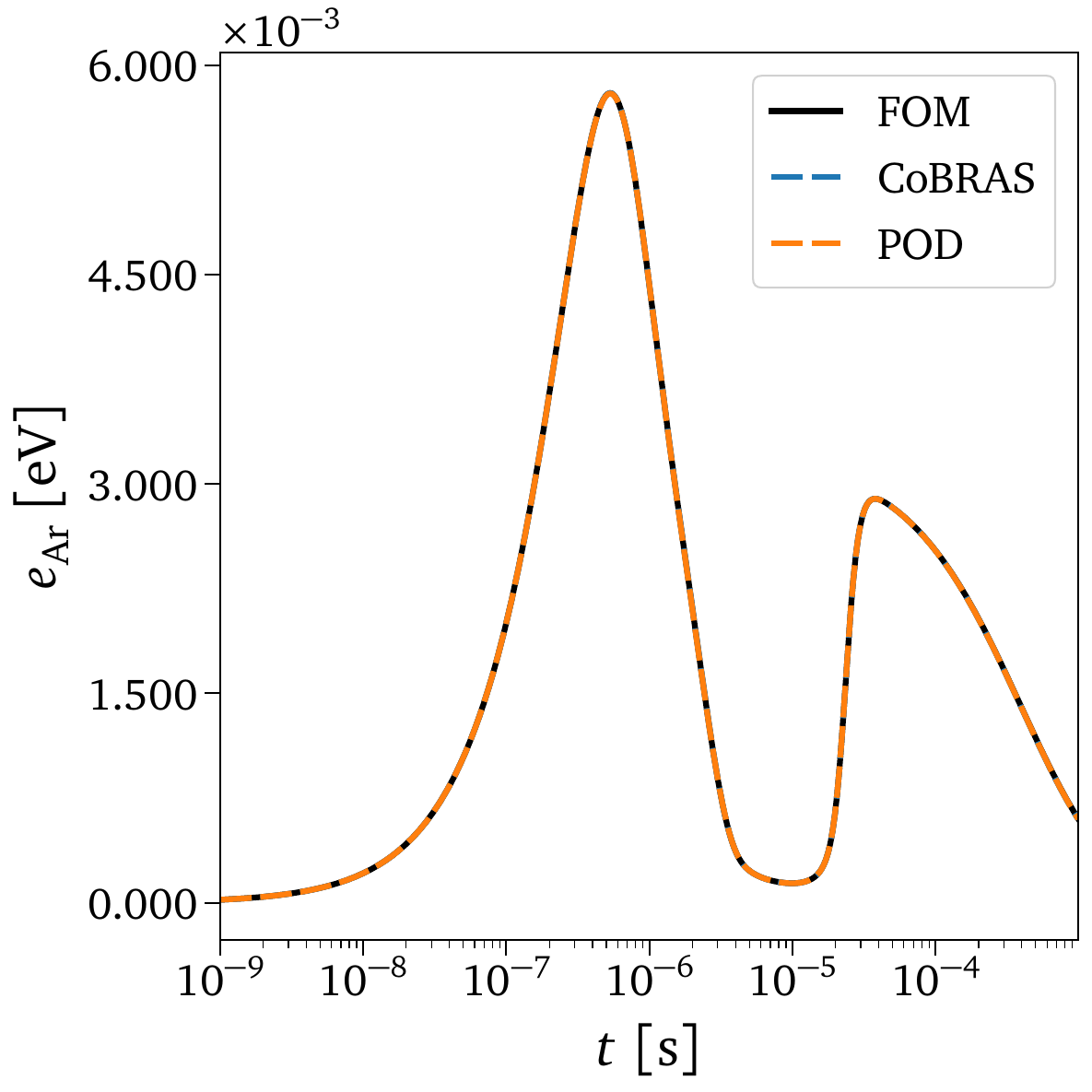}
\end{subfigure}
\begin{subfigure}[htb!]{0.32\textwidth}
    \includegraphics[width=\textwidth]{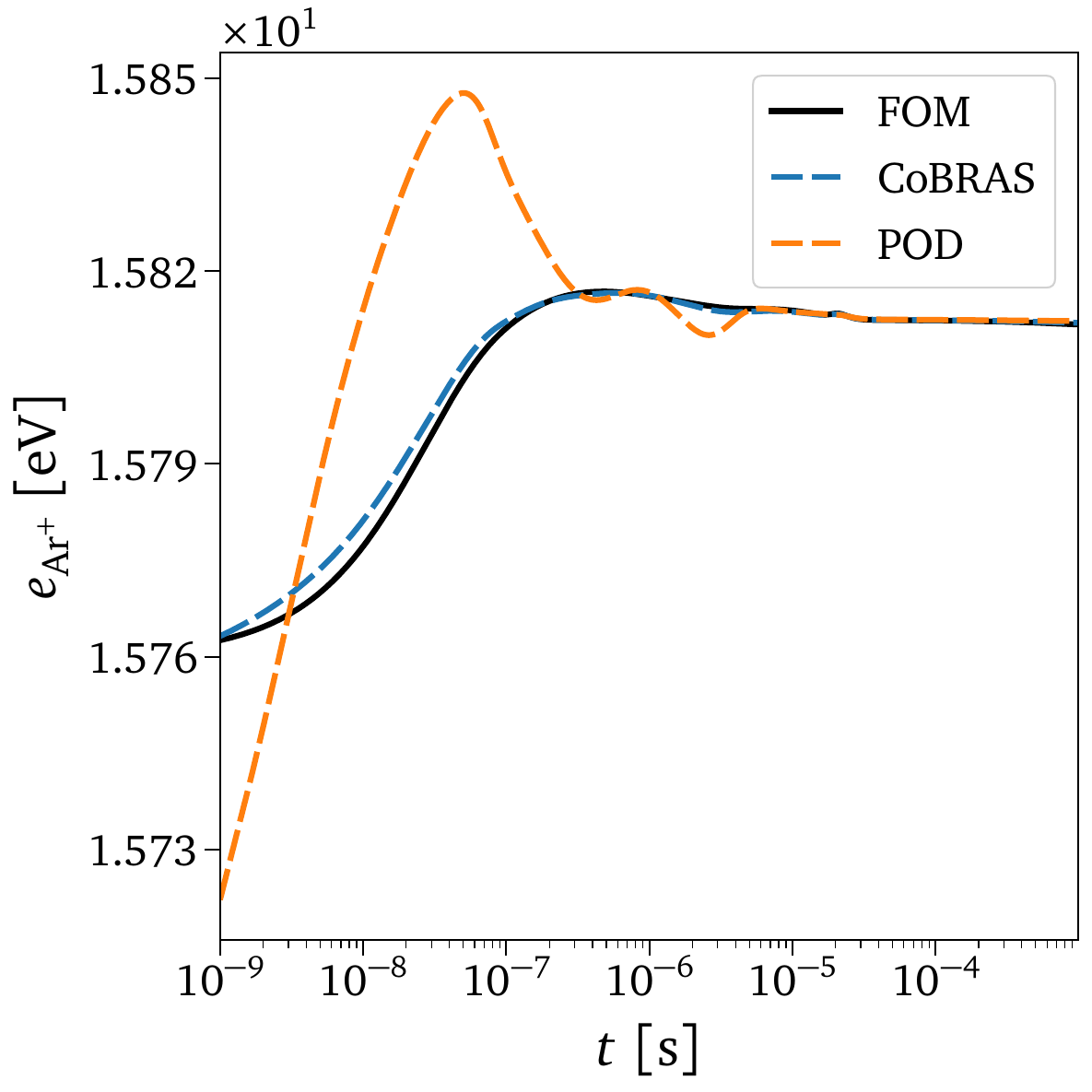}
\end{subfigure}
\begin{subfigure}[htb!]{0.32\textwidth}
    \includegraphics[width=\textwidth]{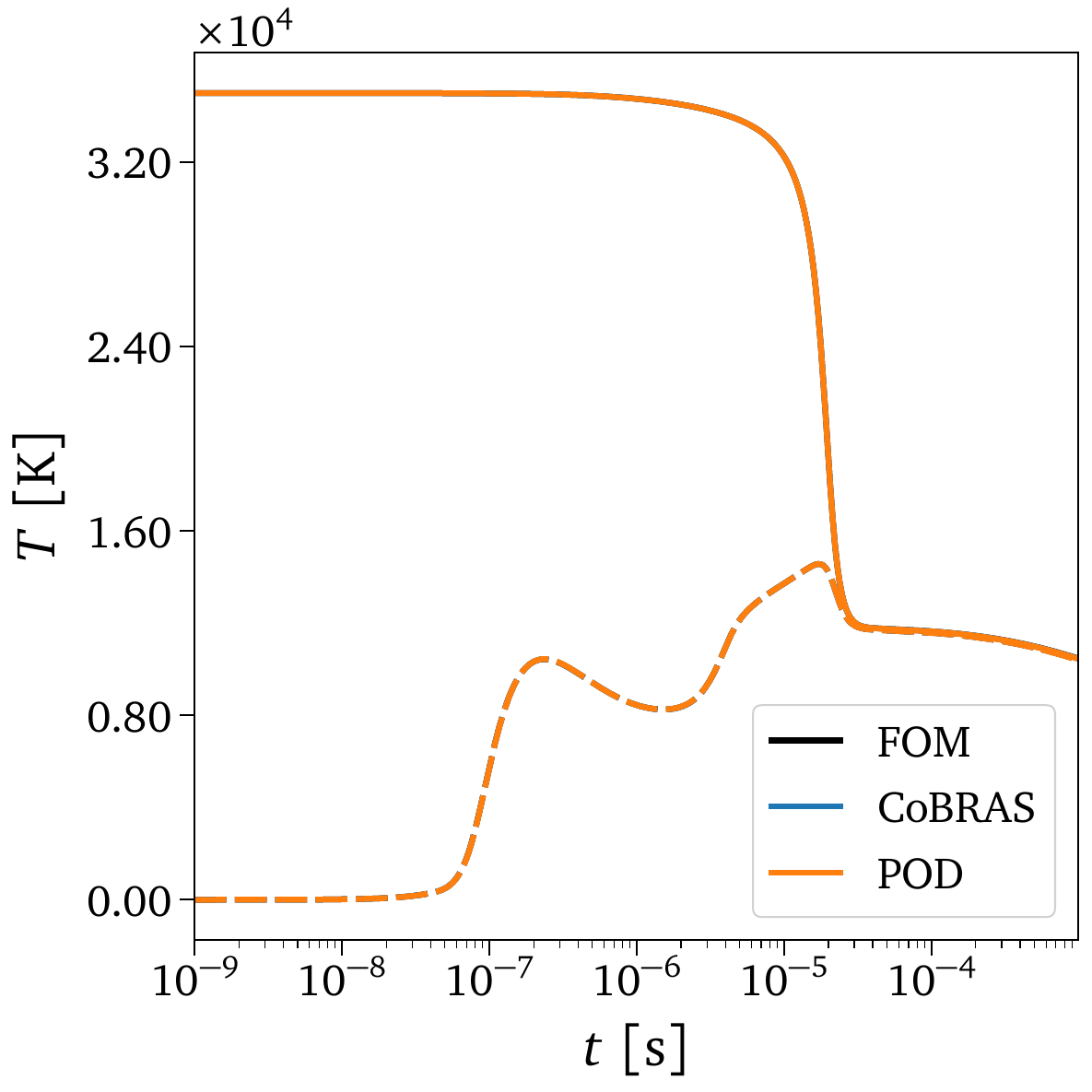}
\end{subfigure}
\caption{\textit{FOM vs. ROMs for 0D simulations: Test case 2.} Time evolution of species zeroth-order moments (molar fractions), first-order moments (internal energies), and temperatures, as predicted by the FOM and by CoBRAS and POD reduced-order models with dimension $r = 8$. Results correspond to test case 2 described in table~\ref{table:testcases}.}
\label{fig:0d.testcase2}
\end{figure}

\clearpage
\section{One‐dimensional simulations}\label{suppl:1d}

\begin{figure}[!htb]
\centering
\includegraphics[width=0.74\textwidth]{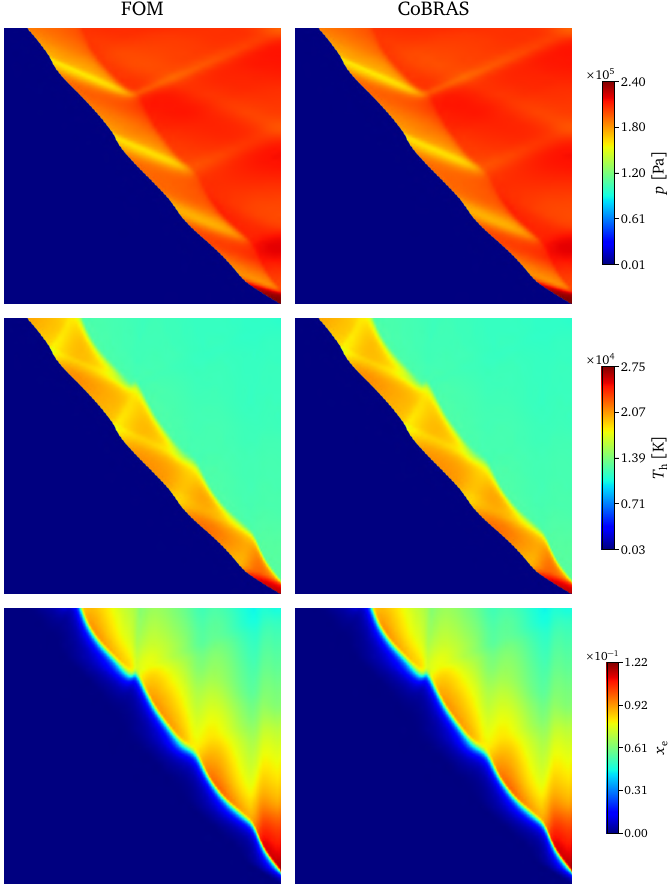}
\caption{\textit{FOM vs. CoBRAS for 1D simulations: Space-time diagrams.} Comparison of FOM and CoBRAS ROM ($r = 9$) for pressure $p$, heavy-particle temperature $T_\mathrm{h}$, and electron molar fraction $x_{\mathrm{e}}$ in space–time ($x$–$y$ plane) diagrams obtained from the 1D simulation. Freestream velocity: $u_\infty = 4\,000$~m/s.}
\label{fig:1d.v3.timespace}
\end{figure}

\begin{figure}[htb!]
\centering
\begin{subfigure}[htb!]{0.4\textwidth}
    \caption*{\hspace{-3mm}\fontfamily{mdbch}\normalsize CoBRAS}
    \includegraphics[width=\textwidth]{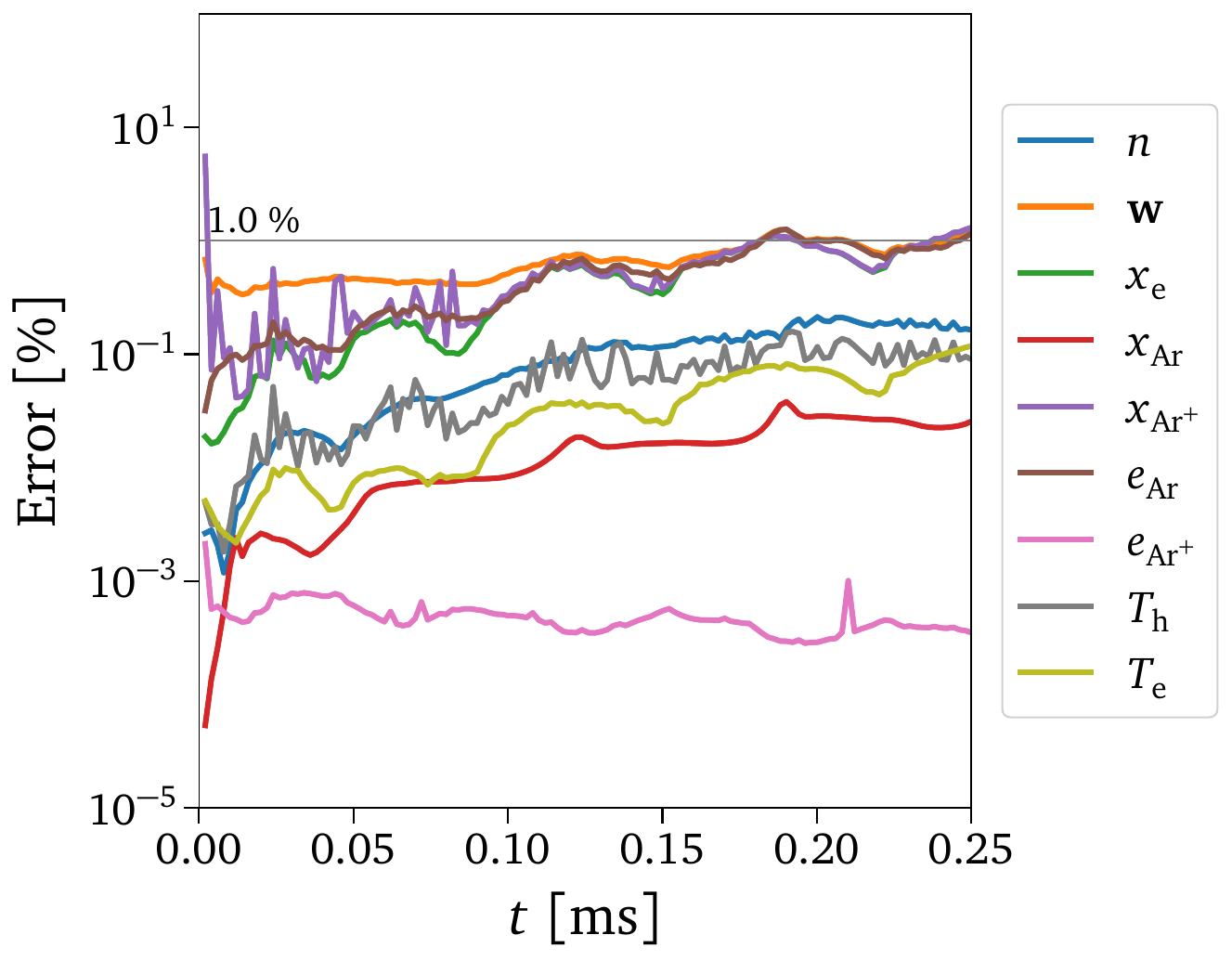}
\end{subfigure}
\quad
\begin{subfigure}[htb!]{0.4\textwidth}
    \caption*{\hspace{-3mm}\fontfamily{mdbch}\normalsize POD}
    \includegraphics[width=\textwidth]{figures/1d/vel1_pod/err_time.pdf}
\end{subfigure}
\caption{\textit{Error evolution of ROMs in 1D simulations.}
Space‐averaged relative errors (\%) for CoBRAS (left panel) and POD (right panel) ROMs with $r = 9$, computed for key quantities of interest. Freestream velocity: $u_\infty = 4\,000$~m/s.}
\label{fig:1d.v3.err_evol}
\end{figure}

\begin{figure}[!htb]
\centering
\includegraphics[width=0.96\textwidth]{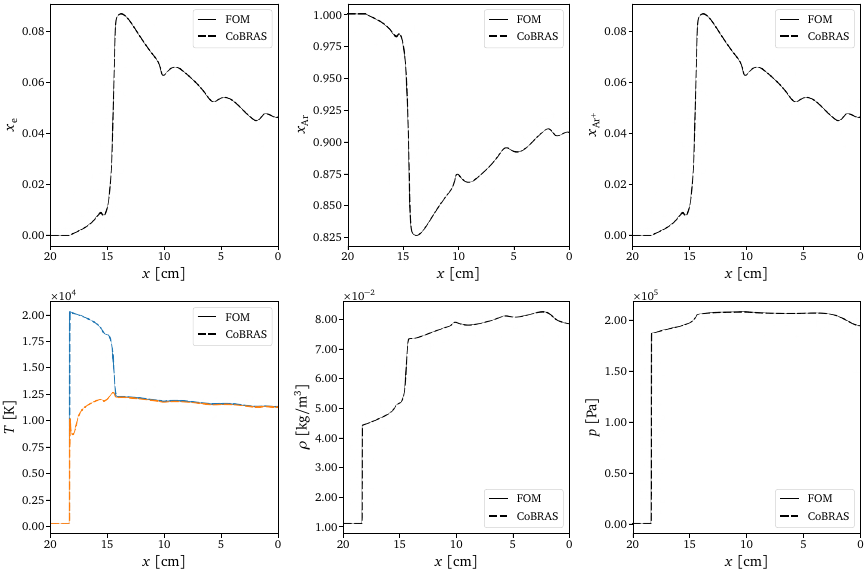}
\caption{\textit{Final-time comparison of FOM and CoBRAS for 1D simulations.} Comparison of FOM and CoBRAS ($r=9$) at the final time $t=2.5\times10^{-4}$ s for the 1D simulation presented in figure~\ref{fig:1d.v1.timespace}. Top row: species molar fractions of e$^-$, Ar, and Ar$^+$. Bottom row: heavy-particle temperature $T_{\mathrm{h}}$ (blue), electron temperature $T_{\mathrm{e}}$ (orange), density $\rho$, and pressure $p$. Freestream velocity: $u_\infty = 4\,000$~m/s.}
\label{fig:1d.v3.snapshots}
\end{figure}

\begin{figure}[!htb]
\centering
\includegraphics[width=0.74\textwidth]{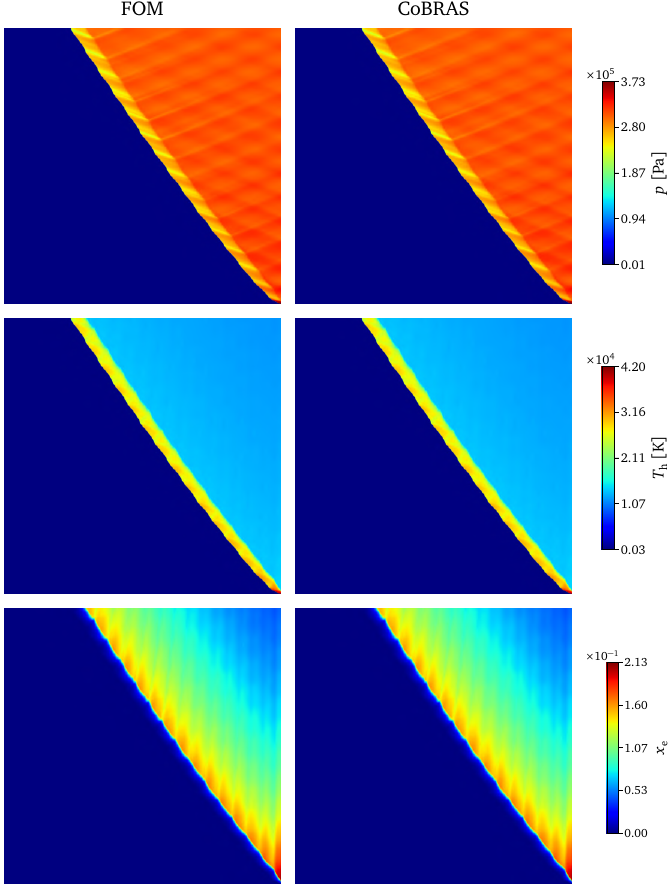}
\caption{\textit{FOM vs. CoBRAS for 1D simulations: Space-time diagrams.} Comparison of FOM and CoBRAS ROM ($r = 9$) for pressure $p$, heavy-particle temperature $T_\mathrm{h}$, and electron molar fraction $x_{\mathrm{e}}$ in space–time ($x$–$y$ plane) diagrams obtained from the 1D simulation. Freestream velocity: $u_\infty = 5\,000$~m/s.}
\label{fig:1d.v2.timespace}
\end{figure}

\begin{figure}[htb!]
\centering
\begin{subfigure}[htb!]{0.4\textwidth}
    \caption*{\hspace{-3mm}\fontfamily{mdbch}\normalsize CoBRAS}
    \includegraphics[width=\textwidth]{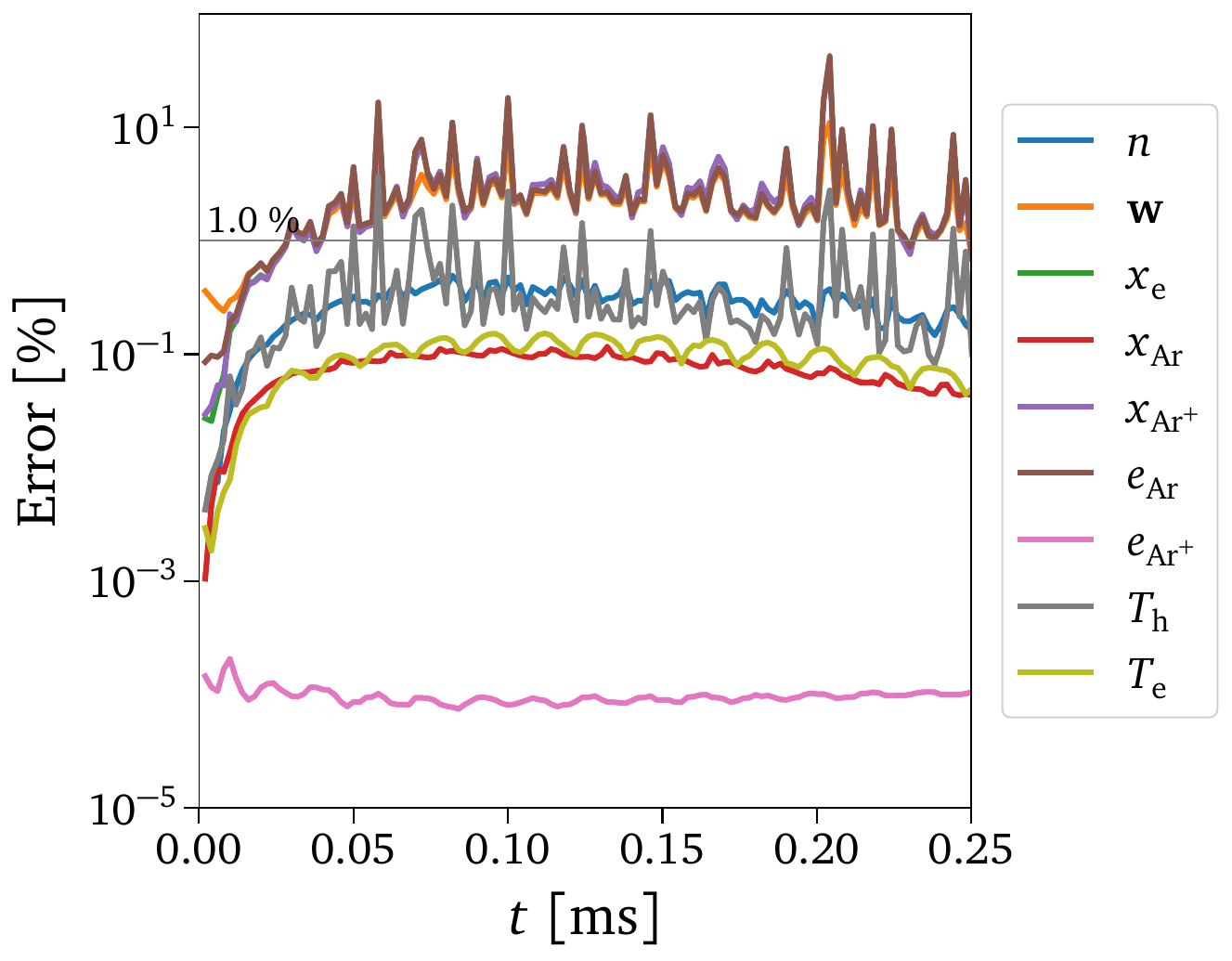}
\end{subfigure}
\quad
\begin{subfigure}[htb!]{0.4\textwidth}
    \caption*{\hspace{-3mm}\fontfamily{mdbch}\normalsize POD}
    \includegraphics[width=\textwidth]{figures/1d/vel1_pod/err_time.pdf}
\end{subfigure}
\caption{\textit{Error evolution of ROMs in 1D simulations.}
Space‐averaged relative errors (\%) for CoBRAS (left panel) and POD (right panel) ROMs with $r = 9$, computed for key quantities of interest. Freestream velocity: $u_\infty = 5\,000$~m/s.}
\label{fig:1d.v2.err_evol}
\end{figure}

\begin{figure}[!htb]
\centering
\includegraphics[width=0.96\textwidth]{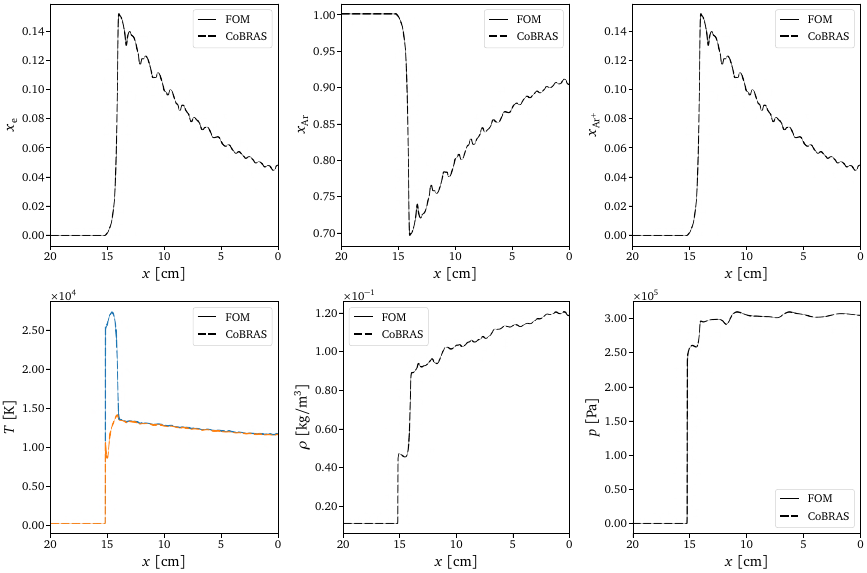}
\caption{\textit{Final-time comparison of FOM and CoBRAS for 1D simulations.} Comparison of FOM and CoBRAS ($r=9$) at the final time $t=2.5\times10^{-4}$ s for the 1D simulation presented in figure~\ref{fig:1d.v1.timespace}. Top row: species molar fractions of e$^-$, Ar, and Ar$^+$. Bottom row: heavy-particle temperature $T_{\mathrm{h}}$ (blue), electron temperature $T_{\mathrm{e}}$ (orange), density $\rho$, and pressure $p$. Freestream velocity: $u_\infty = 5\,000$~m/s.}
\label{fig:1d.v2.snapshots}
\end{figure}

\clearpage
\section{Two‐dimensional simulations}\label{suppl:2d}
\begin{figure}[htb!]
\centering
\includegraphics[width=0.43\textwidth]{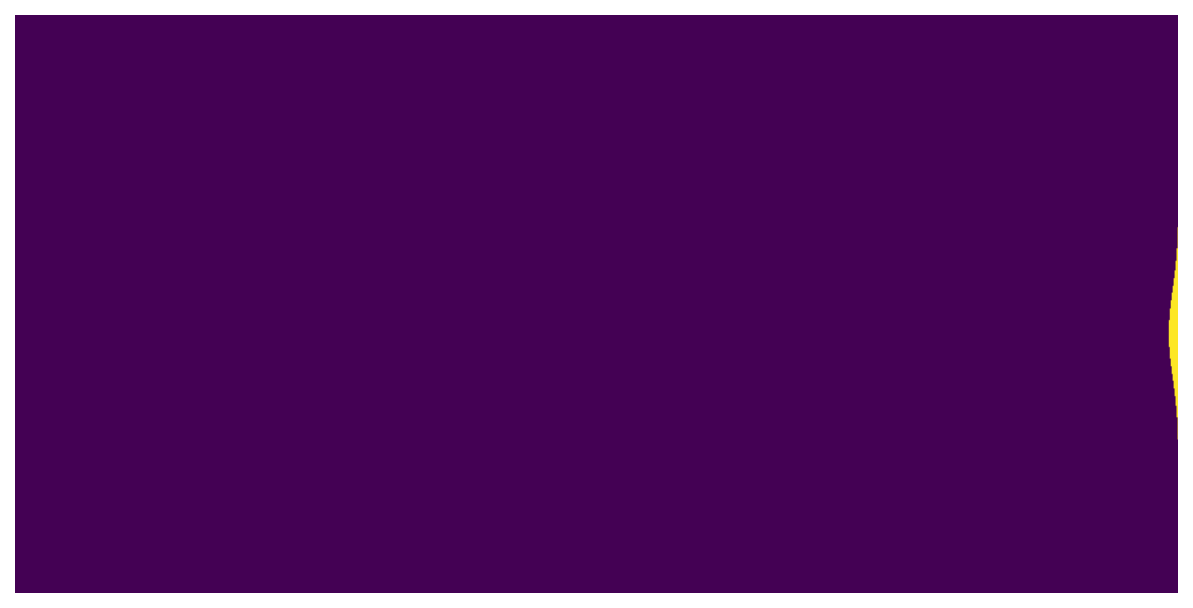}
\caption{\textit{Initial pressure field for 2D simulations.} A localized Gaussian pressure perturbation is introduced near the right wall.}
\label{fig:init_pert}
\end{figure}

\begin{figure}[!htb]
\centering
\includegraphics[width=0.98\textwidth]{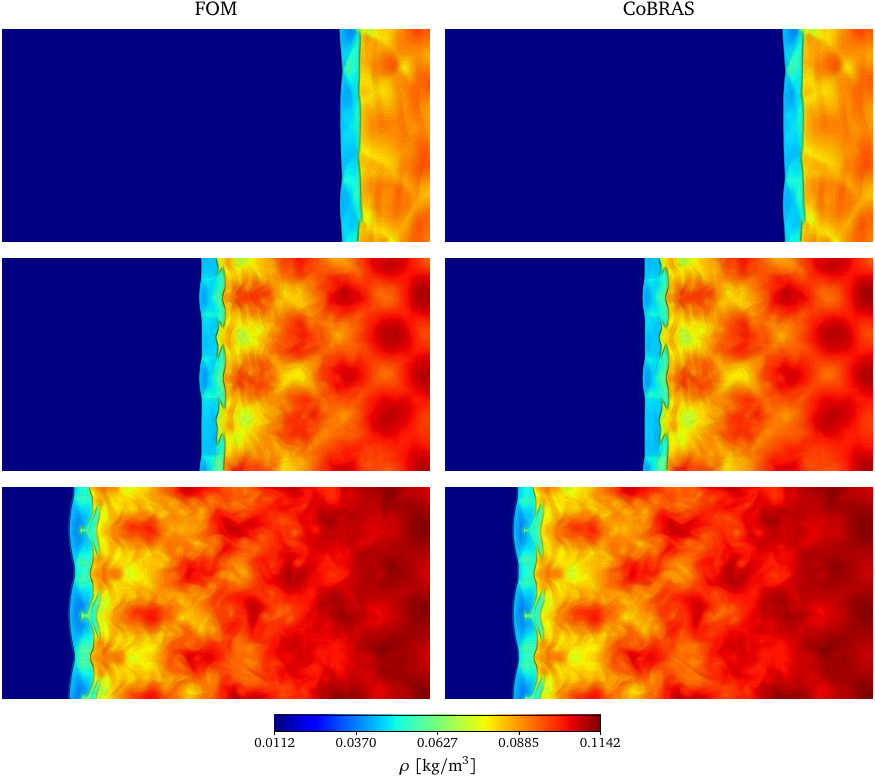}
\vspace{-2mm}
\caption{\textit{FOM vs. CoBRAS for 2D simulations: Density snapshots.} Comparison of FOM and CoBRAS ROM ($r = 9$) at three time instants: $t_i=[1,3,5]\times 10^{-4}$~s for the mass density, $\rho$.}
\label{fig:2d.rho}
\end{figure}

\begin{figure}[!htb]
\centering
\includegraphics[width=0.98\textwidth]{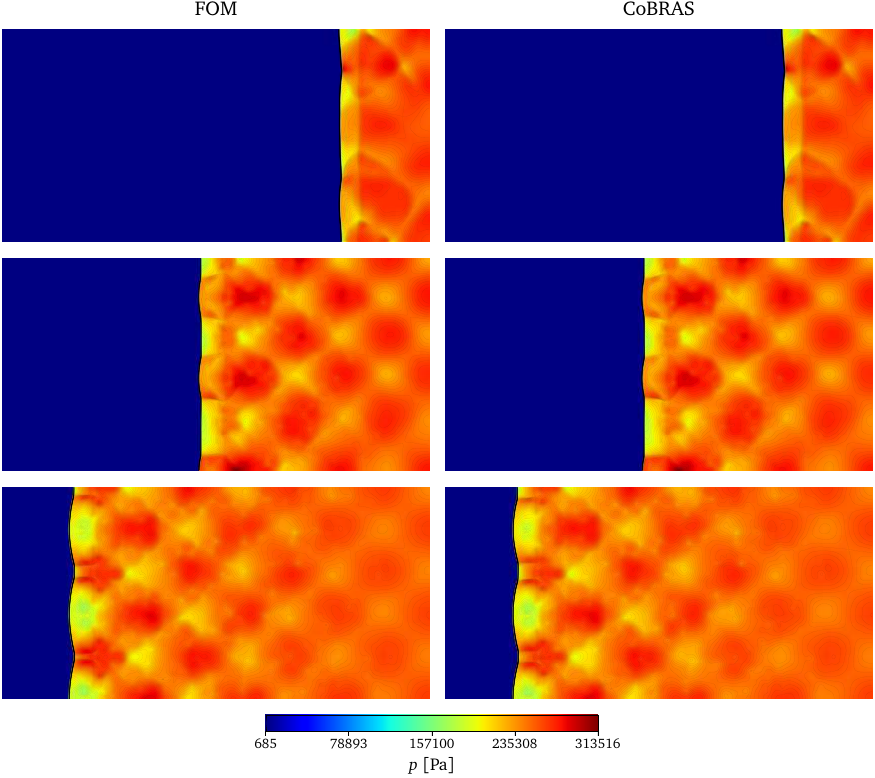}
\vspace{-2mm}
\caption{\textit{FOM vs. CoBRAS for 2D simulations: Pressure snapshots.} Comparison of FOM and CoBRAS ROM ($r = 9$) at three time instants: $t_i=[1,3,5]\times 10^{-4}$~s for the mass density, $p$.}
\label{fig:2d.p}
\end{figure}

\begin{figure}[!htb]
\centering
\includegraphics[width=0.98\textwidth]{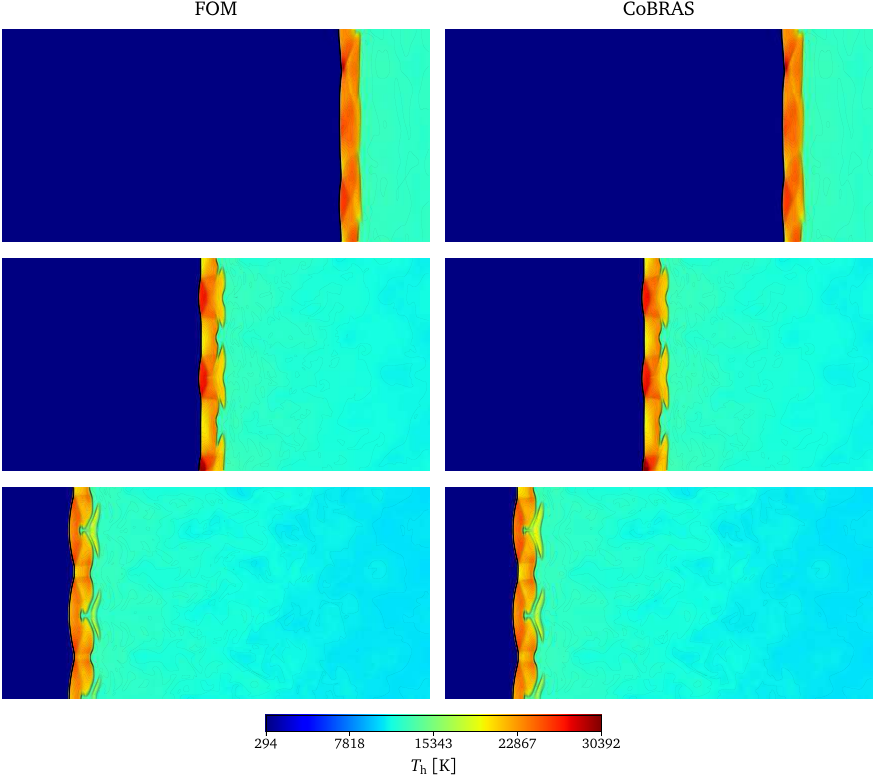}
\vspace{-2mm}
\caption{\textit{FOM vs. CoBRAS for 2D simulations: Heavy-particle temperature snapshots.} Comparison of FOM and CoBRAS ROM ($r = 9$) at three time instants: $t_i=[1,3,5]\times 10^{-4}$~s for the heavy-particle temperature, $T_h$.}
\label{fig:2d.Th}
\end{figure}

\begin{figure}[!htb]
\centering
\includegraphics[width=0.98\textwidth]{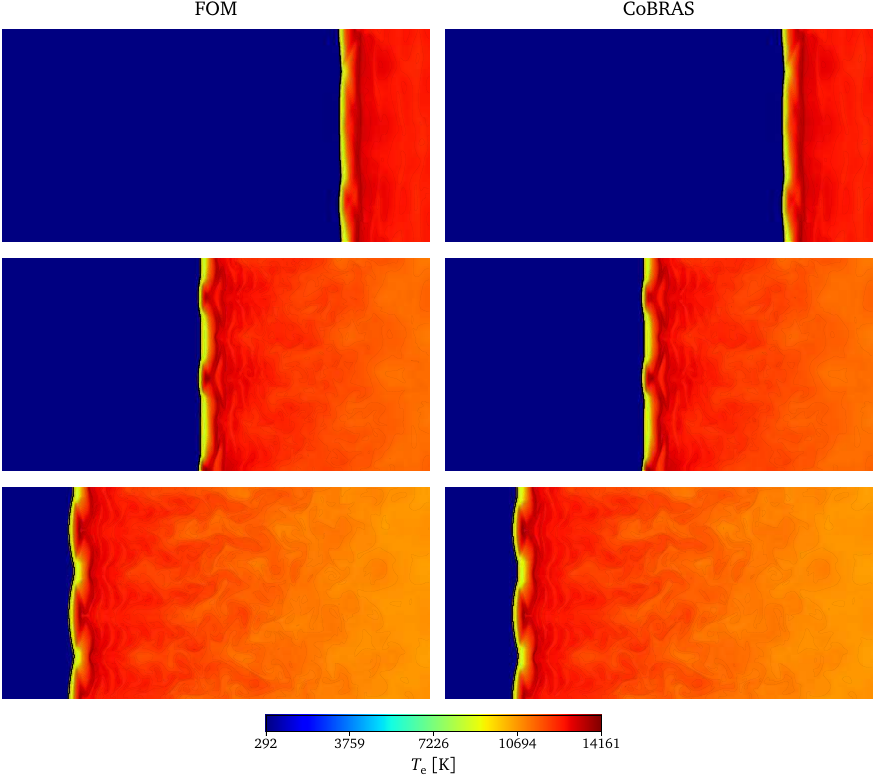}
\vspace{-2mm}
\caption{\textit{FOM vs. CoBRAS for 2D simulations: Electron temperature snapshots.} Comparison of FOM and CoBRAS ROM ($r = 9$) at three time instants: $t_i=[1,3,5]\times 10^{-4}$~s for the electron temperature, $T_e$.}
\label{fig:2d.Te}
\end{figure}

\end{document}